\PassOptionsToPackage{table,dvipsnames}{xcolor}
\PassOptionsToPackage{numbers,sort&compress}{natbib}
\PassOptionsToPackage{bookmarks=false}{hyperref}

\documentclass[manuscript,screen,natbib,nonacm]{acmart}

\usepackage[colorinlistoftodos,prependcaption,textsize=tiny]{todonotes}

\AtBeginDocument{%
  \providecommand\BibTeX{{%
    \normalfont B\kern-0.5em{\scshape i\kern-0.25em b}\kern-0.8em\TeX}}}

\setcopyright{acmcopyright}
\copyrightyear{2024}
\acmYear{2024}
\acmDOI{XXXXXXX.XXXXXXX}

\usepackage[T1]{fontenc}
\usepackage[utf8]{inputenc}
\usepackage[american]{babel}
\usepackage{csquotes}
\usepackage{microtype}
\usepackage{verbatim}

\usepackage{paralist}
\setdefaultenum{(1)}{(a)}{(i)}{A.}
\usepackage{enumitem}
\usepackage{url}
\usepackage{flushend}
\usepackage{etoolbox}
\usepackage{fix-cm}
\usepackage{textpos}
\usepackage{mfirstuc}
\usepackage{titlecaps}
\usepackage[datesep=.,style=ddmmyyyy]{datetime2}
\usepackage{titlesec}
\usepackage[colorinlistoftodos,prependcaption,textsize=tiny]{todonotes}

\usepackage{amsmath}
\usepackage{bm}
\usepackage{bbm}
\usepackage{bbold}
\usepackage{relsize}
\usepackage{siunitx}
\usepackage[super]{nth}
\usepackage{braket}
\usepackage{qcircuit}

\usepackage{booktabs}
\usepackage{multirow}
\usepackage{colortbl}
\usepackage{makecell}
\usepackage{rotating}
\usepackage{threeparttable}
\usepackage{array}
\usepackage{siunitx}
\usepackage{longtable}

\usepackage{float}
\usepackage{graphicx}
\usepackage{xcolor}
\usepackage[table]{xcolor}
\usepackage{subcaption}
\usepackage{tikz}
\usetikzlibrary{positioning, arrows.meta}
\usetikzlibrary{backgrounds}
\usepackage{mdframed}
\usepackage{colortbl} 
\usepackage{tcolorbox}

\usepackage{algorithm}
\usepackage{algpseudocode}
\usepackage{enumitem}
\usepackage{listingsutf8}

\usepackage{xparse}
\usepackage{xspace}

\usepackage[xindy,acronym]{glossaries}
\glsdisablehyper

\usepackage{natbib}

\usepackage{hyperref}
\usepackage[all]{hypcap}

\usepackage[capitalise]{cleveref}
\usepackage{xr}

\definecolor{codegreen}{rgb}{0,0.6,0}
\definecolor{codegray}{rgb}{0.5,0.5,0.5}
\definecolor{codepurple}{rgb}{0.58,0,0.82}
\definecolor{backcolour}{rgb}{0.95,0.95,0.92}
\definecolor{lightgray}{gray}{0.95}

\lstdefinestyle{mystyle}{
    backgroundcolor=\color{backcolour},
    commentstyle=\color{codegreen},
    keywordstyle=\color{magenta},
    numberstyle=\tiny\color{codegray},
    stringstyle=\color{codepurple},
    basicstyle=\ttfamily\footnotesize,
    breakatwhitespace=false,
    breaklines=true,
    captionpos=b,
    keepspaces=true,
    numbers=left,
    numbersep=5pt,       
    xleftmargin=7pt,     
    showspaces=false,
    showstringspaces=false,
    showtabs=false,
    tabsize=2
}

\titleformat{\subsubsection}[runin]{\normalfont\bfseries}{\thesubsubsection}{1em}{}[\\]

\newcommand*\circled[1]{\tikz[baseline=(char.base)]{
            \node[shape=circle,draw,inner sep=2pt] (char) {#1};}}

\newacronym{dps}{DPS}{default program specification}
\newacronym{ex3}{eX\textsuperscript{3}}{Experimental Infrastructure for Exploration of Exascale Computing}
\newacronym{pdo}{PDO}{Probability Distribution Oracle}
\newacronym{ps}{PS}{program specification}
\newacronym{qc}{QC}{quantum computing}
\newacronym{rq}{RQ}{research question}
\newacronym{sdk}{SDK}{software development kit}
\newacronym{sut}{SUT}{system under test}
\newacronym{tc}{TC}{test case}
\newacronym{woo}{WOO}{Wrong Output Oracle}
\newacronym{qst}{QST}{quantum software testing}
\newacronym{iqr}{IQR}{interquartile range}
\newacronym{sd}{SD}{standard deviation}
\newacronym{var}{Var}{Various}
\newacronym{grov}{Grov}{Grover Search}
\newacronym{gs}{Gs}{Graph States}
\newacronym{qwalk}{Qwalk}{Quantum Walk}
\newacronym{mwu}{MWU}{Mann-Whitney U}
\newacronym{avg}{AVG}{average}

\definecolor{lightblue}{HTML}{c6d6e9}
\definecolor{darkblue}{HTML}{9fb1bd}

\tcbset{colback=lightblue, colframe=darkblue, arc=5pt, boxrule=0.75pt}
\definecolor{customlightblue}{RGB}{209, 229, 240}

\newcommand{\summarybox}[2]{
	\begin{tcolorbox}[colback=customlightblue]
		\small{
			\textbf{#1:} #2
		}
	\end{tcolorbox}
}

\begin{document}

\title{Faster and Better Quantum Software Testing through Specification Reduction and Projective Measurements}


\author{Noah H. Oldfield}
\orcid{0000-0002-9059-0694}
\affiliation{%
  \institution{Simula Research Laboratory and University of Oslo}
  \city{Oslo}
  \country{Norway}}
\email{noah@simula.no}

\author{Christoph Laaber}
\orcid{0000-0001-6817-331X}
\affiliation{%
  \institution{Simula Research Laboratory}
  \city{Oslo}
  \country{Norway}}
\email{laaber@simula.no}

\author{Tao Yue}
\orcid{0000-0003-3262-5577}
\affiliation{%
  \institution{Simula Research Laboratory}
  \city{Oslo}
  \country{Norway}}
\email{taoyue@gmail.com}

\author{Shaukat Ali}
\orcid{0000-0002-9979-3519}
\affiliation{%
  \institution{Simula Research Laboratory}
  \city{Oslo}
  \country{Norway}}
\affiliation{%
  \institution{Oslo Metropolitan University}
  \city{Oslo}
  \country{Norway}}
\email{shaukat@simula.no}


\begin{abstract}
\Gls{qc} promises polynomial and exponential speedups in many domains, such as unstructured search and prime number factoring. 
However, quantum programs yield probabilistic outputs from exponentially growing distributions and are vulnerable to quantum-specific faults.
Existing \gls{qst} approaches treat quantum superpositions as classical distributions. 
This leads to two major limitations when applied to quantum programs: (1) an exponentially growing sample space distribution and 
(2) failing to detect quantum-specific faults such as phase flips.
To overcome these limitations, we introduce a \gls{qst} approach, which applies a reduction algorithm to a quantum program specification. 
The reduced specification alleviates the limitations (1) by enabling faster sampling through quantum parallelism and (2) by performing projective measurements in the mixed Hadamard basis.
Our evaluation of 143 quantum programs across four categories demonstrates significant improvements in test runtimes and fault detection with our reduction approach. 
Average test runtimes improved from 169.9s to 11.8s, with notable enhancements in programs with large circuit depths (383.1s to 33.4s) and large program specifications (464.8s to 7.7s). 
Furthermore, our approach increases mutation scores from $54.5\%$ to $74.7\%$, effectively detecting phase flip faults that non-reduced specifications miss. 
These results underline our approach's importance to improve \gls{qst} efficiency and effectiveness.
\end{abstract}



\maketitle

\glsresetall{}

\newcommand{\HLIGHT}[1]{\enquote{#1}}

\section{Introduction}
Quantum computers utilize the principles of quantum mechanics to perform computations at speeds unachievable by classical computers, opening unprecedented possibilities in fields such as quantum chemistry, optimization, machine learning, and cryptography~\citep{feynman2018simulating, QCQIBook, Egger_2020, moll2018quantum, GoogleAILearning, Shor_1997}. 
While the computational power of classical computers scales linearly for each added bit, quantum computational power scales exponentially for each added qubit. 
This exponential power comes from exploiting quantum superposition and entanglement in quantum algorithms, which one implements as quantum programs with a quantum software development kit such as IBM's Qiskit, Rigetti Forest, or Google's Cirq~\citep{QiskitCite,joshua_combes_2019_3455848,cirq_developers_2023_10247207}. 
Quantum programs, similar to classical programs, are also prone to bugs. 
Therefore, their correctness needs to be ensured. 
However, as quantum computing extends classical computing, quantum programs are found to contain not only classic software faults but also quantum-specific faults~\citep{BugsinQCEmpirical,campos2021qbugs,aoun2022bug}.
Therefore, \gls{qst} plays a crucial role in the quantum software development cycle to find these faults, motivating the need for the development of software testing practices that meet the required standards of industrial and academic applications~\citep{zhao2020quantum,ShaukatTaoQST,MiranskyyICSE2020}.

In the evolving landscape of \gls{qst}, research has adapted and applied classical software testing approaches such as input-output coverage, differential, property-based, and metamorphic testing to quantum programs, attempting to incorporate quantum features~\cite{quitoASE21tool,Wang2021QDiffDT,honarvar2020property,QMorph,Muskit}.
These approaches validate quantum programs by statistically sampling the quantum program distribution in the computational basis and comparing a sample distribution with a theoretical distribution using a statistical test, such as the chi-square test. 
This theoretical distribution refers to different concepts depending on the approach. 
For instance, it is often formulated as \gls{ps} in input-output coverage~\citep{quitoASE21tool} and mutation testing~\citep{Muskit}, is considered a property in property-based testing~\citep{honarvar2020property}, or represents the expected statistical relationship between the distributions of a source and follow-up program in metamorphic testing~\citep{QMorph}.
 
We identify two challenges of the current \gls{qst} approaches.
First (1), current \gls{qst} approaches require a considerable number of samples for a scalable validation of a quantum program distribution. 
This scalability issue, however, is not yet noticeable, as current \gls{qst} approaches have only been evaluated on tiny quantum programs having up to 12 qubits and a circuit depth of 30~\citep{QMorph,quitoASE21tool}. 
Scaling to 60 qubits introduces exponentially large search spaces, such as $2^{30}$ elements in Grover Searches, necessitating a high number of measurements~\citep{Preskill_2018,GroversAlgo}.
Similar measurement optimization challenges arise in other applications, such as quantum chemistry, where eigensolvers primarily optimize measurements of a single output value, specifically the expectation value of the 
Hamiltonian~\citep{Garc_a_P_rez_2021}. 
In contrast, quantum software testing frequently requires multiple sets of measurements to validate various outputs across numerous test cases.
Thereby, due to the exponential scaling of quantum program distributions, efficient sampling is paramount for efficient \gls{qst}.
Second (2), quantum programs exhibit a multitude of bugs of which gate faults are a common source~\citep{BugsinQCEmpirical,aoun2022bug,campos2021qbugs}.
Current testing techniques often do not consider such quantum-specific bugs, as these techniques solely sample quantum probability distributions in the computational basis~\citep{QCQIBook}. 
This happens due to phase flip faults manifesting in the program's final state vector as flipped signs in front of one or multiple basis states; thus, counting the basis states from measurements never reveals information about this flipped sign, as it is merely a mathematical artifact in a given basis. 
Only upon performing a projective measurement~\cite{QCQIBook} in the Hadamard basis can we detect phase flip faults because then the fault can be detected as a flipped \emph{bit}.

To tackle challenges (1) and (2), we propose a Greedy reduction-based approach to reduce the \textit{ranks} of quantum program specifications by utilizing projective measurements in mixed Hadamard bases.
According to the state vector postulate of quantum mechanics~\citep{QCQIBook,Sakurai}, all possible information about a quantum system is contained in its state vector. 
Based on this, we define the \gls{ps} as the non-faulty final state vector of the quantum program. 
Further, inspired by the Schmidt rank~\citep{QCQIBook,Sakurai}, we define the rank of the \gls{ps} as the number of basis states, denoting its size.
This reduced program specification enables more efficient sampling of the quantum program distribution due to the smaller program specification rank, tackling challenge (1) and the detection of phase flip faults by projective measurements, alleviating challenge (2).
We evaluate our approach on a program suite consisting of 143 quantum programs in 4 categories.
Given that not all quantum programs yield final state vectors expressible in the eigenstates of mixed Hadamard bases (which we associate with uniform superpositions with real amplitudes), we select four program categories that demonstrate this to varying degrees: Grover search and graph state programs, known for their high uniformity; discrete quantum walks, characterized by their non-uniform final state vectors; and a diverse array of other quantum algorithms to investigate the impact of varying uniformity on our measurements ~\citep{GroversAlgo,Hein_2004,Venegas_Andraca_2012_Quantum_Walk,johnston2019programming}.

In our evaluation, we assess (1) the effectiveness of our reduction approach; (2) the impact of reduced \gls{ps}s on \gls{qst} runtime efficiency; and (3) the impact of reduced \gls{ps}s on effectiveness, as measured through mutation testing.
We find that reduction is generally efficient and effective, achieving reduction rates of $52.7\%$ overall, $27.4\%$ in the low end for quantum walk programs, and $83.7\%$ in the high end for Grover searches.
Our approach significantly improves the test runtimes compared to sampling in the computational basis, yielding a 14-fold improvement on average, from \SI{169.9}{\second} with computational basis sampling to \SI{11.8}{\second} with our approach.
Particularly notable are 11-fold improvements for Grover search programs with large circuit depths and 60-fold improvements for graph state programs with large specification ranks.
In contrast, the categories of quantum walks and various quantum algorithms showed mixed results, with no improvements observed in $12.8\%$ and $9.1\%$ of cases, respectively. 
Nevertheless, quantum walks achieved a threefold improvement, reducing from \SI{2.9}{\second} to \SI{1.0}{\second}, and the category of various algorithms obtained a sixfold improvement, decreasing from \SI{1.1}{\second} to \SI{0.2}{\second}.
Overall, for all program categories, we find that fault detection is enhanced, with mutation scores of $74.7\%$ for various mutations compared to $54.5\%$ using non-reduced specifications.
Notably, non-reduced specifications hardly detect phase flip faults, with a mutation score of only $2.1\%$, while our approach achieves $36.0\%$ overall.

To summarize, the main contributions of our paper are:
\begin{enumerate}
    \item A Greedy rank reduction algorithm to obtain a reduced program specification for projective measurements in a mixed Hadamard basis;
    \item An experimental evaluation of our approach using a suite of 143 quantum programs, including important application programs such as Grover searches, with qubit counts up to $16$, depths up to 684, and program specification ranks up to $2^{14}$ computational basis states;
    \item Empirical evidence of faster sampling of the quantum program distribution through our reduced specifications; and
    \item Empirical evidence of detection of phase flip faults through projective measurements with the mixed Hadamard basis detecting previously undetectable phase flip faults.
\end{enumerate}

We provide a replication package containing the approach's source code, study subjects, experiment scripts, paper results, and an appendix with additional results~\citep{oldfield_2024_11191215}.

\section{Background and Definitions}
\label{sec:background}

This section introduces \gls{qc} and provides important definitions in three parts: \textbf{Qubits and State Vectors}, \textbf{Quantum Gates and Gate Faults}, and \textbf{Quantum Programs and Projective Measurements}.
In the final part, we also provide an illustration in \cref{fig:diagram_background} of how these concepts connect to form a quantum program.

\subsection{\textbf{Qubits and State Vectors}}
\label{sec:state_vectors}

In classical computing, bits represent information in binary as either $0$ or $1$. 
Conversely, quantum computing applies \textit{quantum bits} or \textit{qubits}, which can be in states of $0$ and $1$ simultaneously through \textit{quantum superposition}~\citep{QCQIBook}. Mathematically, we represent a single qubit by the two-component \textit{state vector} $\ket{\psi} \in \mathcal{H}$, where $\mathcal{H}$ denotes the \textit{Hilbert space} of the qubit:

\begin{equation}
\ket{\psi} = \alpha_0 \ket{0} + \alpha_1\ket{1}
\label{eq:single_qubit_state}
\end{equation}

In \cref{eq:single_qubit_state}, the states $\ket{0}$ and $\ket{1}$ are the \textit{computational basis states} for $\ket{\psi}$, corresponding to the possible outcomes after measuring the qubit in the computational basis:

\begin{equation}
\big\{\ket{0},\ket{1}\big\}= \big\{ \begin{pmatrix}
1 \\ 0
\end{pmatrix}, \begin{pmatrix}
0 \\ 1
\end{pmatrix} \big\}
\label{eq:computational_basis}
\end{equation}

Coupled with the basis states are the complex numbers $\alpha_0$ and $\alpha_1$, known as the \textit{probability amplitudes} of the basis states $\ket{0}$ and $\ket{1}$. 
We can write the probability amplitudes as the product $\alpha_m = \sqrt{p_m} e^{i\theta_m}$ consisting of two parameters. 
First, the probability $p_m$ of observing the particular basis state $\ket{m}$ is obtained by computing the squared absolute value of the probability amplitude, i.e., $p_m=|\alpha_m|^2$. 
The set of these probabilities defines the quantum state's probability distribution over the basis states. 
Second, the angle parameter $\theta_m \in [0, 2\pi]$, commonly referred to as the \textit{relative phase}, defines the complex number $e^{i\theta_m}$ for the basis state $\ket{m}$.

While \cref{eq:single_qubit_state} describes single-qubit states, for $n$-qubit states: $q_0,q_1,\cdots, q_{n-1}$, the state vector can be written as a sum of all possible configurations of the $n$-qubit basis states:

\begin{equation}
\label{eq:multi_qubit_state}
    \ket{\psi} = \sum_{j_0=0}^1 \sum_{j_1=0}^1 \cdots \sum_{j_{n-1}}^1 \alpha_{j_0 j_1\cdots j_{n-1}}\ket{j_0 j_1 \cdots j_{n-1}}
\end{equation}

To construct the $n$-qubit basis states in \cref{eq:multi_qubit_state}, we apply the \textit{tensor product} operation $\otimes$~\citep{QCQIBook}, such that each basis state is written as $\ket{j_0 j_1\cdots j_{n-1}} = \ket{j_0}\otimes \ket{j_1}\otimes \cdots \otimes \ket{j_{n-1}}$.

\noindent We can write \cref{eq:multi_qubit_state} more compactly as:

\begin{equation}
   \ket{\psi} = \sum_{j=0}^{N-1} \alpha_j \ket{j}
   \label{eq:multi_qubit_state_compact}
\end{equation}

Here, the states $\ket{j}$ represent the $N=2^n$ basis states from \cref{eq:multi_qubit_state} in compact decimal representation~\citep{QCQIBook}. 
The relation $\ket{j} = \ket{j_0 j_1\cdots j_{n-1}}$ maps between the decimal and binary representations of the state. 
For example, if $j=2$ and $n=3$ then $\ket{2} = \ket{010}$. 
Additionally, we define the probability distribution $\mathcal{P} = \{ p_0, p_1, \dots, p_{N-1} \}$ for a state vector, which must satisfy the normalization condition:
\begin{equation}
 \sum_{j=0}^{N-1} p_j = 1
 \label{eq:normalization_condition}
\end{equation}
In \cref{eq:normalization_condition}, the probabilities in the quantum state distribution sum up to 1. 
A state vector $\ket{\psi_e}$ of a multi-qubit system represents an \textit{entangled state} if and only if it \emph{cannot} be expressed as a tensor product of state vectors from each qubit~\citep{QCQIBook,gurvits2003classical}.
A canonical example of an entangled state is the \textit{Bell state} $ \ket{\beta}= \frac{1}{\sqrt{2}}(\ket{00} + \ket{11})$. 
By contrast, the state $\frac{1}{\sqrt{2}}(\ket{00}+\ket{01})$ \emph{can} be written as a tensor product of the states $\ket{0}$ and $\frac{1}{\sqrt{2}}(\ket{0}+\ket{1})$, and is therefore not an entangled state.
Unlike classical states, entangled states exhibit correlations between qubits that are unique to quantum systems. 
Such correlations have been experimentally confirmed even for quantum states separated by large distances~\citep{van_Leent_2022}.

\paragraph{\textbf{Output Criterion}}

If all basis states of a state vector are equally likely, we call \cref{eq:multi_qubit_state_compact} a \textit{uniform} state vector. 
In addition, if all probability amplitudes are real numbers, we write this class of state vectors as:

\begin{equation}
    \ket{ \psi } = \frac{1}{\sqrt{2^n}}\sum_{j=0}^{N-1} (-1)^{f(j)} \ket{j}
    \label{eq:real_uniform_state_vector}
\end{equation}

In \cref{eq:real_uniform_state_vector}, the function $f$ determines whether the angle parameter is either $\pi$ or $0$, resulting in a sign in front of the state of either $-1$ or $+1$ respectively.
Because these states are simple to study and important for many applications, such as for Grover search and graph states~\citep{GroversAlgo,Hein_2004}, we aim our reduction approach to target these kinds of states. 
Thus, we call \cref{eq:real_uniform_state_vector} the \textit{output criterion} for later reference.

\subsection{\textbf{Quantum Gates and Gate Faults}}
\label{sec:gates}

\begin{table}[tbp]
    \centering
\caption{
Overview of quantum gate operations and their corresponding fault examples based on the running example in \cref{fig:diagram_background}. 
The \textbf{Expected State} column shows the expected quantum state without any faults, while the \textbf{Measured State (Faulty)} column shows the observed state with the introduced fault. 
While not relevant for the other fault examples, in the $R_y(\theta)$ row, the notation (X\%) depicts the probability of observing the expected or the measured state, including the fault's impact. The faults demonstrated here result in bugs in the program’s behavior, which are discussed further in the \textbf{Gate Faults} paragraph of \cref{sec:gates}.}
    \label{tab:background_gates}
    \begin{tabular}{llll}
        \toprule
          &   & \multicolumn{2}{c}{\textbf{Fault Example}} \\
        \cmidrule{3-4}
             \textbf{Gate}          &        \textbf{Operation}            & \textbf{Expected State} & \textbf{Measured State (Faulty)} \\
        \midrule
        X              & Bit Flip           & $\ket{01}$             & $\phantom{-i}\ket{00}$ (Bit flip on $q_2$) \\
        Z              & Phase Flip         & $\ket{01}$             & $\;-\ket{01}$ (Phase flip on $q_2$) \\
        Y              & Bit Flip \& Phase Rotation     & $\ket{01}$             & $-i\ket{00}$ (Bit flip and phase rotation on $q_2$) \\
        H              & Superposition      & $\ket{0+}$             & $\phantom{-i}\ket{00}$ (Undo superposition) \\
        CNOT           & Controlled X       & $\ket{01}$             & $\phantom{-i}\ket{11}$ (Controlled X on $q_1$) \\
        $R_y(\theta)$  & Y-Rotation         & $\ket{01} \, (100\%)$  & $\phantom{-i}\ket{01} \, (50\%)$ (Rotation by $\theta = \pi/2$ on $q_2$) \\
        \bottomrule
    \end{tabular}
\end{table}

As with classical computing, quantum gates transform one computational state to another. 
In \gls{qc}, gates act as unitary matrices on the state vectors to perform computations.
We depict common quantum gates and their corresponding faults in \cref{tab:background_gates}, which we describe in the following.

The $Hadamard$ gate denoted by $H$, performs the following transformations on the computational basis states $\ket{0}$ and $\ket{1}$:

\begin{align}
\label{eq:Hadamard_operations}
 H\ket{0}&=\frac{1}{\sqrt{2}}\Big(\ket{0} + \ket{1}\Big)=\ket{+} & H\ket{1} = \frac{1}{\sqrt{2}}\Big(\ket{0}-\ket{1}\Big)=\ket{-}
\end{align}
In \cref{eq:Hadamard_operations}, the Hadamard gate initiates a uniform superposition with $50\%$ chance of measuring either $\ket{0}$ or $\ket{1}$. 
The quantum equivalent of the classical $NOT$ gate is named the $X$-gate or the bit-flip gate and is denoted by $X$. 
It acts as expected when applied to classical states, i.e., $X\ket{0}=\ket{1}$. 
However, the difference with acting on a quantum state can be shown by applying the $X$-gate to the single qubit state, as depicted in \cref{eq:single_qubit_state}, which performs a bit flip on both terms in the sum. 
The resulting state would thus be $\alpha_1\ket{0} + \alpha_0\ket{1}$, where the probability amplitudes $\alpha_0,\alpha_1$ have exchanged places. 
This is an example of how \gls{qc} differs from classical computing. 
A classical logic gate acts on a single state at a given time, whereas a quantum gate is applied to the state vector of superpositions such that both $\ket{0}$ and $\ket{1}$ are flipped simultaneously. 
The cost, however, is that we can only obtain either $\ket{0}$ or $\ket{1}$ with probabilities $p_0$ or $p_1$ at any given time from observing the state vector~\citep{QCQIBook} by performing a measurement of the state vector to read the output.
We also introduce the \textit{phase flip} gate $Z$, which acts on the computational basis states as $Z\ket{0}=\ket{0}$ and $Z\ket{1}=-\ket{1}$.
The final gate we introduce is the rotation gate $R_y(\theta)=\exp(-i\frac{\theta}{2}Y)$, which applies a rotation of angle $\theta$ to the qubit along the y-axis on the Bloch sphere, where $Y$ is the Pauli Y-gate and $i=\sqrt{-1}$~\citep{QCQIBook}. 
Together, the H, X, and $R_y$ gate parameterize a general single qubit gate~\citep{QCQIBook,Sakurai}.

\paragraph{\textbf{Gate Faults}}
While the gates in \cref{tab:background_gates} exemplify common quantum gate operations applied in most quantum algorithms~\cite{GroversAlgo,Shor_1997,QCQIBook}, an incorrect application of any of these gates results in a gate fault, causing buggy behavior of the quantum program.
In the \textbf{Fault Example} column, we provide examples of faults caused by incorrectly applied gates.
For instance, an incorrect use of an X gate can lead to a bug in a Grover search, where the search locates the wrong element due to a bit flip fault. 
In this case, the basis state representing the correct search element is modified by the bit flip, as demonstrated by the modified state $\ket{00}$ in the \textbf{Measured State (Faulty)} column, when the expected state should be $\ket{01}$ as shown in the \textbf{Expected State} column.
While this type of fault leads to finding the wrong element, other fault types can affect the search quality. 
For example, a faulty $R_y$ gate can alter the probability distribution, reducing the likelihood of finding the correct element. 
This is illustrated by the reduction from 100\% to 50\% probability in the \textbf{Measured State (Faulty)} column for the $R_y(\theta)$ gate.
If a fault from the Z gate occurs, the sign in front of the $\ket{1}$ state flips in a way that it causes a bug in the quantum program.
As the Grover search algorithm works by amplifying states with a negative sign, any modification to this sign, such as in the Z gate fault example, can cause the incorrect element to be amplified, leading to the wrong element being located.
Similarly, faults introduced by the H or CNOT gates can also result in bugs that affect the program's behavior.

\subsection{Quantum Programs and Projective Measurements}
\label{sec:quantum_program}


\begin{figure}
    \centering
    \includegraphics[width=0.85\linewidth]{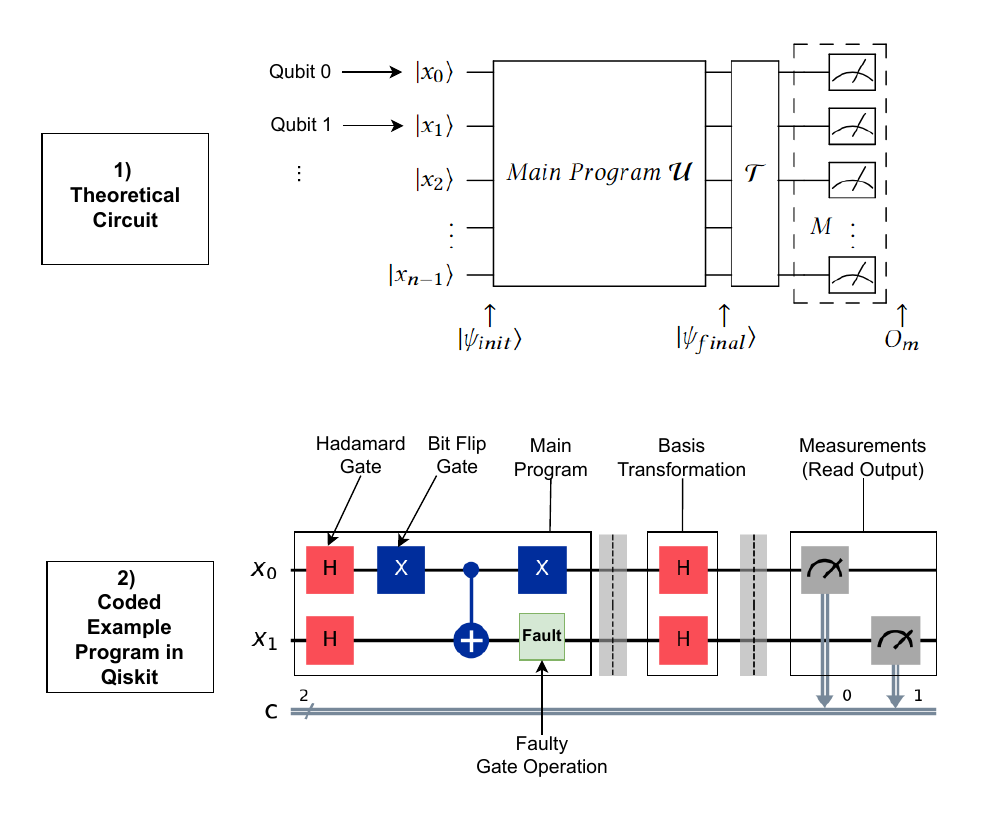}
    \caption{Illustration of the background concepts. In the theoretical circuit 1), we demonstrate the following concepts: (a) state vectors $\ket{\psi_{init}}$ and $\ket{\psi_{final}}$, (b) qubits denoted by $\ket{x_j}$ for $j=0,1,\dots, n-1$, (c) the main program $\mathcal{U}$, (d) the basis transformation $\mathcal{T}$, measurement $M$, and the output $O_m=\ket{m}$ resulting from measurement $M$. In the coded Qiskit example 2), we illustrate additional concepts: (a) quantum gates such as H and X, (b) faults representing incorrectly inserted quantum gates, and (c) the basis transformation with $H$ gates, providing an example of a mixed Hadamard basis in a coded circuit. }
    \label{fig:diagram_background}
\end{figure}

\Cref{fig:diagram_background} illustrates a \textit{quantum} program in two ways: 1) is a theoretical circuit independent of the programming language and particular gates while 2) is a coded quantum program in Qiskit with specific gates, gate faults, basis transformations, and measurements, which we will explain in this section.

A \textit{quantum program}, as illustrated by 1) in \cref{fig:diagram_background}, consists of an initial state vector $\ket{\psi_{init}}$, followed by the main program $\mathcal{U}$ of quantum gates, before performing a measurement of the final state vector $\ket{\psi_{final}}$. 
We have the following relationship  $\ket{\psi_{final}} = \mathcal{U} \ket{\psi_{init}}$, from the evolution postulate of quantum mechanics~\citep{Sakurai,QCQIBook}. 
In order to transition from the final state vector to a definite state of 0s and 1s, one must perform a measurement $M$ as the final operation of the quantum program. 
A measurement is a non-reversible operation and can be represented as a mapping from a state vector onto a specific basis state $\ket{m}$ from the state space, where $\ket{m}$ occurs with probability $p_m$. 
We define the basis state $\ket{m}$ as an \textit{output} $O_m$ of the quantum program resulting from the measurement $M$.

\paragraph{\textbf{Fault Example: Outputs of a 2-Qubit Final State Vector.}}
\label{ex:computational_basis_measurement}

This paragraph demonstrates the coded example program in Qiskit shown by 2) in \cref{fig:diagram_background}.
First, we assume no fault occurs, which means we ignore the inserted faulty gate.
By running the program with an arbitrarily chosen input $\ket{x_0 x_1}=\ket{10}$, we obtain the final state vector: $\ket{\phi} = \frac{1}{\sqrt{4}} ( -\ket{00} + \ket{01}  + \ket{10} + \ket{11} ) $. 
We can identify four possible outputs resulting from a measurement: $\ket{00},\ket{01},\ket{10}$, and $\ket{11}$, where each output occurs with a probability equal to $1/4$.
If we now assume the fault gate has been applied as a common bit flip operation from \cref{tab:background_gates} (which we denote as a \textit{bit flip fault}), we obtain the faulty state vector: $\ket{\phi}_{faulty}=\frac{1}{\sqrt{4}}(\ket{00} - \ket{01} + \ket{10} + \ket{11})$.
We observe that the negative sign (which is supposed to be in front of the $\ket{00}$ state) moves to the $\ket{01}$ state due to the bit flip fault.
To assess the consequences of this fault, we can portray our example in the context of a real-world example. 
Our example program is identical to a sub-operation of a Grover search, where the negative sign acts as a mark on the correct search element, which the search algorithm attempts to amplify the probability of such that it can be measured and located with a high probability~\cite{GroversAlgo}.
Thus, such a bit flip fault results in the algorithm amplifying the incorrect state, namely $\ket{01}$ instead of $\ket{00}$.

In the following, we will further use $\ket{\phi}$ as our running example to demonstrate how basis changes can affect the number of outputs after measurement.

\paragraph{\textbf{Projective Measurements}}

Up to this point, we've only been concerned with the basis states of the computational basis, as shown in \cref{eq:computational_basis}. However, for this paper's approach, we also utilize projective measurements, which represent transformed measurements where we project the quantum state onto a specified subspace depending on the transformation, allowing for a reduced representation of n-qubit state vectors. 
We apply projective measurements that utilize \textit{Hadamard basis} transformations (also known as the \textit{X-basis}, as they are the eigenstates of the Pauli X gate) ~\citep{QCQIBook,Sakurai}:

\begin{equation}
\textbf{H}=\big\{\ket{+},\ket{-}\big\}= \big\{ \begin{pmatrix}
    \frac{1}{\sqrt{2}} \\ \frac{1}{\sqrt{2}}
\end{pmatrix}, \begin{pmatrix}
    \frac{1}{\sqrt{2}} \\ \frac{-1}{\sqrt{2}} 
\end{pmatrix}   \big\}
\label{eq:hadamard_basis}
\end{equation}

\Cref{eq:hadamard_basis} defines the Hadamard basis $\textbf{H}$ for single-qubit states. 
Now, to transform into a \textit{mixed Hadamard basis}, we rewrite the second qubit in the state $\ket{\phi}$, from our running example, in the Hadamard basis as $\ket{\phi}=1/\sqrt{2}(-\ket{0-} + \ket{1+})$. 
From this point on, we will not specify which qubits are transformed, but simply state that the state vector is written in the mixed Hadamard basis. 
The state $\ket{\phi}$ in the mixed Hadamard basis is still mathematically equal to the state in the computational basis and, thus, results in the same outputs after a measurement. 
In order to perform a projective measurement in a quantum program, we must apply the transformation before we perform a normal measurement in the computational basis. 
Thus, we transform $\ket{\phi}$ by the basis transformation $\mathcal{T}=\mathbbm{1}\otimes H$, resulting in the transformed state $\ket{\phi}_{\mathcal{T}} = 1/\sqrt{2}(-\ket{01}+\ket{10})$, by adding the appropriate Hadamard gate to the second qubit of the coded example circuit 2) in \cref{fig:diagram_background}. 
As with the computational basis, we can now identify the two, instead of four, possible outputs $\ket{01}$ and $\ket{11}$ with probabilities $1/2$ after measurement in the new basis. 
We refer to this new basis as the \textit{reduced basis} as it has reduced the number of basis states.
If we apply this basis change to the faulty state vector of the running example, we obtain $\ket{\phi}_{\mathcal{T}, faulty} = 1/\sqrt{2}(\ket{01}+\ket{10})$.
We observe that for the state vector in the reduced basis affected by a bit flip fault, the negative sign in front of the first basis state changes to a positive sign.
We exploit this type of reduction in our approach in \cref{sec:approach}.

Generally, we define a \textit{projective measurement with respect to a particular basis} as the basis transformation $\mathcal{T}$ on the final state vector followed by a measurement $M$ in the computational basis:
\begin{equation}
\label{eq:basis_transform}
 \mathcal{T} = H^{x_0}\otimes H^{x_1}\otimes \cdots \otimes H^{x_{n-1}} 
\end{equation}
\Cref{eq:basis_transform} depicts the mixed Hadamard basis transformations for n-qubit state vectors considered in this paper.
Here, the values $x_m$ for $m=0,1,\cdots,n-1$ take the values 0 or 1 depending on which qubits we transform with a Hadamard gate. 
For instance, if $n=5$, $x_0=1$ means a Hadamard gate will be applied to the $1^{st}$ qubit, while $x_4=0$ means no Hadamard gate is applied to the $5^{th}$ qubit. 
Thus, exponentiating the Hadamard gate to the $0^{th}$ power yields the identity $H^0=\mathbbm{1}$, while the $1^{st}$ power gives the Hadamard gate itself $H^1=H$.
For the basis change performed to our running example state $\ket{\phi}$, we select the transformation $\mathcal{T}= H^0 \otimes H^1= \mathbbm{1}\otimes H$ from \cref{eq:basis_transform} and apply it to the state through adding a Hadamard gate to the second qubit just before we measure.
We introduce a short-hand notation for the transformations in \cref{eq:basis_transform}, where the character $1$ replaces any identity matrix $\mathbbm{1}$, and a lowercase character $h$ replaces any Hadamard gate $H$.
In addition, we combine the characters $1$ and $h$ without the tensor product $\otimes$, such that $H^0 \otimes H^1\rightarrow 1h$.
With this notation, we write the transformed state from the running example as $\ket{\phi}_{1h}=1/\sqrt{2}(-\ket{01}+\ket{10})$.
The notation $1h$ determines that the first qubit is in the computational basis, meaning $0$ and $1$ are read as they are, while the second qubit is in the Hadamard basis, meaning $0$ and $1$ are read as $+$ and $-$, respectively.
In other words, the basis states of $\ket{\phi}_{1h}$ are read as $\ket{0-}$ and $\ket{1+}$, but they are measured as $\ket{01}$ and $ \ket{10}$ in the measurement step in the coded example circuit 2) in \cref{fig:diagram_background}.

\section{Approach}
\label{sec:approach}

In this section, we introduce our reduction approach to \gls{qst}, which consists of two components. 
The first component, called \HLIGHT{Reduction}, introduces a Greedy reduction algorithm designed to reduce a \HLIGHT{Default \gls{ps}} into a \HLIGHT{Reduced \gls{ps}}.
In the second part, called \HLIGHT{Application To Quantum Software Testing}, we describe how we utilize the \HLIGHT{Reduced \gls{ps}} in \gls{qst}.

\newcommand{\mycircled}[1]{%
    \tikz[baseline=(char.base)]{
        \node[shape=circle,draw,inner sep=0.85pt] (char) {#1};}
}

\subsection{Overview}
\label{overview}

\begin{figure}[tbp]
    \centering
    \includegraphics[scale=0.80]{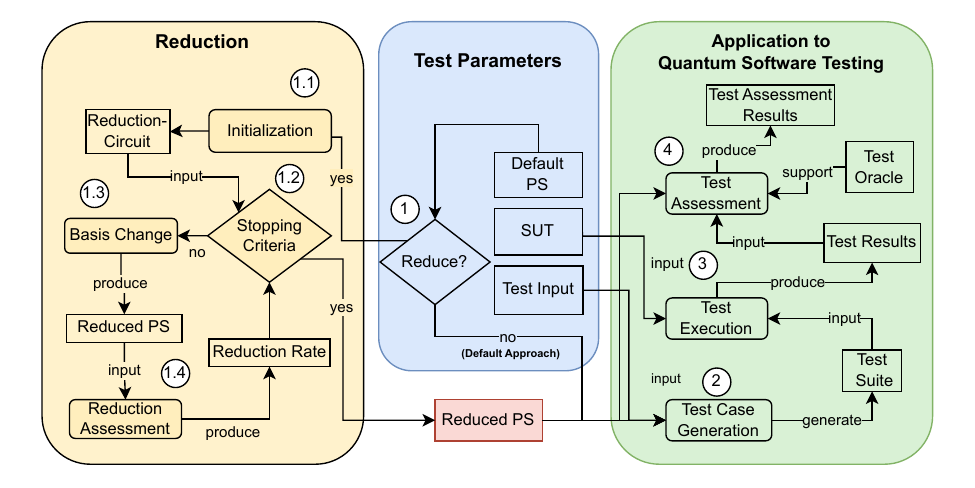}
    \caption{Overview and Application Context of the Reduction Approach.}
    \label{fig:approach_diagram}
\end{figure}

\Cref{fig:approach_diagram} depicts an overview of our approach. 
Initially, we provide the approach's parameters in the \HLIGHT{Test Parameters} step, i.e., a \HLIGHT{Default \gls{ps}}, \HLIGHT{Test Input}, and \gls{sut}. 
Then, if \HLIGHT{Reduce?} yields yes at \circled{1}, the \HLIGHT{Reduction} component applies the reduction algorithm to the provided \HLIGHT{Default PS}.
To initialize reduction in \mycircled{1.1}, the \HLIGHT{Initialization} step constructs a \HLIGHT{Reduction-Circuit}. 
Then, we pass through the \HLIGHT{Stopping Criteria} and enter into a loop at \mycircled{1.2}. 
We iteratively perform \HLIGHT{Basis Change}s in \mycircled{1.3} to the \HLIGHT{Reduction-Circuit} to obtain a \HLIGHT{Reduced \gls{ps}}. 
In each iteration at \mycircled{1.4}, we conduct a \HLIGHT{Reduction Assessment} on the \HLIGHT{Reduced PS}, computing its rank. When the algorithm does not obtain further reductions in the rank, the stopping criteria at \mycircled{1.2} terminates the loop and returns the \HLIGHT{Reduced \gls{ps}}.

In the second stage, \HLIGHT{Application to Quantum Software Testing}, we provide either the \HLIGHT{Default \gls{ps}} or \HLIGHT{Reduced \gls{ps}} as input, depending on whether or not we perform \HLIGHT{Reduction}. 
The test process begins with the \textit{Test Case Generation} at \circled{2} constructing a \HLIGHT{Test Case}. Then, in the \HLIGHT{Test Execution} step \circled{3}, the distribution of the quantum program is sampled by multiple executions, each time with a specific input state vector \HLIGHT{Test Input} provided in the \HLIGHT{Test Parameters} step. Finally, the \HLIGHT{Test Assessment} at \circled{4} evaluates these distributions, represented as \HLIGHT{Test Results}, using \HLIGHT{Test Oracles} yielding \HLIGHT{Test Assessment Results}.

\subsection{Test Parameters}
\label{sec:test_parameters}

Here, we describe the initial step \HLIGHT{Test Parameters} component of \cref{fig:approach_diagram} consisting of the \HLIGHT{Default \gls{ps}}, \gls{sut}, and \HLIGHT{Test Input}. 
These parameters provide the initial setup for our approach.

\subsubsection{Program Specification}
\label{sec:program_specifications}

In \gls{qst}, a \gls{ps} plays a crucial role in validating the correctness of quantum computations by providing the expected behavior of a quantum program~\citep{ShaukatTaoQST, wille2022decision,quitoASE21tool,Proq}. 
From the state vector postulate in quantum mechanics~\citep{Sakurai,QCQIBook}, the state vector provides the complete description of the quantum program. 
Thus, we define the \gls{ps} of a quantum program as its non-faulty final state vector $\ket{PS}_{\mathcal{T}}$ with respect to the basis $\mathcal{T}$ as:

\begin{equation}
    \ket{PS}_{\mathcal{T}} = \sum_{j\in \{ x | \alpha_x \neq 0 \} } \alpha_j \ket{j}_{\mathcal{T}}
    \label{eq:program_specification_state_vector}
\end{equation}

In \cref{eq:program_specification_state_vector}, $\{x | P_x \neq 0\}$ is the set of indices that correspond to the basis states $\ket{j}$ with non-zero probability amplitudes. 
As the \gls{ps} reflects the correct final state vector of a given quantum program, we define the set of probabilities of the \gls{ps} as the \emph{Theoretical Probability Distribution}.
The expression of the \gls{ps} in \cref{eq:program_specification_state_vector} is a convenience rewrite of the general state vector form \cref{eq:multi_qubit_state_compact}, such that we can define $N_{ps} \equiv |\{x | \alpha_x \neq 0\}| $ as the number of basis states in the sum of \cref{eq:program_specification_state_vector}, which we call the \textit{rank} of the \gls{ps}. 
Thereby, the number of basis states in the \gls{ps} may be less than the number of basis states in the state space $N$, i.e., $N_{ps}\leq N$.

For a non-faulty quantum program, we define two types of \gls{ps}: (1) the \HLIGHT{Default \gls{ps}}, which is defined by the unaltered quantum program where $\mathcal{T}=\mathbbm{1}$ and (2) the \HLIGHT{Reduced \gls{ps}}, where the \gls{ps} is reduced by a mixed Hadamard basis transformation of the type in \cref{eq:basis_transform}, such that $\ket{PS}_{\mathcal{T}}=\mathcal{T}\ket{PS}_{\mathbbm{1}}$. 

\textbf{Note, when $\mathcal{T}=\mathbbm{1}$ we will omit the basis notation, such that $\ket{PS}=\ket{PS}_{\mathbbm{1}}$ refers to the \HLIGHT{Default \gls{ps}}}.

In practice, we specify a \HLIGHT{Default \gls{ps}} in the \HLIGHT{Test Parameters} step obtained through a non-faulty version of the \gls{sut}. 
We provide an extensive discussion of the practical and theoretical aspects of obtaining a \HLIGHT{Default \gls{ps}} in the discussion \cref{sec:discussion}.

\subsubsection{\gls{sut} and Test Input}
\label{sec:sut}

The \gls{sut} consists of a main program $\mathcal{U}$ (see \cref{sec:quantum_program}) and a projective measurement with respect to the basis $\mathcal{T}$. 
We compare the \HLIGHT{Default \gls{ps}} to measurements of the \gls{sut} performed with respect to the computational basis, while the \HLIGHT{Reduced \gls{ps}} describes measurements on the \gls{sut} performed with respect to a mixed Hadamard basis. 
To run the \gls{sut}, we provide the initial state vector $\ket{\psi_{init}}$, called the \HLIGHT{Test Input}. 

\subsection{Reduction \label{sec:reduction_component}}

In this section, we present the reduction algorithm of the \HLIGHT{Reduction} component in \cref{fig:approach_diagram}.
We define a Greedy reduction algorithm to obtain a reduced program specification, providing a simple and effective way to handle the search space consisting of the basis transformations in \cref{eq:basis_transform}. 
The simplicity of the search space allows for the efficient application of Greedy, making it an appropriate choice without requiring more advanced search algorithms. 
While Greedy may not perform as well in more complex scenarios, where, for instance, other gates besides Hadamard are included in \cref{eq:basis_transform}, its low computational overhead makes it ideal for the mixed Hadamard basis.

Our approach is not limited to any particular search algorithm, such as Greedy. 
More advanced search algorithms could be applied in future work when dealing with larger or more intricate search spaces. 
However, this paper focuses on evaluating the reduction approach itself rather than comparing different search algorithms. 
We further discuss the rationale and the tradeoff for using Greedy in \cref{sec:discussion}.

Thereby, to assess reductions, we apply the rank $N_{ps}$ of the \gls{ps}, defined in \cref{sec:program_specifications}, as the objective function. 


\paragraph{\textbf{Running Example}}
Throughout this section, we consider the following 3-qubit \gls{ps} as a running example:

\begin{equation}
\ket{\text{Default \gls{ps}}} = \frac{1}{\sqrt{8}} ( \ket{000} + \ket{001} + \ket{010} - \ket{011} + \ket{100} + \ket{101} - \ket{110} + \ket{111})
\label{eq:approach_running_example}
\end{equation}

We chose this example state vector because it allows us to demonstrate two key properties of the basis search space that affect the runtime requirements of our algorithm design. 
First, it shows that a \HLIGHT{Reduced \gls{ps}} is not unique in general, so we might ask how much search space exploration is necessary for a sufficient rank reduction. 
Second, two different \HLIGHT{Reduced \gls{ps}}s may give the same degree of reduction, so how do we select them? 
For instance, given the three basis transformations $\mathbbm{1}\otimes H \otimes \mathbbm{1}$, $\mathbbm{1}\otimes \mathbbm{1} \otimes H$ and $H\otimes \mathbbm{1} \otimes H$, they transform \cref{eq:approach_running_example} into three possible reduced states:

\begin{equation}
    \ket{\text{Reduced \gls{ps}}}_{ihi} = \frac{1}{2} \Bigg( \underbrace{\ket{0+0} + \ket{0-1} + \ket{1-0} + \ket{1-1}}_{N_{ps} = 4} \Bigg)
\label{eq:running_example_definition_example1}
\end{equation}

\begin{equation}
    \ket{\text{Reduced \gls{ps}}}_{iih} = \frac{1}{2} \Bigg( \underbrace{\ket{00+} + \ket{01-} + \ket{10+} - \ket{11-}}_{N_{ps} = 4} \Bigg)
\label{eq:running_example_definition_example3}
\end{equation}

\begin{equation}
    \ket{\text{Reduced \gls{ps}}}_{hih} = \frac{1}{\sqrt{2}} \Bigg( \underbrace{\ket{+0+} + \ket{-1-}}_{N_{ps} = 2} \Bigg)
\label{eq:running_example_definition_example2}
\end{equation}

In the two \HLIGHT{Reduced \gls{ps}} in \cref{eq:running_example_definition_example1,eq:running_example_definition_example3}, the rank
is the same, from $N_{ps}=8$ to $N_{ps}=4$ in the \HLIGHT{Default \gls{ps}}, while in \cref{eq:running_example_definition_example2} it is reduced to $N_{ps}=2$. 
We consider the best reduction of the three to be the latter as it achieved the smallest rank $N_{ps}$. 
Thereby, we treat the other two bases in \cref{eq:running_example_definition_example1,eq:running_example_definition_example3} as indistinguishable. 
We discuss the implications of this assumption in \cref{sec:discussion}.
In terms of runtime considerations, to obtain the better result of \cref{eq:running_example_definition_example2}, two Hadamard gates are required, compared to only one in \cref{eq:running_example_definition_example1} or \cref{eq:running_example_definition_example3}. 
Thus, this illustrates that a larger reduction in the rank may require more search time to find additional Hadamard gates.

\hrulefill 

In the remainder of this section, we define our Greedy reduction algorithm, outlined in \cref{alg:reduction_algorithm}, which consists of three steps: \HLIGHT{Initialization}, \HLIGHT{Basis Change}, and \HLIGHT{Reduction Assessment}. 
We also demonstrate its application by using the running example.

\newcommand{\Input}{\Require}
\newcommand{\Output}{\Ensure}

\algrenewcommand\algorithmicrequire{\textbf{Input:}}
\algrenewcommand\algorithmicensure{\textbf{Output:}}


\begin{algorithm}
\caption{Reduction Algorithm}
\label{alg:reduction_algorithm}
\begin{algorithmic}[1]
\Input \HLIGHT{Default \gls{ps}} $\ket{PS}$, Number of Qubits \(n\)
\Output \HLIGHT{Reduced \gls{ps}} $\ket{PS}_{\mathcal{T}}$
\Statex
\Statex \boxed{\textbf{Initialization Step:}}
\Statex
\State \( \ket{RC_{\texttt{Initial}}} \leftarrow \ket{PS} \) \hspace{5.32cm} /* Reduction-Circuit */ 
\State \( \mathcal{T} \leftarrow \{ \; \mathbbm{1} \; | \;  \ell = 0, \; 1, \; \dots, \; n-1 \; \} \) \hspace{3.75cm} /* Basis Transformation with Identity Operators */ 
\State \( N_{\texttt{minPrevious}} \leftarrow \texttt{CountBasisStates}\big(\ket{PS}\big) \) 
\State \( N_{\texttt{minCurrent}} \leftarrow  N_{\texttt{minPrevious}} \)
\State \( \texttt{search\_space} \leftarrow \{ \; \ell \; | \; \ell = 0, \; 1, \; \dots,\; n-1 \;\} \)
\State \(  N_{\texttt{psList}} \leftarrow \{\; 0 \;|\;  \ell = 0, \; 1, \;\dots,\; n-1 \;\} \) \hspace{3.05cm} /* Set of 0s to Store Number of Basis States */
\Statex
\While{\( \neg \left( \textbf{STOPPING\_CRITERIA} \Big(N_{\texttt{minCurrent}} \geq N_{\texttt{minPrevious}}\Big) \right) \)}
    \Statex
    \State \boxed{\textbf{Basis Change Step:}}
    \Statex
    \For{ \texttt{j} in  \texttt{search\_space} }
        \State \( \mathcal{T}[\texttt{j}] := H \)    
        \State \( \ket{RC_{\texttt{Reduced}}} \leftarrow \texttt{ApplyBasisTransformation}\big( \ket{RC_{\texttt{Initial}}}, \mathcal{T}\big)\)
        \State \( N_{\texttt{psReduced}} \leftarrow \texttt{CountBasisStates}\big( \ket{RC_{\texttt{Reduced}}} \big) \)
        \State \( N_{\texttt{psList}}[\texttt{j}] := N_{\texttt{psReduced}} \)  
        \State \( \mathcal{T}[\texttt{j}] := \mathbbm{1} \)  
    \EndFor
    \Statex
    \State \boxed{\textbf{Reduction Assessment Step:}}
    \Statex
    \State \( N_{\texttt{minCurrent}} = \texttt{min}\big( N_{\texttt{psList}} \big) \)  \hspace{3.9cm} /* Find Minimum Value */
        \State \( N_{\texttt{minList}} \leftarrow \) \texttt{FindMinIndices}(\(N_{\texttt{psList}}\), \(N_{\texttt{minCurrent}})\)  \hspace{0.75cm} /* Indices in \( N_{\texttt{psList}} \) Equal to \(N_{\texttt{min}}\) */
        \State \( \texttt{k} \leftarrow \texttt{RandomChoice}(N_{\texttt{minList}}) \) \hspace{3.55cm} /* Randomly Select a Minimum Index */
        \State \( \mathcal{T}[\texttt{k}] := H \)   \hspace{6.15cm}  /* Fix the k'th Gate */
        \State \texttt{search\_space.RemoveIndex(k)}
        \State \( N_{\texttt{minPrevious}} \leftarrow  N_{\texttt{minCurrent}} \)
    \EndWhile
\Statex
\Return \( \mathcal{T} \) 
\end{algorithmic}
\end{algorithm}


\subsubsection{Initialization}

We first create a quantum circuit, called the \textit{reduction-circuit}, for storing the \HLIGHT{Default \gls{ps}} in the circuit and for performing basis transformations. 
We denote the reduction-circuit by \( \ket{RC_{\texttt{Initial}}} \) and initialize it to the state vector \( \ket{PS} \), representing the \HLIGHT{Default \gls{ps}}.
Next, on Line 2 in \cref{alg:reduction_algorithm}, we initialize the basis transformation array \( \mathcal{T} \) with \( n \) identity operators \( \mathbbm{1} \) indicating no initial transformations on the qubits. 
This array stores and applies transformations 
throughout the algorithm.
Following this on Line 3, we compute the rank $N_{ps}$ of the initial \gls{ps}, storing the result in \( N_{\texttt{psPrevious}} \), using the \texttt{CountBasisStates} function. 
We use this initial rank to determine the \HLIGHT{Stopping Criteria}. 
Subsequently, Lines 4 and 5 define the \texttt{search\_space} and \( N_{\texttt{psList}} \) arrays. 
The \texttt{search\_space} represents the indices of the qubit register. 
We initialize the array \( N_{\texttt{psList}} \) with \( n \) zeros, which stores the ranks of the \HLIGHT{Reduced \gls{ps}} for each transformed bases throughout the algorithm, keeping track of the reductions in each iteration.

\paragraph{\textbf{Running Example (Initialization Step):}}
\label{sec:initialization_running_example}
We declare the reduction-circuit, $\ket{RC_{\texttt{Initial}}}$, that holds the state vector in \cref{eq:approach_running_example}. 
Now, the final state vector of the reduction-circuit is identical to the running example state vector, i.e., $\ket{RC_{\texttt{Initial}}}=\ket{\text{Default \gls{ps}}}$. 
The remaining variables for the initialization step for our running example are specified as follows.
We start by initializing the basis transformation array, \(\mathcal{T} = [\mathbbm{1}, \mathbbm{1}, \mathbbm{1}]\), to the computational basis. 
Next, we compute the rank of the initial \gls{ps}, setting \(N_{\texttt{psPrevious}} = 8\). 
The array of qubit indices available for Hadamard transformations is defined as \(\texttt{search\_space} = [0,1,2]\). 
Finally, \(N_{\texttt{psList}} = [0,0,0]\) is used to store the ranks resulting from the transformations, maintaining the minimum rank at the end of each iteration.

\subsubsection{Basis Change}

After the initialization step, the \HLIGHT{Basis Change} step is performed, which includes a while loop on line 7. This loop continues until satisfying the \HLIGHT{Stopping Criteria}.
The stopping criteria assess whether the previous iteration resulted in a reduction, and consequently, the algorithm terminates and returns the \HLIGHT{Reduced \gls{ps}}. 
Inside the while loop, we initiate the basis change by iterating over the qubit indices in the \texttt{search\_space} array. Line 10 activates a single Hadamard gate at the index $j$ of $\mathcal{T}$, and we provide $\mathcal{T}$ and $\ket{RC_{\texttt{Initial}}}$ to \texttt{ApplyBasisTransformation} on line 11. \texttt{ApplyBasisTransformation} converts $\mathcal{T}$ into a basis transformation according to \cref{eq:basis_transform} and applies it to $\ket{RC_{\texttt{Initial}}}$, yielding the transformed state $\ket{RC_{\texttt{Reduced}}}$. 
We then compute the rank of $\ket{RC_{\texttt{Reduced}}}$ and store it at index $j$ in $N_{\texttt{psList}}$ on line 13. 
Line 14 resets the Hadamard gate of the current iteration back to an identity operator at $\mathcal{T}[\texttt{j}]$. 
This process constitutes the core operation of the \HLIGHT{Basis Change}: we activate single Hadamard gates based on the indices from \texttt{search\_space}, perform the basis transformation on the reduction-circuit, compute the rank of the transformed reduction-circuit, and deactivate the Hadamard gate again. 
The for-loop ends when we have performed all basis transformations specified by the indices in \texttt{search\_space} and the corresponding ranks are stored in $N_{\texttt{psList}}$.

\subsubsection{Reduction Assessment}
\label{sec:reduction_assessment}

After we try all the basis transformations in \texttt{search\_space}, we move to the \HLIGHT{Reduction Assessment} step. 
Line 17 calculates the minimum of \(N_{\texttt{psList}}\), which points to the basis with the largest reduction.
If there are multiple equal minimum values, we pick one randomly on lines 18 and 19.
On line 20, we select an index $k$ randomly from $N_{\texttt{minList}}$ and fix the $k$\textsuperscript{th} Hadamard gate in $\mathcal{T}$ on line 21 by removing $k$ from \texttt{search\_space}.
The final stage of the \HLIGHT{Reduction Assessment} step is to update $N_{\texttt{minPrevious}}$ by assigning it the value of $N_{\texttt{minCurrent}}$ such that the \HLIGHT{Stopping Criteria} in the next iteration compares the next current minimum value found against the previously one.

\paragraph{\textbf{Running Example (Basis Change \& Reduction Assessment):}}
\label{sec:basis_change_running_example_Greedy}

\begin{figure}[tbp]             
    \includegraphics[scale=0.80]{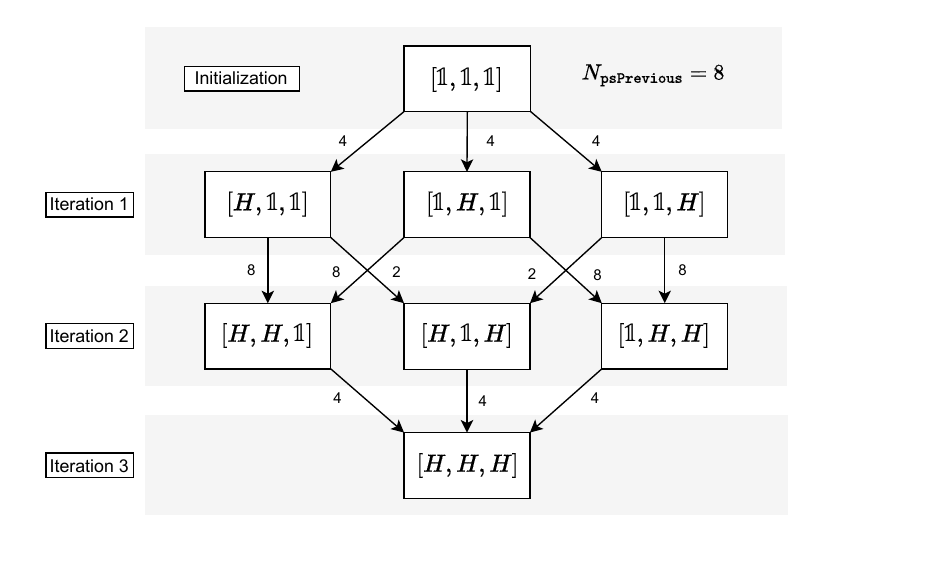}
    \caption{Directional graph representation of the Greedy reduction \cref{alg:reduction_algorithm} for the running example in \cref{eq:approach_running_example}.}
    \label{fig:graph_examples}
\end{figure}

Our example state has \num{8} possible basis transformations:

\begin{align}
    \label{eq:approach_basis_transforms_regular}
    \begin{aligned}
        & \mathbbm{1}\otimes \mathbbm{1} \otimes \mathbbm{1} \hspace{1.5cm}& 
        & \mathbbm{1}\otimes \mathbbm{1} \otimes H \hspace{1.5cm}& 
        & \mathbbm{1}\otimes H \otimes \mathbbm{1} \hspace{1.5cm}&
        & \mathbbm{1}\otimes H \otimes H \\    
        & H\otimes \mathbbm{1} \otimes \mathbbm{1} \hspace{1.5cm}& 
        & H\otimes \mathbbm{1} \otimes H \hspace{1.5cm}& 
        & H\otimes H \otimes \mathbbm{1} \hspace{1.5cm}&
        & H \otimes H \otimes H.
    \end{aligned}
\end{align}

We perform three iterations of the while-loop. 
In iteration 1 (illustrated in \cref{fig:graph_examples}), we perform the basis transformations: $[H, \mathbbm{1},\mathbbm{1}], [\mathbbm{1},H,\mathbbm{1}]$ and $[\mathbbm{1},\mathbbm{1},H]$ and compute the ranks (Line 11--12 in \cref{alg:reduction_algorithm}). 
Recall that our initial state had $N_{ps}=8$; these three transformations all lead to a reduction in the rank by $4$, which we can see from the transformed states:

\begin{equation}
    \begin{aligned}
        \ket{PS}_{h11} &= \frac{1}{2}\Bigg( \underbrace{\ket{+00} + \ket{+01} + \ket{-10} -\ket{-11}}_{N_{ps}=4} \Bigg)\\
        \ket{PS}_{1h1} &= \frac{1}{2}\Bigg( \underbrace{\ket{0+0} + \ket{0-1} + \ket{1-0} +\ket{1-1}}_{N_{ps}=4} \Bigg)\\
        \ket{PS}_{11h} &= \frac{1}{2}\Bigg( \underbrace{\ket{00+} + \ket{01-} + \ket{10+} -\ket{11-}}_{N_{ps}=4} \Bigg)
    \end{aligned}
    \label{eq:running_example_reductions_from_algo}
\end{equation}

Thus, $N_{\texttt{psList}}=[4,4,4]$, which means we select one of the transformations at random on Line 19 in \cref{alg:reduction_algorithm}.
If we choose $[H,\mathbbm{1},\mathbbm{1}]$, we remove index $0$ from \texttt{search\_space}:=\texttt{search\_space}=[1,2], such that in the next iteration we only iterate over qubits 1 and 2. 
Then, in iteration 2 (see \cref{fig:graph_examples}), we apply the basis transformations $[H,H,\mathbbm{1}]$ and $[H,\mathbbm{1},H]$. 
The first choice leads to an increase in rank, back to $8$ basis states, while the latter leads to the state $1/\sqrt{2}(\ket{+0+}+\ket{-1-})$ with $N_{ps}=2$. 
Thereby, we obtain $N_{\texttt{psList}}=[4,8,2]$ and select the second index, i.e., the minimum, corresponding to the basis $[H,\mathbbm{1}, H]$. 
Despite this being the global minimum, the algorithm continues to iteration 3 and terminates on the \HLIGHT{Stopping Criteria} as the basis $[H, H, H]$ leads to an increase in the rank to $N_{ps}=4$. 
Thus \cref{alg:reduction_algorithm} returns the \HLIGHT{Reduced \gls{ps}} $[H,\mathbbm{1},H]$.
Let's explore a couple of other possibilities.
If we had chosen $[\mathbbm{1},\mathbbm{1}, H]$ at the end of iteration 1 instead, we also find the reduction to $2$ (which we see from \cref{fig:graph_examples}).
However, if we instead choose $[\mathbbm{1}, H,\mathbbm{1}]$ with a reduction to the rank $4$ state $1/2(\ket{0+0}+\ket{0-1}+\ket{1-0}+\ket{1+1})$, then we are stuck at a local minimum, as the next transformations in iteration 2 are either $[H, H,\mathbbm{1}]$ or $[\mathbbm{1}, H, H]$, which both lead to an increase back to $8$ basis states, terminating the \HLIGHT{Stopping Criteria} and returning the local minimum basis $[\mathbbm{1}, H,\mathbbm{1}]$.
All possible reduction paths can be studied in \cref{fig:graph_examples}. 

As we reduce the basis search space by a Hadamard gate after each iteration, our reduction algorithm finds a local or global minimum with $\mathcal{O}(n^2)$ number of basis transformations, which we show in our appendix~\citep{oldfield_2024_11191215}.

Completing the \HLIGHT{Reduction} component, we then provide the \HLIGHT{Reduced \gls{ps}} as input to the \HLIGHT{Application to Quantum Software Testing} component.

\subsection{Application to Quantum Software Testing}

In this section, we present the \HLIGHT{Application to Quantum Software Testing} component in \cref{fig:approach_diagram}, which applies \HLIGHT{Reduction} to \gls{qst} and contains the following steps: \HLIGHT{Test Case Generation} and \HLIGHT{Test Execution}.

\subsubsection{Basis Dependent Test Cases}
\label{sec:test_cases}

To test a quantum program, we compare its theoretical probability distribution to a sample distribution resulting from multiple executions of the program. 
In order to account for measurements in different bases, we define a test case as a function of the basis $\mathcal{T}$:

\begin{equation}
    TC(\mathcal{T}) = \Big\{\ket{\psi_{init}}, \mathcal{P}_{ps}(\mathcal{T}), Outputs \Big\}
    \label{eq:test_case_definition}
\end{equation}

In \cref{eq:test_case_definition}, \( \ket{\psi_{init}} \) represents the input state vector, while \( \mathcal{P}_{ps}(\mathcal{T}) \) is the theoretical probability distribution of the quantum program (see \cref{sec:program_specifications}), defined as the set of probabilities of each output basis state, but extended to include measurement with respect to the basis $\mathcal{T}$:

\begin{equation}
\mathcal{P}_{ps}(\mathcal{T}) = \left\{ p_{0}(\mathcal{T}), p_{1}(\mathcal{T}), \cdots ,p_{N-1}(\mathcal{T}) \right\}
\label{eq:theoretical_distribution}
\end{equation}

We compute the $k$\textsuperscript{th} probability in \cref{eq:theoretical_distribution} by applying the basis transformation $\mathcal{T}$ to the \HLIGHT{Default \gls{ps}}, then computing the squared norm of the probability amplitudes:

\begin{equation}
p_k(\mathcal{T}) = \Big| \alpha_k(\mathcal{T}) \Big|^2
\label{eq:basis_dependent_probabilities}
\end{equation}

Where the probability amplitudes in \cref{eq:basis_dependent_probabilities} are the amplitudes for the basis states in the \HLIGHT{Reduced \gls{ps}}.

Finally, we define \textit{Outputs} as the set of possible basis states in $\ket{\text{Reduced \gls{ps}}}$:

\begin{equation}
    Outputs = \Big\{ \ket{j} | \text{if $j$ is a basis state of $\ket{\text{Reduced \gls{ps}}}$ } \Big\}
    \label{eq:output_set}
\end{equation}

\paragraph{\textbf{Test Case and Suite Generation:}}
We follow these steps to generate a test case:
\begin{enumerate}
    \item Specify an input $\ket{\psi_{init}}$ and $\ket{\text{Default \gls{ps}}}$, then obtain $\mathcal{T}$ from \cref{alg:reduction_algorithm}
    \item Compute $\ket{\text{Reduced \gls{ps}}}=\mathcal{T}\ket{\text{Default \gls{ps}}}$
    \item Calculate the probability distribution and outputs of $\ket{\text{Reduced \gls{ps}}}$ with \cref{eq:basis_dependent_probabilities} and \cref{eq:output_set}
\end{enumerate}

When steps (1) to (3) are performed, the pair $\ket{\psi_{init}}$ with $\mathcal{P}_{ps}(\mathcal{T})$ is a test case from \cref{eq:test_case_definition}.

\paragraph{\textbf{Running Example (Basis Dependent Test Cases):}}

Continuing from \cref{sec:reduction_assessment},
we obtain $\ket{\text{Default \gls{ps}}}$ from \cref{eq:approach_running_example} along with the arbitrarily chosen input $\ket{\psi_{init}}=\ket{000}$.
To complete steps (1) and (2), we can assume that an execution of \cref{alg:reduction_algorithm}
resulted in the basis $[H,\mathbbm{1}, \mathbbm{1}]$, corresponding to the reduced state $\ket{\text{\gls{ps}}}_{h11}$ in \cref{eq:running_example_reductions_from_algo}.
Now, we can complete step (3) by computing the set of possible outputs, along with its theoretical probability distribution.
To compute $Outputs$, we list the possible basis states by inspecting the state $\ket{\text{\gls{ps}}}_{h11}$ in \cref{eq:running_example_reductions_from_algo}, yielding:
\begin{equation}
    Outputs_{Reduced} = \Big\{ \ket{+00},\ket{+01},\ket{-10},\ket{-11}\Big\} \Leftrightarrow  \Big\{ \ket{000},\ket{001},\ket{110},\ket{111}\Big\}
    \label{eq:running_example_outputs}
\end{equation}

\Cref{eq:running_example_outputs} lists the basis states of $\ket{\text{\gls{ps}}}_{h11}$ and additionally demonstrates its equivalence with the set to the right by showing that $+$ is read as $0$ and $-$ is read as $1$, which is due to the basis choice with the Hadamard gate applied to the first qubit.
Now, to complete step (3), we calculate the theoretical probability distribution of $\ket{\text{\gls{ps}}}_{h11}$ with \cref{eq:basis_dependent_probabilities}, yielding:

\begin{equation}
    \mathcal{P}_{ps}(\mathcal{T}) = \Big\{ 0.25, 0.25, 0, 0, 0, 0, 0.25, 0.25 \Big\}
\label{eq:running_example_reduced_distribution}
\end{equation}

\Cref{eq:running_example_reduced_distribution} emphasizes that the index position of the probabilities in the list corresponds to the particular basis states.
For instance, the probability of $\ket{000}$ should be placed in the first position of the list, as $000$ is the first basis state according to its binary value.
Thus, \cref{eq:running_example_reduced_distribution}, along with the input $\ket{\psi_{init}}=\ket{000}$ forms a test case according to our definition in \cref{eq:test_case_definition}.
We generate a test suite by repeating these steps.

Now, to illustrate the reduction, we can repeat these steps for the $\ket{\text{Default \gls{ps}}}$ to obtain:

\begin{equation}
    Outputs_{Default} = \Big\{ \ket{000},\ket{001},\ket{010},\ket{011}, \ket{100}, \ket{101}, \ket{110}, \ket{111}\Big\}
    \label{eq:running_example_default_output}
\end{equation}

Along with the default theoretical distribution:

\begin{equation}
    \mathcal{P}_{ps}(\mathbbm{1}) = \Big\{ 0.125, 0.125, 0.125, 0.125, 0.125, 0.125, 0.125, 0.125 \Big\}
\label{eq:running_example_reduced_distribution_default}
\end{equation}

We observe that the reduced output list and reduced theoretical distribution of \cref{eq:running_example_outputs} and \cref{eq:running_example_reduced_distribution} contains half of the possible basis states and probabilities compared to the default in \cref{eq:running_example_default_output} and \cref{eq:running_example_reduced_distribution_default}.
Our approach exploits this to improve efficiency and effectiveness.

\subsubsection{Test Execution with Projective Measurements}
\label{sec:test_execution}

The \HLIGHT{Test Execution} step takes the previously generated \HLIGHT{Test Case} and the \gls{sut} as inputs. 
We execute the test cases against the \gls{sut} and sample the quantum program with $N_{E}$ program executions, acquiring the sample distribution $\Lambda_{sample}$, which constitutes the \HLIGHT{Test Results}  in \cref{fig:approach_diagram}. 
We now sample the quantum program $\mathcal{U}$, depicted as 1) in \cref{fig:diagram_background}, with projective measurements $\mathcal{M}$ in the mixed Hadamard basis $\mathcal{T}$.
Thus, we define the \textit{sample distribution} for a given test case as a function of the basis $\mathcal{T}$:

\begin{equation}
\Lambda_{sample}(\mathcal{T}, N_E) = \big\{\lambda_0(\mathcal{T}, N_E), \lambda_1(\mathcal{T}, N_E),\cdots, \lambda_{N-1}(\mathcal{T}, N_E)\big\}. 
\label{eq:sample_distribution}
\end{equation}

In \cref{eq:sample_distribution}, the sample distribution quantifies the relative frequencies $\lambda_m(\mathcal{T}, N_E)$ of each output basis state from the quantum program measured in $\mathcal{T}$:

\begin{equation}
    \lambda_m(\mathcal{T}, N_E) = \frac{C_m(\mathcal{T})}{N_E(\mathcal{T})}
    \label{eq:lambda_m}
\end{equation}

Thus, $C_m(\mathcal{T})$ in \cref{eq:lambda_m}, denotes the number of occurrences of the output basis state $\ket{m}_{\mathcal{T}}$, and $N_{E}$ is the total quantum program executions for a given test case. 
As programs with more qubits require higher sample sizes compared to programs with fewer qubits, we set the sample size $N_{E}$ to be proportional to the number of basis states in the \gls{ps}:

\begin{equation}
    N_E = \gamma N_{ps}
\label{eq:sample_size}
\end{equation}

In \cref{eq:sample_size}, $\gamma\in \mathbbm{N}$ determines the \HLIGHT{Test Execution} step's sample size relative to the rank $N_{ps}$ of the \gls{ps}. 

Thus, our approach offers two potential improvements to \gls{qst}. First, defining a test case in \cref{sec:test_cases} as a function of the basis reduces the number of output states in the program specification, which in turn decreases the size of the test case.
Second, in \cref{eq:sample_size}, by defining the sample size as a function of the number of basis states, a smaller test case will yield a reduction in the number of quantum program executions required for obtaining the sample distribution in \cref{eq:sample_distribution}. 

\paragraph{\textbf{Test Execution Procedure:}}
We define the procedure for sampling the quantum program and measuring in $\mathcal{T}$ as:
\begin{enumerate}
    \item Prepare the quantum program $\mathcal{U}$ with the input state $\ket{\psi_{init}}$ from the test case
    \item Apply the basis transformation $\mathcal{T}$ to $\mathcal{U}\ket{\psi_{init}}$
    \item Measure in the computational basis $M$, resulting in the projective measurement $\mathcal{M}=M\mathcal{T}$ applied to $\mathcal{U}\ket{\psi_{init}}$
    \item An execution of the program $M\mathcal{T}\mathcal{U}\ket{\psi_{init}}$ results in an output basis state $O_m$
    \item Update the sample distribution $\Lambda_{\mathcal{T}}$
\end{enumerate}

We repeat steps (1) to (5) according to the sample size $N_{E}$.

\paragraph{\textbf{Running Example (Test Execution with Projective Measurements):}}

In step (1), we initialize the quantum program with the  $\ket{\text{Default \gls{ps}}}$ from \cref{eq:approach_running_example} with the input state $\ket{000}$.
Then, assuming the same basis as in the running example of \cref{sec:test_cases}, we complete step (2) by applying a Hadamard gate to the end of our circuit before the measurement operation, according to \cref{sec:background}.
Now, with the basis transformation gates applied, we read all outputs from the program according to the basis transformation.
We always obtain states consisting of 0 and 1, such as $\ket{001}$ or $\ket{1100}$. 
However, since the basis is now $[H, \mathbbm{1}, \mathbbm{1}]$, we interpret 0 in the first qubit as $+$ and 1 as $-$, as introduced in \cref{sec:background}. 
Thus, when we measure the state $\ket{010}$, we interpret it as $\ket{+10}$.
In step (3), we perform the measurement, and in step (4), we obtain a result, such as $\ket{010}$, which we interpret as $\ket{+10}$. Finally, in step (5), we update the sample distribution. 
The sample distributions have the same form as the theoretical distributions but use sampled frequencies instead of ideal probabilities. 
We perform multiple samples and record how often each result occurs, which should approximate the theoretical distribution more closely as the sample size increases. 
For example, if we perform $N_E = 40$ samples after applying the basis transformation, we obtain the following randomly-generated sample distribution:

\begin{equation}
    \Lambda_{\mathcal{T}} = \Big\{0.175,0.30, 0, 0, 0, 0, 0.275, 0.25     \Big\}
\label{eq:running_example_reduced_sample_distribution}
\end{equation}

We then compare this sample distribution in \cref{eq:running_example_reduced_sample_distribution} with the theoretical distribution in \cref{eq:running_example_reduced_distribution}. 
Similarly, we perform the same sampling for the quantum program without the basis transformation, running it in the computational basis. 
In that case, we compare the sample distribution with the theoretical distribution in \cref{eq:running_example_reduced_distribution_default}.

\section{Experiment Design}
\label{sec:experiment_design}

This section presents the experiment design for the evaluation of our approach.

\subsection{Research Questions}
\label{sec:research_questions}

We pose three research questions: RQ1, RQ2, and RQ3:

\begin{enumerate}[label=\textbf{RQ\arabic*:},leftmargin=*]
    \item How efficiently and effectively do we obtain reduced program specifications through reduction?
    \item What impact does applying reduced program specifications have on the efficiency of \gls{qst}?
    \begin{enumerate}[label=\textbf{\alph*)}]
        \item How efficient is \gls{qst} with reduced program specifications?
        \item How does the reduction correlate with the \gls{qst} efficiency?
    \end{enumerate}
    \item What effect does applying reduced program specifications have on the effectiveness of \gls{qst}?
    \begin{enumerate}[label=\textbf{\alph*)}]
        \item How effective is \gls{qst} with reduced program specifications?
        \item How does the reduction correlate with the \gls{qst} effectiveness?
    \end{enumerate}
\end{enumerate}

\subsection{Overview of the Experiment Design}
\label{sec:experiment_overview}

Following the nomenclature of \citet{abcSoftwareResearch}, we conduct two laboratory experiments, illustrated in the overview \cref{tab:evaluation_overview}. 
The first (Experiment 1) applies the \HLIGHT{Reduction} component from the approach in \cref{fig:approach_diagram}, while the second (Experiment 2) applies the \HLIGHT{Application to Quantum Software Testing} component on a suite of 143 quantum programs. 
In Experiment 1, we run \cref{alg:reduction_algorithm} multiple times on the same quantum program, obtaining a distribution of the reduction rate metric and the runtime. 
We answer RQ1 with statistical assessments of the reduction rate distributions and the time cost of running \cref{alg:reduction_algorithm}.
In Experiment 2, we run the \HLIGHT{Application to Quantum Software Testing} component of \cref{fig:approach_diagram}, where we subject the \gls{sut} to mutation testing. 
We conduct two testing scenarios: one with a \HLIGHT{Reduced \gls{ps}}, derived from the median reduction rate resulting from Experiment 1, and another with a \HLIGHT{Default \gls{ps}}. 
The data gathered, consisting of the metrics of testing runtime and mutation score for each scenario, are used to evaluate the efficiency posed in RQ2 and the effectiveness in RQ3, respectively.

\begin{table}[tbp]
    \centering
    \caption{Overview table of the experimental evaluation.}
    \label{tab:evaluation_overview}
    \begin{tabular}{lllll}
        \toprule
       Research Question  & Approach Component       & Experiment ID         & Metric                  & Statistic  \\
        \midrule
        RQ1                        & Reduction    &   1          & Reduction Rate, Runtime & Mean, Standard Deviation  \\
        RQ2                        & Testing     &     2         & Speedup, Slowdown                 & Mean, Standard Deviation  \\
        RQ3                        & Testing     &     2         & Mutation Score          & Mean, Standard Deviation  \\
    \bottomrule
    \end{tabular}
\end{table}

\subsection{Random Baseline}

In our evaluation, we compare our Greedy algorithm to a random search baseline, which we call Random reduction. 
For a given quantum program, we first evaluate using the Greedy approach, resulting in a number of executions of the reduction-circuit. 
Our random baseline algorithm then also performs the same number of executions but with random samples of the exhaustive basis search space. 
In the end, Random selects the basis that minimizes the rank (i.e., maximizes the reduction). 
We then compare the results of Greedy to Random in both Experiment 1 and Experiment 2.

\subsection{Study Subjects (Quantum Programs)}
\label{sec:study_subjects}

Our approach, which obtains specification reductions through basis transformations, is independent of any specific basis. 
However, by selecting mixed Hadamard bases, we expect some quantum algorithms to perform well with reductions while others may not.
To avoid bias toward programs expected to perform well with our approach, we select study subjects that satisfy the output criterion in \cref{eq:real_uniform_state_vector} to a varying extent.
These are programs where the final state vectors are uniform with real amplitudes, approximating eigenstates of the mixed Hadamard bases.
Thus, different quantum algorithms satisfy the output criterion to varying extents based on their gate composition and semantics. 
Graph states, consisting only of Hadamard and controlled Z gates, meet this criterion often~\citep{Hein_2004}. 
In contrast, while Quantum Fourier transforms give uniform superpositions, they yield complex amplitudes, and quantum walks have the opposite problem~\citep{QCQIBook,Venegas_Andraca_2012_Quantum_Walk,gerhardt2003continuoustime}. 

Additionally, since large reductions increase runtime overhead due to the additional Hadamard gates in the circuit for the projective measurements, we aim to test our approach against programs varying in depth and in the number of output states (\(N_{ps}\)).

Based on these criteria, we select 145 quantum programs divided into four categories: 
\begin{itemize}
    \item \gls{grov} with 48 programs, noted for their high program depth.
    \item \gls{qwalk} with 39 programs, characterized by a lower satisfaction of the output criterion.
    \item \gls{gs} comprising 14 programs, marked by their high number of output states.
    \item \gls{var} with 44 programs, characterized by representing multiple quantum algorithms.
\end{itemize}
These categories, detailed in \cref{tab:study_subjects_programs}, source from \citet{johnston2019programming, quetschlich2022mqtbench, Hein_2004}. 
In our selection of these four categories, we aimed to obtain a representative sample, following the nomenclature of \citet{baltes2021sampling}, of algorithms where our output criterion is satisfied to varying extents. 
This allows us to evaluate our approach across both cases where it may be highly effective and those where it may work to a lesser extent. 
While we exclude optimization algorithms like VQE and QAOA, as their final state vectors yield single output values that do not satisfy the output criterion, the general applicability of our approach to other programs depends on the choice of gates in the search space. 
These gates determine the possible basis transformations applied to reduce program specifications.

In general, and in our evaluation, the runtime cost of classical simulations of quantum programs scales exponentially with the number of qubits.
Therefore, to maximize the depth of \gls{grov} programs and the number of outputs from \gls{gs} programs, we set upper thresholds for depth and qubit counts, as detailed in the following.
We select our thresholds based on the points where adding one qubit doubles the runtime and exceeds our budget.
This budget sets a maximum evaluation runtime of one week, aiming to use as many or more qubits and depths than other \gls{qst} evaluations in related work~\citep{quitoASE21tool, Muskit, QMorph, honarvar2020property}.

\begin{table}[tbp]
    \centering
    \caption{Overview table of the study subjects in our evaluation.}
    \label{tab:study_subjects_programs}
    \begin{tabular}{lllll}
        \toprule
       \textbf{Categories}  & \textbf{\gls{grov}} & \textbf{\gls{gs}} & \textbf{\gls{qwalk}} & \textbf{\gls{var}} \\
        \midrule
        \textbf{\#Programs} & 48 & 14 & 39 & 44 \\
        \textbf{Qubit-Range} & [6, 9] & [3, 16] & [3, 5] & [2, 8] \\
        \textbf{Depth-Range} & [2, 684] & [5, 18] & [1, 154] & [3, 70] \\
        \textbf{Rank $N_{ps}$-Range} & [32, 256] & [2, 16384] & [1, 16] & [4, 16] \\
        \textbf{Program Variations} & Random Oracles & Ring Graphs & \# Walks & Inputs 0-3 \\
    \bottomrule
    \end{tabular}
\end{table}

\paragraph{\textbf{Grover Search (\gls{grov})}}

The Grover Search program takes a set of specified output states in a search oracle, then carries out iterations of Grover operations, marking and amplifying the probability amplitudes of the specified set of output states in the oracle~\citep{johnston2019programming}.

We establish our upper threshold by considering the empirical trial of the nine qubit \gls{grov} programs with depths of $684$ and runtime of $9$ hours.
To include multiple variations of these programs, going beyond this qubit count and depth becomes challenging, as runtimes double from this point for each added qubit. 
Thus, we use the depth of $700$ and $9$ qubits as the suitable upper thresholds for \gls{grov}.
With respect to these thresholds, we generate programs incrementally from 6 to 9 qubits and generate \num{360} program variations per qubit with random oracles from \num{256} possible oracles.
For each variation, we randomly select sets and perform Grover operations until non-input state amplitudes are below $10^{-4}$. 
In order to achieve this precision, we allow a non-optimal number of Grover iterations.
This allows our sample to simulate large depths, where many iterations are needed for the optimal case. 
Excluding two variations for exceeding depth $700$, we obtain $46$ study subjects from the $\gls{grov}$ category.

\paragraph{\textbf{Graph States (\gls{gs})}}

These programs represent a quantum state encoding of a graph $G=(V, E)$, with vertices $V$ and edges $E$~\citep{Hein_2004, PhysRevA.73.022334}.
This process yields low-depth \gls{gs} programs with consistent $2^{n-2}$ output states, making them ideal as large-rank programs.
In our selection, we focus on ring graphs for their edge set $E=\{(0,1), (1,2), \cdots, (n-1, 0)\}$, given a qubit count $n$. 

As with \gls{grov}, we use the empirical trial for the \num{16} qubit \gls{gs} program with a runtime of $4937s$.
Thus, from the qubit threshold of \num{16}, we generate programs incrementally from \num{3} to \num{16} qubits.

\paragraph{\textbf{Quantum Walk (\gls{qwalk})}}

We aim to include programs in our evaluation that do not necessarily meet the output criteria specified in \cref{eq:real_uniform_state_vector}. Discrete quantum walks, the quantum equivalents of classical random walks, are particularly suitable for this purpose as they typically do not converge to uniform superpositions~\citep{Venegas_Andraca_2012_Quantum_Walk, gerhardt2003continuoustime}.

Unlike the \gls{grov} and \gls{gs} programs, maximization of qubit counts and depths is not required for \gls{qwalk} programs. 
Our goal is to evaluate our approach using programs that may not satisfy our predefined output criteria. Consequently, we have chosen to include \gls{qwalk} programs with qubit counts ranging from 5 to 8, generating 12 variations per qubit count and varying the number of walks from 1 to 15 to ensure a diverse sample of program variations.

\paragraph{\textbf{Various (\gls{var})}}

In the \gls{var} category, we include programs from multiple quantum algorithms to evaluate our approach. 
We analyze final state vectors using the QCengine API~\citep{johnston2019programming} to select these programs. 
Initially, we consider 52 programs, but exclude simple demonstrations (e.g., Examples: 2-1, 2-2, 2-3, 3-1), programs with non-reducible single outputs (e.g., Examples: 3-4, 5-6, 10-2, 12-2, 12-4, 14-GT, 14-BV, 14-S), and any exceeding 16 qubits (e.g., Examples: 4-2, 11-2, 11-4, 11-6, 12-1). 
We detail the selection in \cref{tab:study_subjects_var}. 
For each program, we generate four manually verified valid input states from 0 to 3 to ensure adequate sample diversity.

\begin{table}[tbp]
    \centering
    \caption{Overview of the \gls{var} category study subjects used for our evaluation, sourced from~\citep{QCBookGentle}.}
    \label{tab:study_subjects_var}
    \begin{tabular}{llrrr}
        \toprule
       \textbf{Example id}  & \textbf{Program} & \textbf{\#Qubits} & \textbf{Depth} & $\bm{N_{ps}}$ \\
        \midrule
        \textbf{3-3} & Phase kickback & 3 & 3 & 4 \\
        \textbf{3-5} & Custom conditional-phase & 2 & 6 & 4 \\
        \textbf{4-1} & Basic teleportation & 3 & 4 & 8 \\
        \textbf{5-2} & Adding two quantum integers & 6 & 8 & 4 \\
        \textbf{5-3} & Add-squared & 6 & 14 & 4 \\
        \textbf{5-4} & Quantum conditional execution & 6 & 11 & 4 \\
        \textbf{5-5} & Quantum conditional phase flip & 5 & 11 & 8 \\
        \textbf{6-3} & Multiple flipped entries & 4 & 69 & 4 \\
        \textbf{10-1} & Phase Logic 1 & 4 & 11 & 8 \\
        \textbf{10-4} & Unsatisfiable 3-Sat & 7 & 41 & 8 \\
        \textbf{11-3} & Drawing into small tiles & 8 & 17 & 16 \\
        \textbf{12-3} & Shor step-by-step & 8 & 13 & 16 \\
    \bottomrule
    \end{tabular}
\end{table}

\subsection{Experiment Setup}
\label{sec:experiment_setup}

This section details the experimental setup for evaluating RQ1--RQ3, according to the overview in \cref{tab:evaluation_overview}.

\subsubsection{Experiment 1 for RQ1 Evaluation}
\label{sec:experiment_ID_1_eval}

In Experiment 1, we conduct $r_1$ repetitions of the reduction algorithm (\cref{alg:reduction_algorithm}) on each study subject. 
We first perform Experiment 1 with the Greedy approach, storing the number of objective function calls. 
Then, we perform Experiment 1 with the Random baseline, providing the number of objective function calls from the Greedy run as input. 

\paragraph{ \textbf{Metric RQ1 -- Reduction Rate}:}

At the end of each repetition of Experiment 1, we calculate the \textit{Reduction Rate} $R$, defined as the ratio between the ranks $\Tilde{N}_{ps}$ and $N_{ps}$ of the \HLIGHT{Reduced \gls{ps}} and \HLIGHT{Default \gls{ps}}, respectively.

\begin{equation}
    R = 100(1 - \frac{\Tilde{N}_{ps}}{N_{ps}})
    \label{eq:metric_reduction_rate}
\end{equation}

While \citet{ArcuriBriand} recommends 30 repetitions as a rule of thumb, the efficiency of our approach allows us to exceed this by performing $r_1 = 100$ repetitions of Experiment 1 for each program. 
This is feasible because our approach requires at most $\mathcal{O}(n^2)$ objective function calls.
After conducting $r_1$ repetitions for each of the 143 study subjects across the four program categories, we collect four sets of \textit{Reduction Rate Distributions} to evaluate the effectiveness of our approach.

\paragraph{\textbf{Statistical Analyses (Experiment 1):}}
\label{sec:statistical_analysis_experiment1}

We apply the Mann-Whitney U test to statistically analyze the significant differences between the Greedy approach and the Random baseline. 
The null hypothesis $H_0$ states that the Greedy and Random distributions are the same, while the alternative hypothesis $H_1$ states that the Greedy distribution is stochastically different from the Random distribution. 
To assess the strength of statistical significance, we use the Vargha-Delanay effect size $\hat{A}_{12}$ \cite{ArcuriBriand, VarghaDelanay}. 
If $\hat{A_{12}}=0.5$, there is no difference between Greedy and Random, while $\hat{A}_{12} > 0.5$ favors the Greedy approach. 
For a given effect size, we define four nominal magnitude categories based on the scaled value $\hat{A}_{12}^{scaled}=2(\hat{A}_{12} - 1/2)$~\citep{HessRobustConfidence,laaber2024evaluating}.
The nominal magnitude of the effect sizes are: 

\begin{enumerate}
    \item \textit{Negligible} (N) for $|\hat{A}_{12}^{scaled}|<0.147$
    \item \textit{Small} (S) for $0.147 \leq |\hat{A}_{12}^{scaled}|\leq 0.33$
    \item \textit{Medium} (M) for $ 0.33 \leq |\hat{A}_{12}^{scaled}|<0.474$
    \item \textit{Large} (L) for $|\hat{A}_{12}^{scaled}|\geq 0.474$
\end{enumerate}
We consider the results statistically significant if the p-value $\leq 0.05$ and the effect size $\hat{A}_{12}$ is greater than the magnitude (N).

\subsubsection{Experiment 2 for RQ2--RQ3 Evaluation}

\label{sec:experiment_ID_2_eval}

Current benchmark suites of quantum programs are still in their infancy, only containing a limited number of faulty programs~\citep{zhao2020quantum}.
Therefore, and similar to other quantum software testing works~\citep{quitoASE21tool,Muskit,honarvar2020property}, we seed gate faults representing typical quantum program operations, such as bit flips, phase flips, and rotations around the y-axis (as introduced in \cref{tab:background_gates}). 
Faults of this kind can cause failures in the expected functionality of a quantum program, resulting in a bug. 

In Experiment 2, we apply mutation testing to obtain a faulty program suite using three distinct single qubit mutation operators: a bit flip ($X$-gate), a phase flip ($Z$-gate), and a probability perturbation along the y-axis ($R_y$-gate). 
These operators enable us to simulate all possible single qubit faults by covering the Bloch sphere~\citep{QCQIBook}. 
We manually generate mutations because, to our knowledge, no existing mutation tools support the projective measurements required by our program specifications~\citep{Muskit,QMutPy}.
We insert these mutation operators between the main quantum program $\mathcal{U}$ (see \cref{fig:diagram_background}) and the projective measurement $\mathcal{M}$ to simulate single gate faults at the end of the circuit.
To ensure a comprehensive representation of gate insertions across the 16 potential locations (considering the maximum qubit count of 16 in our study subjects),
we create \num{90} mutants at random locations for each mutation operator.
For $R_y$, we also need to include a representative sample of the angle $\theta$.
We achieve this by \num{30} randomly assigned values of the angle $\theta$.
Consequently, we generate 2145 mutants (15 mutants for each of the 143 programs).
For each mutant, we perform $r_2$ repetitions of the \HLIGHT{Application to Quantum Software Testing} component in \cref{fig:approach_diagram} for each type of \gls{ps}, default and reduced.
For the \HLIGHT{Reduced \gls{ps}}, we apply our basis-dependent test case generation and execution procedures in \cref{sec:test_cases} and \cref{sec:test_execution}. 
Thereby obtaining sample distributions ready for the \HLIGHT{Test Assessment} step. 
Following the best practice by \citet{ArcuriBriand}, we perform $r_2=30$ repetitions of Experiment 2 for all mutants, obtaining $30 \times 2145$ test results for the three cases of \HLIGHT{Default \gls{ps}} and \HLIGHT{Reduced \gls{ps}} with both Greedy and random baseline. 
Furthermore, we aim for a minimal sample size rule that is comparable across program specification sizes. 
Inspired by the 10 times rule~\citep{10timesrule}, we set $\gamma=10$ from \cref{eq:sample_size}, performing 10-fold samples of the size of the \gls{ps}.

Finally, as a technical note, Qiskit optimizes circuit simulations by classically sampling the same state vector using the \textit{shots} variable~\citep{QiskitCite} after performing matrix multiplications. 
While this is an effective way to simulate quantum programs, it does not allow for an accurate evaluation of our approach, as we must measure the realistic circuit runtime and not the runtime of classical probability sampling. 
Therefore, we run each circuit execution with a single shot to accurately measure the runtime.

\paragraph{\textbf{Test Assessment:}}

We apply the following two test oracles from related work~\citep{quitoASE21tool, Muskit,QUSBT,honarvar2020property,QMorph,Wang2021QDiffDT}, incorporating projective measurements:

\textbf{\gls{woo}.} The first test oracle, $f_{woo}$, validates whether an output basis state $\ket{m}$, obtained from a test execution, is a basis state of the \gls{ps} or not:
\begin{equation}
    f_{woo}\Big[\ket{m}, TC(\mathcal{T})\Big] =\begin{cases}
        1 & \text{if } \ket{m} \notin Outputs  \\
        0 & \text{else}
    \end{cases}
    \label{eq:woo_definition}
\end{equation}
In \cref{eq:woo_definition}, $f_{woo}$ takes the observed output state $\ket{m}$ and a test case (see \cref{eq:test_case_definition}) as inputs and fails ($f_{woo}=1$) if the output cannot be found among the output states of the \gls{ps}.

\textbf{\gls{pdo}.}
The second oracle, \gls{pdo}, conducts a chi-square hypothesis test to compare the sample distribution $\Lambda_{sample}$ against the theoretical distribution $\mathcal{P}_{ps}$ from the \gls{ps}~\citep{Pearson1900}:
\begin{equation}
    f_{pdo}\Big[\mathcal{P}_{ps}(\mathcal{T}), \Lambda_{sample}(\mathcal{T}, N_E)\Big]=\begin{cases}
        1 & \text{if } p_{value} < \alpha  \\
        0 & \text{otherwise}
    \end{cases}
    \label{eq:oracle_pdo_definition}
\end{equation}
In \cref{eq:oracle_pdo_definition}, the null hypothesis states that there is no significant difference between the sample and theoretical distributions, while the alternative hypothesis suggests that such a difference is present according to the significance level $\alpha=0.05$. 

Now, after applying the two test oracles (\cref{eq:woo_definition} and \cref{eq:oracle_pdo_definition}), we obtain the \HLIGHT{Test Assessment Result}.

\paragraph{ \textbf{Metric RQ2 -- Runtime Speedup \& Slowdown:}}
We define the \textit{Test Runtime} as the time required to run the approach, including the \HLIGHT{Test Execution} and \HLIGHT{Test Assessment} steps (whether the WOO or PDO oracle fails). 
For the \HLIGHT{Default \gls{ps}} case, we only run the latter two steps, as no \HLIGHT{Reduction} is performed.
Thus, we define the \textit{Default Test Runtime} ($T_{Def}$) and the \textit{Reduction Test Runtime} ($T_{Red}$), from which we calculate the runtime metrics \textit{Speedup} and \textit{Slowdown} to evaluate RQ2:
\begin{align}
Speedup &= \frac{T_{Def}}{T_{Red}} & Slowdown &= -\frac{T_{Red}}{T_{Def}}
\label{metric:speedupslowdown}
\end{align}
In \cref{metric:speedupslowdown}, we define the \textit{Speedup} (\%) as the ratio between the default and reduction test runtime. 
Similarly, we define the \textit{Slowdown} (\%) as the ratio between the reduction and default test runtime:
If $T_{Def}>T_{Red}$, we obtain a Speedup in test runtime from using the \HLIGHT{Reduced \gls{ps}}; otherwise, the \HLIGHT{Reduced \gls{ps}} results in a Slowdown of test runtime.

\paragraph{ \textbf{Metric RQ3 -- Mutation Score}:}

To evaluate RQ3, we define the \textit{Mutation Score}~\citep{OriginalMutationTestingPaper,Muskit,QMutPy} as the percentage of killed mutants for a given quantum program.
\begin{equation}
    MUT = 100\frac{\# Killed\; Mutants}{15}
\label{eq:mutation_score}
\end{equation}
In \cref{eq:mutation_score}, we divide the number of killed mutants by $15$, as we have $15$ mutants per quantum program.

\paragraph{\textbf{Statistical Analyses (Experiment 2):}}
\label{sec:statistical_analysis_experiment2}

In addition to the statistical analyses, p-value, and effect size defined in \cref{sec:statistical_analysis_experiment1}, we employ the Kruskal-Wallis test~\citep{Kruskal} for comparing more than two approaches.
When Kruskal-Wallis finds significant differences between the distributions, we employ pairwise \gls{mwu} tests along with effect size measures to assess which distributions differ.
We compute the Spearman rank coefficient to determine correlation between reduction rate and runtime efficiency or mutation score effectiveness, whose value is between -1 and 1, where -1 indicates no monotonic relationship and 1 indicates the strongest relationship. 
We adopt the following magnitude categories for correlation in order to draw categorical conclusions about our data~\citep{schober2018correlation}:
\begin{enumerate}
    \item \textit{Negligible} if $r_s \in [0.00,0.10]$
    \item \textit{Weak} if $r_s \in [0.10,0.39]$
    \item \textit{Moderate} if $r_s \in [0.40, 0.69]$
    \item \textit{Strong} if $r_s \in [0.70, 0.89]$
    \item \textit{Very Strong} if $r_s \in [0.90, 1.0]$
\end{enumerate}

\subsection{Threats to Validity}
\label{sec:threats_to_validity}

We structure the threats to validity along the types: construct, internal, and external~\citep{Threats_to_Validity_Dag}.

\paragraph{\textbf{Construct Validity:}}
The biggest threat to the construct validity is whether our metrics represent efficiency and effectiveness.
While runtime is a straightforward measure of efficiency, it might not capture all dimensions of testing effectiveness, such as the time taken to set up a test environment. 
This includes obtaining a \HLIGHT{Default \gls{ps}} for the correct program by experimental or mathematical methods~\citep{Efficient_State_Tomography,zhou2019applied}. 
This may be alleviated by constructing efficient software tools and processes for ease of application. 
We discuss the challenge of obtaining a \gls{ps} in a later section.
In addition, although mutation score provides a direct measure of the test case's ability to identify errors, our mutation operator set of the X gate, Z gate, and $R_y$ may not represent all types of faults. 
Although the set forms a universal single-qubit gate, we may not detect certain two-qubit gate faults with our reduction approach, which are essential for entanglement in quantum computation~\citep{QCQIBook}. 

\paragraph{\textbf{Internal Validity:}}
When it comes to internal validity, a crucial threat comes from potential new faults being introduced from the reduction basis being added onto the end of the circuit in order to perform the projective measurement. 
This threat may cause systematic faults that are introduced by the reduction approach and not by controlled experimental design, leading to a falsely high mutation score for the reduction approach. 
This can occur by the oracles becoming over-sensitive. 
We mitigated this threat by calibrating the experiment by performing a fault-free run of the Default and the reduction approach and observing negligible failure rates.
In addition, the ten times rule, which inspired our choice of sample size, has been found to give inadequate sample estimates~\citep{10timesrulecriticism}.  
While our calibration run helps mitigate this threat, we still may fail to detect subtle distribution perturbations caused by \(R_y\) faults through inadequate sample sizes.

\paragraph{\textbf{External Validity:}}

Our approach is limited to specific program categories in addition to our choices for qubit and depth ranges. 
In addition, although we conduct our evaluation using Qiskit, the generalizability of our results to other quantum software development kits like CirQ or Forest is not significantly threatened. 
This is because our approach primarily requires gate operations and computational basis measurements, which facilitate the projective measurements essential to our methodology. 
However, these SDKs must be gate-based to ensure the generality of our results.

As real quantum devices in the current NISQ era contain noise, a crucial threat to external validity is our usage of an ideal simulator. 
Thus, if noise were present in our experiment, this noise would also cause faulty operations in the circuit, which we would treat as faults of the quantum programs. 
To effectively eliminate this threat, we see two mitigation methods. 
The first is by modifying the oracles to accept noise tolerance thresholds as inputs, then using a known noise model representing the backend to be tested. 
Or by using a single \gls{pdo} type oracle modified to accept all possible output states, effectively eliminating the \gls{woo} oracle. 
The second is to apply noise mitigation techniques~\citep{Cai_2023} and, in the future, error correction~\citep{ShorErrorCorr}. 
For our experiment, we assume an ideal simulator and discuss the former cases in our future work.
While our results are specific to the program categories in our experiment, our evaluated program categories represent real-world quantum programs, such as Grover searches, projected to be a crucial quantum algorithm for the quadratic speedup of unstructured search~\citep{Khadiev_2023,mcinroy2024benchmarking,wu2024efficiency,fernández2024implementing,zhao2024gqhan}.
However, a common challenge in Grover search is the construction of the search oracle, which is domain specific~\citep{seidel2021automatic,Sinitsyn_2023}. 
While our results hold for the specific search oracle construction applied in our experiment, we cannot claim that our results generalize to all types of search oracles.
Given that Grover search is a type of quantum walk on a bipartite graph~\citep{santos2016szegedys}, we suggest our reduction approach can be applied to certain types of discrete quantum walks that satisfy our output criterion, such as applied in hypercube quantum search~\citep{Pillin_2023}, where the output states are uniform superpositions.

Next to consider is the generality of our results to programs with large specification sizes, such as our \gls{gs} programs. 
Graph states, including cluster and complete graphs, often meet our output constraints effectively by design~\citep{gori2024gaussian,kaldenbach2023mapping,Li_2024}. 
Therefore, our evaluation of ring graphs provides a strong basis for extending our results to both complete and cluster graphs.

\section{Results and Analyses}
\label{sec:results_and_analyses}

In this section, we present the experimental results for each research question.

\subsection{RQ1: Reduction Efficiency and Effectiveness}
\label{sec:results_rq1}

\begin{table}[tbp]
    \centering
\caption{
The gray columns \textbf{Greedy [\%]} and \textbf{Random [\%]} show the overall average reduction rates and standard deviations. 
The \textbf{Different (X) [\%]} columns show the success rates of \gls{mwu} tests and effect size categories (S, M or L) defined in \cref{sec:statistical_analysis_experiment1}, depicting the percentages where we find Greedy and Random are significantly different, in the format of x/y, where x is the percentage where Greedy outperformed Random, and y is the reverse. \textbf{Different (NN) [\%]} shows the sum of the success rates for significant effect size categories (S), (M) or (L). }
\label{tab:rq1_reduction_rate_summary}
\resizebox{\columnwidth}{!}{%
\begin{tabular}{lll>{\columncolor{lightgray}}r>{\columncolor{lightgray}}rrrrrrr}
\toprule
\textbf{Category} & \textbf{\#Qubits} & \textbf{Depth} & \textbf{Greedy [\%]} & \textbf{Random [\%]} & \textbf{Different (NN) [\%]} & \textbf{Different (S) [\%]} & \textbf{Different (M) [\%]} & \textbf{Different (L) [\%]} & \textbf{Equal [\%]} \\
\midrule
              All &           [2, 15] &       [3, 684] &         52.7 $\pm$ 28.2 &      46.7 $\pm$ 28.5 &        62.2/2.8 &       16.8/2.1 &        9.1/0.7 &       36.4/0.0 &                    35.0 \\
             \gls{grov} &            [6, 9] &      [30, 684] &          66.8 $\pm$ 8.4 &       62.4 $\pm$ 8.1 &        82.6/0.0 &        8.7/0.0 &       13.0/0.0 &       60.9/0.0 &                    17.4 \\
            \gls{qwalk} &            [3, 5] &      [14, 154] &         27.4 $\pm$ 23.1 &      22.5 $\pm$ 22.8 &        35.9/5.1 &       12.8/2.6 &        7.7/2.6 &       15.4/0.0 &                    59.0 \\
              \gls{var} &            [2, 8] &        [3, 70] &         49.8 $\pm$ 29.8 &      42.0 $\pm$ 30.2 &        54.5/4.5 &       27.3/4.5 &        6.8/0.0 &       20.5/0.0 &                    40.9 \\
               \gls{gs} &           [3, 15] &        [5, 17] &         83.7 $\pm$ 16.9 &      75.1 $\pm$ 22.3 &        92.9/0.0 &       21.4/0.0 &        7.1/0.0 &       64.3/0.0 &                     7.1 \\
\bottomrule
\end{tabular}
}
\end{table}


In the \textbf{Greedy} and \textbf{Random} columns in \cref{tab:rq1_reduction_rate_summary}, we observe that the Greedy approach achieved $52.7\%$ average reduction rate, while random achieved $46.7\%$. 
The \gls{gs} category exhibits the largest reductions at $83.7\%$ for Greedy and $75.1\%$ for Random.
The worst performing category is the \gls{qwalk} with $27.4\%$ for Greedy and $22.5\%$ for Random. 
Reduction rates vary widely across program categories, as indicated by the large standard deviations. 
This can be explained due to the wide qubit ranges. 
For example, the average runtime for the 16 qubit programs of \gls{gs} is \num{44088}s while only \num{65}s for the \num{6} qubit programs. 
More detailed average results by qubit count are found in the appendix~\citep{oldfield_2024_11191215}.
The Greedy method typically outperforms Random, with the largest gap at \num{8.6} (pp) in the \gls{gs} category between the approaches, favoring Greedy. 
We find the smallest gap in the \gls{grov} category, showing more similar performances. 
The lowest average reduction rates are in the \gls{qwalk} category, with $27.4\%$ for Greedy and $22.5\%$ for Random.


We compare the Greedy and Random approaches using the statistical tests shown in the last five columns. 
For the \textbf{Different (NN) [\%]} column in \cref{tab:rq1_reduction_rate_summary}, we find that for all categories, Greedy is better than Random in $62.2\%$ of cases, while Random is only better than Greedy $2.8\%$ of the time. 
By category, we find that Greedy outperformed Random with a large margin for \gls{grov} and \gls{gs} in $82.6\%$ and $92.9\%$ of cases, with Random never being better. 
While for \gls{qwalk} and \gls{var}, Greedy is better for $54.5\%$ and $35.9\%$ of cases, with Random being better for only $5.1\%$ and $4.5\%$. 
Moving into the \textbf{Different (X) [\%]} columns where (X) is the effect size magnitude category. 
Here, we find that more than $60\%$ of the magnitude categories for \gls{grov} and \gls{gs} are of the type (L). 
For \gls{qwalk} and \gls{var}, however, the magnitudes are more evenly spread out between (S) and (L), with a slightly lower occurrence of (M). 
The last column, \textbf{Equal [\%]}, indicates the occurrence of tests where Greedy and Random performed the same.
Here, we find that for all categories, Greedy and Random are equal in $35.0\%$ of cases.
By category, we observe, in accordance with the results in the \textbf{Different (X)} columns, that Greedy and Random performed equally in $7.1\%$ of cases for \gls{grov} and $17.4\%$ of cases for \gls{gs}. 
In addition, we find equal performance for a clear majority of the cases at $59\%$ for \gls{qwalk} and $40.9\%$ for \gls{var}. 

\begin{table}[tbp]
    \centering
\caption{Summary of runtime results. The gray columns \textbf{Greedy [ms]}  and \textbf{Random [ms]} show the average reduction runtime and standard deviations. The remaining columns to the right are described as in \cref{tab:rq1_reduction_rate_summary}. }
\label{tab:rq1_runtime_summary}
\resizebox{\columnwidth}{!}{%
\begin{tabular}{lll>{\columncolor{lightgray}}r>{\columncolor{lightgray}}rrrrrrr}
\toprule
\textbf{Category} & \textbf{\#Qubits} & \textbf{Depth} & \textbf{Greedy [ms]} & \textbf{Random [ms]} & \textbf{Different (NN) [\%]} & \textbf{Different (S) [\%]} & \textbf{Different (M) [\%]} & \textbf{Different (L) [\%]} & \textbf{ Equal [\%]} \\
\midrule
              All &           [2, 15] &       [3, 684] &      732.2 $\pm$ 4013.7 &   743.9 $\pm$ 4083.0 &       18.2/68.5 &        6.3/9.8 &       4.2/14.7 &       7.7/44.1 &                    13.3 \\
             \gls{grov} &            [6, 9] &      [30, 684] &       484.8 $\pm$ 600.2 &    471.2 $\pm$ 587.3 &       19.6/63.0 &       8.7/15.2 &       4.3/26.1 &       6.5/21.7 &                    17.4 \\
            \gls{qwalk} &            [3, 5] &      [14, 154] &         70.8 $\pm$ 30.2 &      71.9 $\pm$ 30.0 &        7.7/87.2 &       2.6/10.3 &        5.1/5.1 &       0.0/71.8 &                     5.1 \\
              \gls{var} &            [2, 8] &        [3, 70] &         53.8 $\pm$ 32.3 &      56.3 $\pm$ 33.7 &       29.5/54.5 &        6.8/0.0 &        4.5/9.1 &      18.2/45.5 &                    15.9 \\
               gs &           [3, 15] &        [5, 17] &    5555.3 $\pm$ 11813.5 & 5711.7 $\pm$ 11995.4 &        7.1/78.6 &       7.1/21.4 &       0.0/21.4 &       0.0/35.7 &                    14.3 \\
\bottomrule
\end{tabular}
}
\end{table}

The runtime comparison in \cref{tab:rq1_runtime_summary} indicates that Greedy and Random methods have similar average runtimes across all programs, from the \textbf{Greedy [ms]} and \textbf{Random [ms]} columns, displaying runtime in milliseconds. 
This is expected, as described in \cref{sec:experiment_ID_1_eval}, we design Experiment 1 such that Greedy and Random perform the same number of objective function calls. 
However, the statistical tests reveal that Greedy still shows lower average runtimes overall in the \textbf{Different (NN) [\%]} column, signifying Greedy slightly outperforming Random in $68.5\%$ of the cases, while Random is better in $18.2\%$ of the cases. 
We also find these results to be mostly consistent by category and evenly spread out across magnitude categories.

\summarybox{RQ1 Summary}{
We find the reduction approach to be generally efficient and effective, with the Greedy approach outperforming the Random one in average reduction rate, runtime, and consistency. 
However, the results vary by category: \gls{grov} and \gls{gs} categories achieve the highest and most consistent reduction rates, while \gls{qwalk} and \gls{var} exhibit significantly lower performance. 
Runtimes across categories favor the Greedy over the Random approach.
}

\subsection{RQ2: Impact of Reduction on \gls{qst} Efficiency}
\label{sec:results_rq2}

Here, we present the results for RQ2a and RQ2b.

\subsubsection{RQ2a: Test Efficiency}
\label{sec:results_rq2a}


\begin{table}[tbp]
    \centering
    \caption{Summary of testing runtime results for all program categories. To the right of the gray columns, the headers \textbf{Default--Greedy}, \textbf{Default--Random} and \textbf{Greedy--Random} show the results of the statistical tests in the \textbf{Different (X)} and \textbf{Equal} columns for the respective pair of approaches. 
For each approach comparison, we show the success rate of pairwise statistical tests in the same column format as for RQ1 with \textbf{Different} and \textbf{Equal} columns, but only including the (NN) effect sizes.}
    \label{tab:rq2_summary_test_runtime_table}
\resizebox{\columnwidth}{!}{%
\begin{tabular}{lll>{\columncolor{lightgray}}r>{\columncolor{lightgray}}r>{\columncolor{lightgray}}rrrrrrrrrrrrrrr}
\toprule
 & &  & &  & & \multicolumn{2}{c}{\textbf{Default--Greedy}} & \multicolumn{2}{c}{\textbf{Default--Random}} & \multicolumn{2}{c}{\textbf{Greedy--Random}} \\
\cmidrule(l{1pt}r{1pt}){7-8}
\cmidrule(l{1pt}r{1pt}){9-10}
\cmidrule(l{1pt}r{1pt}){11-12}
\textbf{Category} & \textbf{\#Qubits} & \textbf{Depth} & \textbf{Default [s]} & \textbf{Greedy [s]} & \textbf{Random [s]} & \textbf{Different (NN) [\%]} & \textbf{Equal [\%]} & \textbf{Different (NN) [\%]} & \textbf{Equal [\%]} & \textbf{Different (NN) [\%]} & \textbf{Equal [\%]} \\
\midrule
              All &           [2, 15] &       [3, 684] &    169.9 $\pm$ 473.7 &        11.8 $\pm$ 61.2 &     15.6 $\pm$ 73.8 &                  93.7/0.0 &                 6.3 &                  93.0/0.0 &                 7.0 &                   3.5/0.0 &                96.5 \\
             \gls{grov} &            [6, 9] &      [30, 684] &    383.1 $\pm$ 503.3 &       33.4 $\pm$ 104.2 &    36.6 $\pm$ 113.5 &                 100.0/0.0 &                 0.0 &                 100.0/0.0 &                 0.0 &                   2.2/0.0 &                97.8 \\
            \gls{qwalk} &            [3, 5] &      [14, 154] &        2.9 $\pm$ 4.2 &          1.0 $\pm$ 2.0 &       1.0 $\pm$ 2.1 &                  87.2/0.0 &                12.8 &                  89.7/0.0 &                10.3 &                   5.1/0.0 &                94.9 \\
              \gls{var} &            [2, 8] &        [3, 70] &        1.1 $\pm$ 2.1 &          0.2 $\pm$ 0.3 &       0.3 $\pm$ 0.5 &                  90.9/0.0 &                 9.1 &                  86.4/0.0 &                13.6 &                   0.0/0.0 &               100.0 \\
               \gls{gs} &           [3, 15] &        [5, 17] &   464.8 $\pm$ 1027.7 &         7.7 $\pm$ 16.7 &    35.1 $\pm$ 100.4 &                 100.0/0.0 &                 0.0 &                 100.0/0.0 &                 0.0 &                  14.3/0.0 &                85.7 \\
\bottomrule
\end{tabular}%
}
\end{table}

In \cref{tab:rq2_summary_test_runtime_table}, we present the average test runtimes in seconds by program category in the gray columns \textbf{Default}, \textbf{Greedy} and \textbf{Random}.

\paragraph{\textbf{Test Runtimes.}} 
Overall, we find that the Default approach obtained an average test runtime of \num{169.9}s.
Then, after we apply the reduction, we achieve an average test runtime of \num{11.8}s for Greedy and \num{15.6}s for random.
By category, we find the same patterns. 
The Default approach generally requires a considerably higher runtime to achieve a test result when compared to the reduction approaches.

\paragraph{\textbf{Default--Greedy Comparison:}}
We consider the statistical tests, comparing Default to Greedy. 
In the \textbf{Different (NN)} column, we observe that the Default test runtimes are lower in 93.7\% of the cases overall. By category, we find that Default is slower for 100\% of the cases for \gls{grov} and \gls{gs}, while 90.9\% and 87.2\% for \gls{var} and \gls{qwalk}.
Conversely, we find no occurrences where Greedy is slower than Default.
From the \textbf{Equal} column, we find that the approaches are equally fast in $6.3\%$ of cases overall. by category, Greedy is always faster for \gls{grov} and \gls{gs} and equal to Random for $12.8\%$ and $9.1\%$ of cases for \gls{qwalk} and \gls{var}. 

\paragraph{\textbf{Default--Random Comparison:}}
For the comparison between Default and Random, we find very similar results to the comparison of Default with Greedy. 

\paragraph{\textbf{Greedy--Random Comparison:}}
In comparing Greedy to Random directly in the \textbf{Different (NN)} column, we find that Greedy is, for the most part, similar to Random. However, while Greedy is faster than Random only $3.5\%$ overall, by category, we find that Greedy is faster for $14.3\%$ of cases for \gls{gs}.

\begin{figure}[tbp]
    \centering
    \begin{subfigure}{0.48\textwidth}
        \centering
        \includegraphics[scale=0.39]{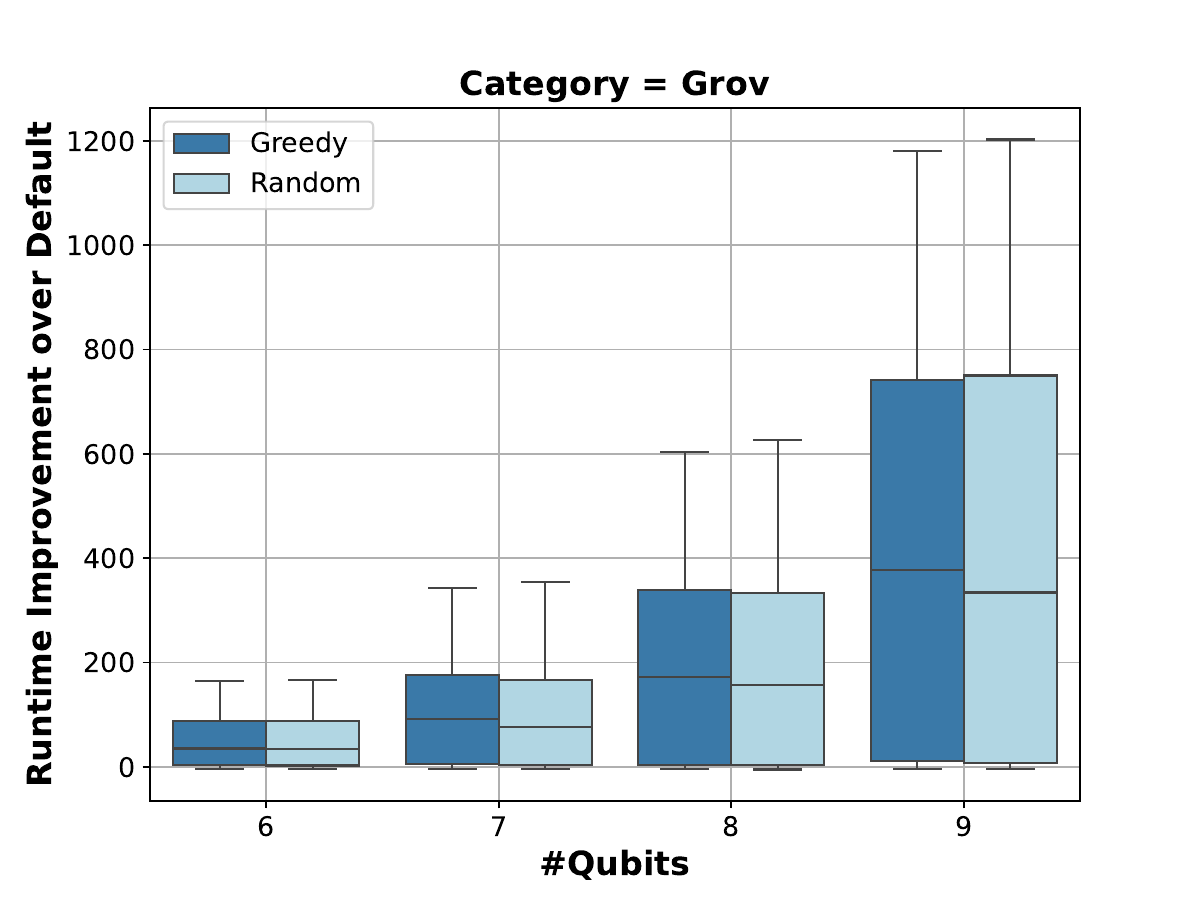}
        \caption{}
        \label{fig:rq2_boxplot_grov}
    \end{subfigure}
    \hfill 
    \begin{subfigure}{0.48\textwidth}
        \centering 
        \includegraphics[scale=0.39]{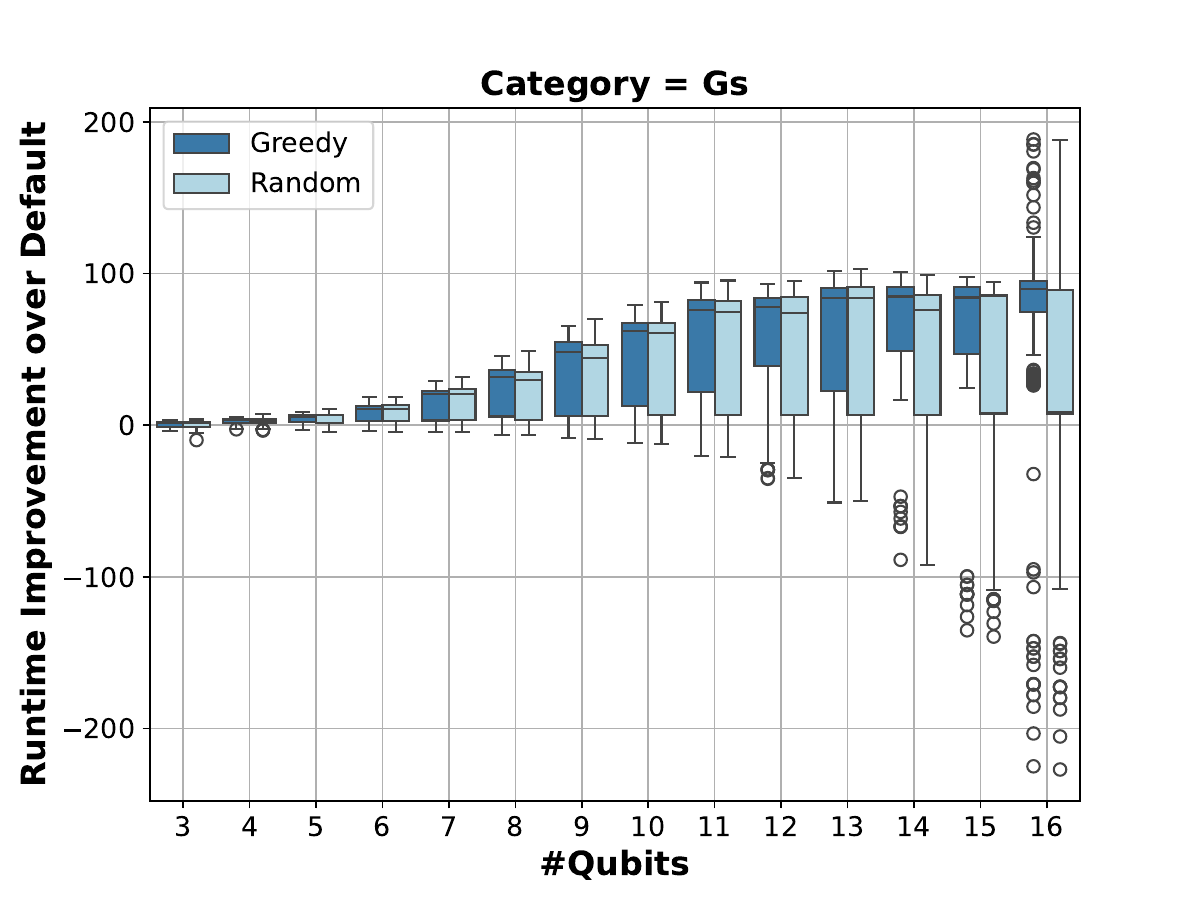}
        \caption{}
        \label{fig:rq2_boxplot_gs}
    \end{subfigure}
    
    \begin{subfigure}{0.48\textwidth}
        \centering
        \includegraphics[scale=0.39]{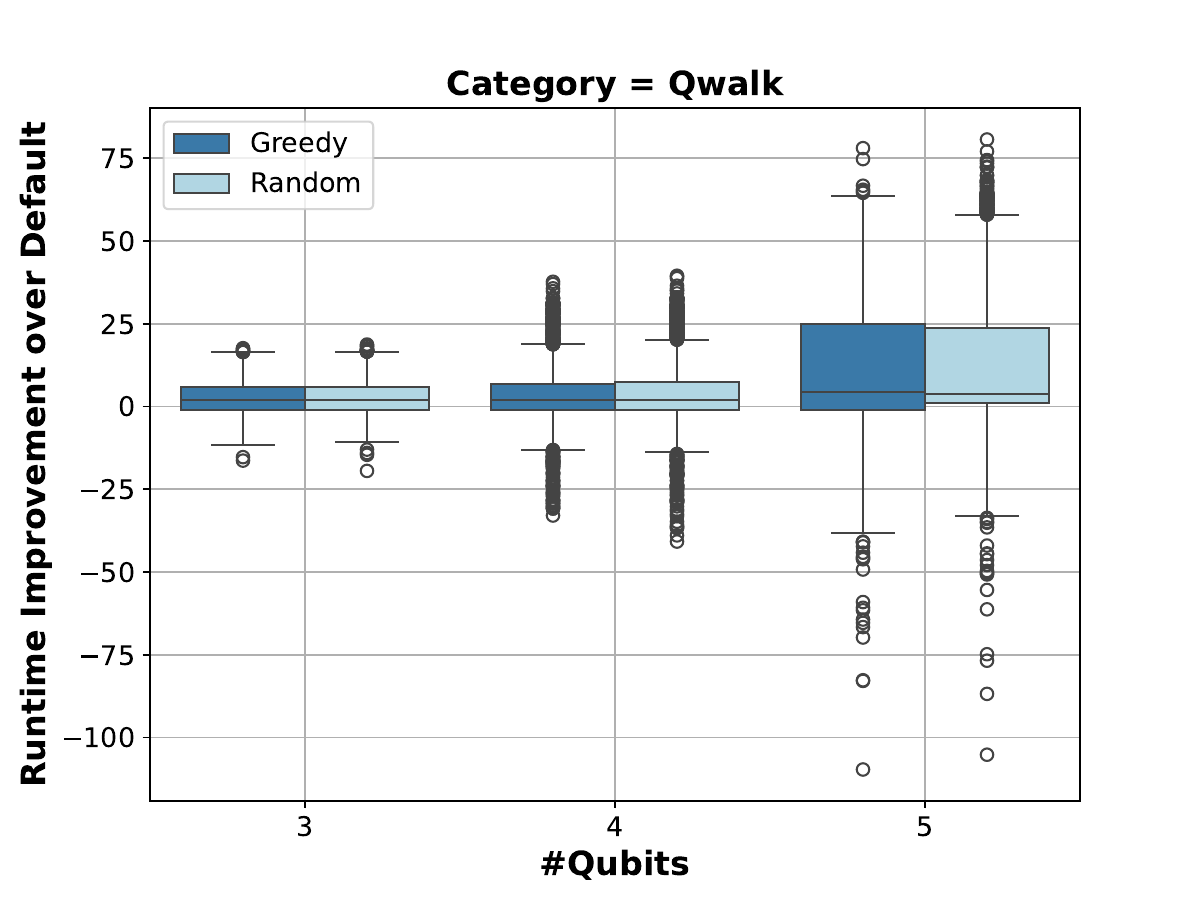}
        \caption{}
        \label{fig:rq2_boxplot_qwalk}
    \end{subfigure}
    \hfill 
    \begin{subfigure}{0.48\textwidth}
        \centering
        \includegraphics[scale=0.39]{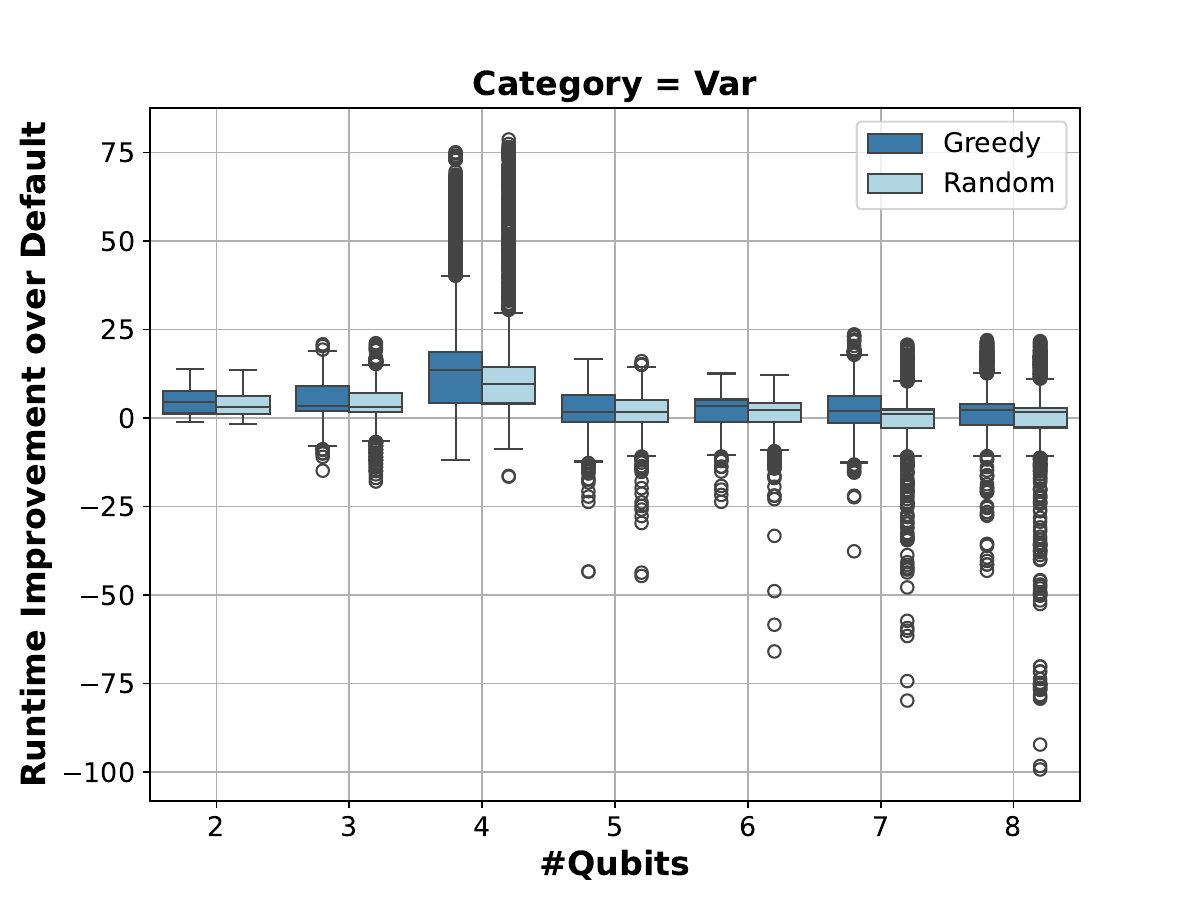}
        \caption{}
        \label{fig:rq2_boxplot_var}
    \end{subfigure}

    \caption{Boxplots of runtime improvement over Default when we apply reduction to \gls{qst} by the number of qubits. Speedups are above the x-axis, and Slowdowns are below.} 
    \label{fig:runtime_performance_plots} 
\end{figure}

\paragraph{\textbf{Relative Runtime improvement over Default:}}
Similar to RQ1, the relatively large standard deviations stem from the fact that the average overall runtime includes runs from various qubit counts and depths, resulting in large test time differences. 
To account for differences in qubit count and depth, we compute the \textbf{Runtime Improvement over Default}, (denoting the metrics Speedup or Slowdown defined in \cref{metric:speedupslowdown}), between the Default test runtime and the test runtime of the reduction approach Greedy or Random. 
We compute these ratios for each test result and depict the results as the boxplots in \cref{fig:rq2_boxplot_grov,fig:rq2_boxplot_qwalk,fig:rq2_boxplot_var,fig:rq2_boxplot_gs} by category.
First, from visual inspection, we observe a clear divide in improvement between the \gls{grov} and \gls{gs} categories and the \gls{qwalk} and \gls{var} categories. 
For the latter categories, we observe that all boxes have their lower whiskers reaching into the negative lower half of the y-axis, indicating that the reduction approaches performed worse than the Default for a significant number of cases. 
Still, however, the boxes stay above the y-axis, indicating that most ($72.5\%$ for \gls{qwalk} and $81.6\%$ for \gls{var}) experienced an increase in runtime due to reduction while the remaining $27.5\%$ for \gls{qwalk} and $18.4\%$ for \gls{var} experienced a slowdown. 
On the other hand, we find only $2.8\%$ and $6.9\%$ slowdowns \gls{grov} and \gls{gs}.
Furthermore, the runtime improvement over Default for \gls{grov} reaches median values of just below \num{400} for Greedy, followed by Random at closer to \num{300}. 
Given the same qubit range in \gls{gs}, i.e., from \num{6} to \num{9} qubits, the maximum median improvement over Default is around \num{50} for both Greedy and Random.

For the reductions of \gls{gs}, we can make two observations regarding the differences between the Greedy and Random approaches in \cref{fig:rq2_boxplot_gs}. 
First, the runtime improvements over Default for the reduction approaches seem to plateau at below 100. 
Second, the runtime improvements for Greedy become more consistently higher for higher qubit counts. 
We see this from the first quartile of Greedy increasing consistently for higher qubit counts, while the first quartile for Random stays constant. 

\subsubsection{RQ2b: Correlation Between Reduction and Efficiency}
\label{sec:results_rq2b}

\begin{table}[tbp]
    \centering
\caption{Spearman correlation $r_s$ between reduction rate and runtime improvement over Default.}
\label{tab:rq2_correlation_table}
\resizebox{\columnwidth}{!}{%
\begin{tabular}{lll>{\columncolor{lightgray}}r>{\columncolor{lightgray}}r>{\columncolor{lightgray}}r>{\columncolor{lightgray}}rll}
\toprule
 & &  & \multicolumn{2}{c}{\textbf{Correlation $r_s$}} & \multicolumn{2}{c}{\textbf{p-value}} & \multicolumn{2}{c}{\textbf{Correlation Magnitude}}  \\
\cmidrule(l{1pt}r{1pt}){4-5}
\cmidrule(l{1pt}r{1pt}){6-7}
\cmidrule(l{1pt}r{1pt}){8-9}
\textbf{Category} & \textbf{\#Qubits} & \textbf{Depth} & \textbf{Greedy} & \textbf{Random} & \textbf{Greedy} & \textbf{Random} & \textbf{Greedy} & \textbf{Random} \\
\midrule
              All &           [2, 15] &       [3, 684] &                        0.43442 &                     0.46043 &                    0.0e+00 &                 0.0e+00  & Moderate & Moderate \\
             \gls{grov} &            [6, 9] &      [30, 684] &                        0.08752 &                     0.06189 &                    1.7e-36 &                 5.0e-19  &  Negligible & Negligible\\
            \gls{qwalk} &            [3, 5] &      [14, 154] &                        0.29152 &                     0.29155 &                    0.0e+00 &                 0.0e+00 & Weak & Weak\\
              \gls{var} &            [2, 8] &        [3, 70] &                         0.3276 &                      0.3714 &                    0.0e+00 &                 0.0e+00 & Weak & Weak\\
               \gls{gs} &           [3, 15] &        [5, 17] &                        0.77997 &                     0.60156 &                    0.0e+00 &                 0.0e+00  & Strong & Moderate\\
\bottomrule
\end{tabular}
}
\end{table}

Next, we investigate the relationship between reduction and testing efficiency. 
Thus, in \cref{tab:rq2_correlation_table}, we display the Spearman correlations $r_s$ between the reduction rate and the runtime improvement over Default along with the correlation magnitude categories from \cref{sec:statistical_analysis_experiment1}.

Overall, we find a Moderate correlation of \num{0.43442} for Greedy and \num{0.46043} for Random. 
This indicates that overall, a significant relationship exists between a reduction rate and obtaining an improved test runtime over Default.
However, by category, we notice that \gls{grov}, which achieved the fastest runtimes after reduction, has a Negligible correlation.
Conversely, \gls{gs} achieved the strongest correlation.
Thus, given the very high depths of \gls{grov} programs compared to \gls{gs}, this means that high depth, paired with reduction, could also be important to achieve runtime improvements.
We also note for \gls{gs} that we find a Strong correlation when applying the Greedy approach, followed by Random with a Moderate correlation.
The categories \gls{qwalk} and \gls{var} exhibit Weak correlations. 
We see from the \textbf{p-value} column that all correlations are statistically significant with p-values $\sim 10^{-19}$.

\summarybox{RQ2 Summary}{
Our results show that our reduction approach improves \gls{qst} efficiency from \num{169.9}s with the Default approach to \num{11.8}s with reduction, but the impact varies by program characteristics. 
Both the Greedy and Random approaches generally improve on the Default approach, with Greedy favored in \gls{gs} programs for large specification sizes.
We also observe the largest improvements for \gls{grov} programs with high circuit depth. 
While \gls{qwalk}, and \gls{var} also experienced improvements to efficiency, we find no improvements to these categories in $27.5\%$ and $18.4\%$ of cases respectively, compared to $2.8\%$ for \gls{grov} and $6.9\%$ for \gls{gs}.
The correlation between reduction and runtime improvements is Moderate overall.
}

\subsection{RQ3: Impact of Reduction on \gls{qst} Effectiveness}
\label{sec:results_rq3}

Here, we present the results for RQ3a and RQ3b.

\subsubsection{RQ3a: Test Effectiveness}

\begin{table}[tbp]
    \centering
\caption{We show the average mutation score and standard deviations in the gray columns \textbf{Default [\%]}, \textbf{Greedy [\%]} and \textbf{Random [\%]}. In the \textbf{DGR} column, we show the p-value result from a Kruskal-Wallis test between the Default, Greedy and Random approaches.
To the right, we depict the results of the pairwise statistical tests between the respective approaches. }
\label{tab:rq3_summary_mutation_score}
\resizebox{\columnwidth}{!}{%
\begin{tabular}{llll>{\columncolor{lightgray}}r>{\columncolor{lightgray}}r>{\columncolor{lightgray}}rrrrlrrlrrl}
\toprule
& & & & & & & \textbf{DGR} & \multicolumn{3}{c}{\textbf{Default-Greedy}} & \multicolumn{3}{c}{\textbf{Default-Random}} & \multicolumn{3}{c}{\textbf{Greedy-Random}} \\
\cmidrule(l{1pt}r{1pt}){8-8}
\cmidrule(l{1pt}r{1pt}){9-11}
\cmidrule(l{1pt}r{1pt}){12-14}
\cmidrule(l{1pt}r{1pt}){15-17}
\textbf{Category} & \textbf{Mutant Type} & \textbf{\#Qubits} & \textbf{Depth} & \textbf{Default [\%]} & \textbf{Greedy [\%]} & \textbf{Random [\%]} & \textbf{p-value}  & \textbf{p-value}  & $\bf{\hat{A}_{12}}$  & \textbf{Magnitude}  & \textbf{p-value}  & $\bf{\hat{A}_{12}}$  & \textbf{Magnitude}  & \textbf{p-value}  & $\bf{\hat{A}_{12}}$  & \textbf{Magnitude}  \\
\midrule
              All &         all &           [2, 15] &       [3, 684] &       54.5 $\pm$ 22.5 &          74.7 $\pm$ 9.7 &       74.0 $\pm$ 9.9 &              0.0e+00 &             0.0e+00 &                      0.206 &                   (L) &             0.0e+00 &                      0.219 &                   (L) &             0.00129 &                       0.52 &                   (N) \\
              All &           X &           [2, 15] &       [3, 684] &       66.7 $\pm$ 42.1 &         65.5 $\pm$ 32.5 &      64.8 $\pm$ 32.9 &              1.8e-15 &             3.2e-12 &                      0.541 &                   (N) &             1.7e-13 &                      0.543 &                   (N) &              0.3887 &                      0.505 &                   (N) \\
              All &           Z &           [2, 15] &       [3, 684] &         2.1 $\pm$ 9.0 &         36.0 $\pm$ 32.4 &      33.1 $\pm$ 31.6 &              0.0e+00 &             0.0e+00 &                      0.192 &                   (L) &             0.0e+00 &                       0.21 &                   (L) &             5.9e-05 &                      0.524 &                   (N) \\
              All &       $R_y$ &           [2, 15] &       [3, 684] &       67.9 $\pm$ 29.1 &         90.7 $\pm$ 10.8 &      90.7 $\pm$ 10.8 &              0.0e+00 &             0.0e+00 &                      0.255 &                   (L) &             0.0e+00 &                      0.256 &                   (L) &             0.90403 &                      0.501 &                   (N) \\
              \midrule
             \gls{grov} &         all &            [6, 9] &      [30, 684] &       72.5 $\pm$ 10.5 &          77.4 $\pm$ 9.6 &       77.3 $\pm$ 9.7 &              1.3e-35 &             1.8e-28 &                      0.381 &                   (S) &             8.6e-28 &                      0.383 &                   (S) &              0.8789 &                      0.502 &                   (N) \\
             \gls{grov} &           X &            [6, 9] &      [30, 684] &       96.9 $\pm$ 15.8 &         58.7 $\pm$ 30.1 &      62.0 $\pm$ 29.5 &              0.0e+00 &            3.0e-302 &                      0.861 &                   (L) &            1.6e-278 &                      0.842 &                   (L) &             0.00657 &                      0.472 &                   (N) \\
             \gls{grov} &           Z &            [6, 9] &      [30, 684] &        3.9 $\pm$ 12.9 &         43.8 $\pm$ 30.1 &      39.3 $\pm$ 29.5 &              0.0e+00 &            1.9e-294 &                      0.136 &                   (L) &            1.2e-262 &                      0.161 &                   (L) &             7.3e-05 &                      0.541 &                   (N) \\
             \gls{grov} &       $R_y$ &            [6, 9] &      [30, 684] &       87.3 $\pm$ 15.1 &          94.8 $\pm$ 8.0 &       95.1 $\pm$ 7.8 &              1.8e-61 &             3.5e-43 &                      0.363 &                   (S) &             4.5e-48 &                      0.355 &                   (S) &             0.30569 &                      0.491 &                   (N) \\
                           \midrule
            \gls{qwalk} &         all &            [3, 5] &      [14, 154] &       51.5 $\pm$ 16.1 &          73.1 $\pm$ 8.9 &       73.1 $\pm$ 9.1 &              0.0e+00 &            8.3e-236 &                      0.112 &                   (L) &            2.5e-234 &                      0.114 &                   (L) &             0.98696 &                        0.5 &                   (N) \\
            \gls{qwalk} &           X &            [3, 5] &      [14, 154] &       84.0 $\pm$ 31.2 &         75.9 $\pm$ 30.0 &      74.9 $\pm$ 31.1 &              1.9e-26 &             9.5e-22 &                      0.597 &                   (S) &             1.4e-22 &                      0.599 &                   (S) &             0.61458 &                      0.505 &                   (N) \\
            \gls{qwalk} &           Z &            [3, 5] &      [14, 154] &         1.4 $\pm$ 6.6 &         23.3 $\pm$ 28.0 &      24.8 $\pm$ 30.0 &             5.0e-148 &            9.8e-132 &                      0.276 &                   (M) &            3.0e-132 &                      0.275 &                   (M) &             0.43133 &                      0.491 &                   (N) \\
            \gls{qwalk} &       $R_y$ &            [3, 5] &      [14, 154] &       57.4 $\pm$ 22.8 &         88.8 $\pm$ 11.1 &      88.7 $\pm$ 11.1 &              0.0e+00 &            5.2e-242 &                      0.109 &                   (L) &            9.8e-241 &                       0.11 &                   (L) &             0.62748 &                      0.505 &                   (N) \\
                          \midrule
              \gls{var} &         all &            [2, 8] &        [3, 70] &       38.1 $\pm$ 25.0 &          72.6 $\pm$ 9.6 &       71.6 $\pm$ 9.9 &              0.0e+00 &            1.7e-297 &                      0.089 &                   (L) &            5.8e-282 &                        0.1 &                   (L) &             0.01587 &                      0.526 &                   (N) \\
              \gls{var} &           X &            [2, 8] &        [3, 70] &       31.0 $\pm$ 38.1 &         63.3 $\pm$ 36.0 &      60.9 $\pm$ 36.4 &             2.0e-117 &             2.7e-95 &                      0.276 &                   (M) &             2.2e-84 &                      0.289 &                   (M) &             0.08552 &                      0.518 &                   (N) \\
              \gls{var} &           Z &            [2, 8] &        [3, 70] &         1.1 $\pm$ 6.0 &         38.7 $\pm$ 36.2 &      35.8 $\pm$ 34.8 &             1.8e-261 &            3.1e-236 &                      0.191 &                   (L) &            1.4e-220 &                      0.206 &                   (L) &             0.05576 &                      0.521 &                   (N) \\
              \gls{var} &       $R_y$ &            [2, 8] &        [3, 70] &       52.8 $\pm$ 33.1 &         87.0 $\pm$ 11.6 &      87.1 $\pm$ 11.7 &             7.0e-258 &            4.4e-192 &                      0.173 &                   (L) &            9.2e-193 &                      0.172 &                   (L) &             0.80537 &                      0.497 &                   (N) \\
                            \midrule
               \gls{gs} &         all &           [3, 15] &        [5, 17] &       55.0 $\pm$ 12.8 &          77.0 $\pm$ 9.8 &      73.0 $\pm$ 10.0 &             1.9e-126 &            1.2e-103 &                      0.074 &                   (L) &             9.2e-83 &                      0.121 &                   (L) &             8.4e-09 &                      0.612 &                   (S) \\
               \gls{gs} &           X &           [3, 15] &        [5, 17] &       30.9 $\pm$ 30.3 &         65.6 $\pm$ 27.5 &      58.3 $\pm$ 31.3 &              2.2e-55 &             8.6e-51 &                      0.212 &                   (L) &             1.2e-32 &                      0.272 &                   (M) &             0.00106 &                      0.562 &                   (N) \\
               \gls{gs} &           Z &           [3, 15] &        [5, 17] &         0.9 $\pm$ 5.3 &         37.3 $\pm$ 28.4 &      27.6 $\pm$ 26.4 &             1.1e-102 &             5.4e-98 &                      0.136 &                   (L) &             5.6e-73 &                      0.205 &                   (L) &             4.6e-07 &                      0.594 &                   (S) \\
               \gls{gs} &       $R_y$ &           [3, 15] &        [5, 17] &       81.1 $\pm$ 23.8 &          94.0 $\pm$ 9.2 &      93.0 $\pm$ 10.1 &              1.4e-12 &             1.8e-11 &                      0.379 &                   (S) &             7.3e-09 &                      0.395 &                   (S) &             0.17699 &                      0.524 &                   (N) \\
\bottomrule
\end{tabular}
}
\end{table}

In \cref{tab:rq3_summary_mutation_score}, we show the summary of the mutation score results, including statistical tests between all the pairs of approaches. 
The table is divided into five sections for each program category, with four rows in each section for the mutant types X, Z $R_y$, and All, which consists of all three mutant types.

First, we observe that the average mutation score for all program categories and all mutant types is $54.5\%$ for the Default approach, with the reduction approaches Greedy at $74.7\%$ and Random at $74.0\%$. 
We see from the statistical tests in the columns \textbf{Default-Greedy} and \textbf{Default-Random} that there is no difference in killing X mutants overall when applying either reduction approach. 
However, the mutation score for the Z mutants is close to zero for Default, while $36.0\%$ for Greedy and $33.1\%$ for Random.
The $R_y$ mutants are also killed at a higher rate when a reduction is applied, at $90.0\%$ for Greedy and Random, an increase from $67.9\%$ for Default. 

By category, we first observe for \gls{grov} a statistically significant increase in overall mutation score, but the effect is small, as we can see from the \textbf{Default-Greedy} and \textbf{Default-Random} columns where the effect size is (S) in favor of the reduction approaches. 
Interestingly, we observe that Default has a higher mutation score for X mutants at $96.9\%$, while the reduction approaches obtain scores of $58.7\%$ for Greedy and $62.0\%$ for Random. 
For $R_y$, we also see an increase when a reduction is applied. 
Thus, fewer X mutants are killed for \gls{grov} when a reduction is applied, but more Z and $R_y$ mutants are killed.
For \gls{qwalk} and \gls{var}, we observe that both achieve a large increase in overall mutation scores for all mutant types. However, the X mutants for \gls{qwalk} are killed at a statistically significant higher rate for Default with effect size (S). 
All other mutants are killed at a higher rate with a large or medium effect size when applying a reduction. 
Lastly, for the \gls{gs} programs, we make two distinct observations. 
First, we see that a statistically large increase in mutation score is obtained when applying either reduction approach, particularly due to the killing of Z mutants. 
Second, we see from the \textbf{Greedy-Random} column that there are only negligible (N) differences between the Greedy and random reduction approaches, except for the \gls{gs} category, which exhibits statistically significant differences overall and for the Z mutants with small effect sizes.

\subsubsection{RQ3b: Correlation Between Reduction and Effectiveness}

\begin{table}[tbp]
    \centering
\caption{Spearman correlation $r_s$ between reduction rate and mutation score. We use N as a shorthand for the correlation magnitude Negligible.}
\label{tab:rq3_correlation_table}
\resizebox{\columnwidth}{!}{%
\begin{tabular}{llll>{\columncolor{lightgray}}r>{\columncolor{lightgray}}r>{\columncolor{lightgray}}r>{\columncolor{lightgray}}rll}
\toprule
 & &  &  & \multicolumn{2}{c}{\textbf{Correlation $r_s$}} & \multicolumn{2}{c}{\textbf{p-value}} & \multicolumn{2}{c}{\textbf{Correlation Magnitude}}\\
\cmidrule(l{1pt}r{1pt}){5-6}
\cmidrule(l{1pt}r{1pt}){7-8}
\cmidrule(l{1pt}r{1pt}){9-10}
\textbf{Category} & \textbf{Mutant Type} & \textbf{\#Qubits} & \textbf{Depth} & \textbf{Greedy} & \textbf{Random} & \textbf{Greedy} & \textbf{Random} & \textbf{Greedy} & \textbf{Random} \\
\midrule
              All &                  All &           [2, 15] &       [3, 684] &                     0.05689 &                     0.01056 &                 1.9e-04 &                 0.48917 &  N & N \\
              All &                    X &           [2, 15] &       [3, 684] &                    -0.46222 &                    -0.46551 &                4.2e-226 &                1.0e-229 &  Moderate(-) & Moderate(-)\\
              All &                    Z &           [2, 15] &       [3, 684] &                     0.46749 &                     0.40918 &                6.3e-232 &                7.2e-173 & Moderate & Moderate \\
              All &                $R_y$ &           [2, 15] &       [3, 684] &                     0.08801 &                     0.11216 &                 7.7e-09 &                 1.7e-13 & N & N \\
              \midrule
             \gls{grov} &                  All &            [6, 9] &      [30, 684] &                    -0.00749 &                    -0.03106 &                 0.78099 &                 0.24896 &  N & N\\
             \gls{grov} &                    X &            [6, 9] &      [30, 684] &                    -0.09993 &                    -0.11574 &                 2.0e-04 &                 1.6e-05 & N & N\\
             \gls{grov} &                    Z &            [6, 9] &      [30, 684] &                     0.12547 &                     0.10869 &                 2.9e-06 &                 5.2e-05 & N & N\\
             \gls{grov} &                $R_y$ &            [6, 9] &      [30, 684] &                    -0.03745 &                    -0.03895 &                 0.16435 &                 0.14811 & N & N \\
                           \midrule
            \gls{qwalk} &                  All &            [3, 5] &      [14, 154] &                    -0.05194 &                    -0.02963 &                 0.07576 &                 0.31116 & N & N\\
            \gls{qwalk} &                    X &            [3, 5] &      [14, 154] &                    -0.63187 &                    -0.61523 &                2.1e-131 &                8.8e-123 & Moderate(-) & Moderate(-) \\
            \gls{qwalk} &                    Z &            [3, 5] &      [14, 154] &                      0.5946 &                     0.58487 &                8.8e-113 &                2.6e-108 & Moderate & Moderate \\
            \gls{qwalk} &                $R_y$ &            [3, 5] &      [14, 154] &                    -0.00965 &                    -0.01322 &                 0.74169 &                 0.65145 & N & N \\
            \midrule
              \gls{var} &                  All &            [2, 8] &        [3, 70] &                    -0.18416 &                    -0.19856 &                 1.6e-11 &                 3.3e-13 & N & N\\
              \gls{var} &                    X &            [2, 8] &        [3, 70] &                    -0.64931 &                    -0.65898 &                6.9e-159 &                3.2e-165 & Moderate(-) & Moderate(-)\\
              \gls{var} &                    Z &            [2, 8] &        [3, 70] &                     0.61412 &                     0.63563 &                1.1e-137 &                2.6e-150 & Moderate & Moderate\\
              \gls{var} &                $R_y$ &            [2, 8] &        [3, 70] &                     -0.2136 &                    -0.20875 &                 4.4e-15 &                 1.8e-14 & Weak(-) & Weak(-)\\
                            \midrule
               \gls{gs} &                  All &           [3, 15] &        [5, 17] &                     0.30426 &                     0.05494 &                 1.9e-10 &                 0.26125 & Weak & N \\
               \gls{gs} &                    X &           [3, 15] &        [5, 17] &                     -0.1179 &                    -0.26432 &                 0.01563 &                 3.8e-08  & N & Weak(-) \\
               \gls{gs} &                    Z &           [3, 15] &        [5, 17] &                     0.17611 &                    -0.03972 &                 2.9e-04 &                 0.41681 & N & N\\
               \gls{gs} &                $R_y$ &           [3, 15] &        [5, 17] &                     0.51672 &                     0.44351 &                 4.8e-30 &                 1.1e-21  & Moderate & Moderate\\
\bottomrule
\end{tabular}
}
\end{table}

Finally, we investigate the correlation between the reduction and approach effectiveness. 
In \cref{tab:rq3_correlation_table}, we show the Spearman correlation between the reduction rate and mutation score by category.
For all program categories and mutant types in the first row, we observe a Negligible correlation between the Greedy and Random approaches. 
For the X mutants, we can see that there is a Moderate negative correlation for both approaches, while the Z mutants have a Moderate positive correlation for both approaches. 
The X and Z mutant correlation coefficients are similar in absolute values, showing opposite correlations. 
Finally, the $R_y$ mutants show a Negligible correlation across the program categories.
Regarding the individual program categories, we can make two observations. 
First, for the All categories except \gls{gs}, we observe either a 
Negligible correlation. 
Thus, \gls{gs} is the only program category where we observe a Weak positive correlation for the Greedy approach, while Negligible for the Random. 
The main contributions to this correlation stem from correlations for the Z and $R_y$ mutants.
The second observation is that the X mutants are negatively correlated for all the program categories with significant magnitudes for the correlation coefficients, while the Z mutants have similar magnitude coefficients but are positively correlated. 

\summarybox{RQ3 Summary}{
The reduction approaches significantly improve mutation scores to $74.7\%$ with the Greedy approach, from $54.5\%$ with the Default approach, primarily driven by an increase in Z and $R_y$ mutant kills.
There are no significant differences between the reduction approaches, except in the \gls{gs} category, where the Greedy approach shows slightly better performance, mainly due to the more effective killing of Z mutants.
The overall correlation between reduction effectiveness and mutation score is Negligible for both approaches.
}

\section{Discussions}
\label{sec:discussion}

Our discussion of the reduction approaches in \gls{qst} highlights three key themes:
\begin{inparaenum}
    \item \textbf{Program Sensitivity}, 
    \item \textbf{Mutant Sensitivity}, 
    and
    \item \textbf{Greedy vs. Random}. 
\end{inparaenum}
These points guide our discussion on optimizing reduction strategies for diverse \gls{qst} applications. 
After these points are addressed, we discuss the practical implications of our findings to the field of \gls{qst}.

\subsection{\textbf{Program Sensitivity}}

Although we find that reduction significantly improved both \gls{qst} efficiency and effectiveness, this is not true for all types of programs. 

\paragraph{ \textbf{Higher Depth Programs Benefit More From Reduction. } }

First, we discuss the observation that the programs from \gls{grov} experienced a very high runtime improvement from reduction compared to the other programs. 
Longer-depth circuits take more time to run than short-depth circuits. 
Thus, by requiring fewer circuit executions due to our approach, we save considerably more \emph{longer} circuit executions, especially for high-depth circuits, such as \gls{grov}.
This is not always the case; however, as we also find a considerable number of slowdowns in the moderate depth category, such as \gls{qwalk}, we argue that higher depth alone does not lead to positive benefits in \gls{qst}. 
In addition, a high satisfiability of the output criterion is required, such as achieved by the uniform and real amplitudes of the states of \gls{grov} and \gls{gs}.

\paragraph{ \textbf{Large Program Specifications Benefit More From Reduction: } }

For the programs with large specification sizes (ranks), e.g., \gls{gs}, we observe a tendency of both reduction approaches to plateau below a runtime improvement over the Default of 100.
This is likely due to the combination of many output states and low depth, which leads to the runtime overhead from the Hadamard gates required for larger reductions competing with the program's depth. 
To achieve larger reductions, we need more Hadamard gates, which for low-depth circuits like \gls{gs} programs may cause either a stagnation of runtime improvement or even a slowdown after a certain point. 
However, our data doesn't provide conclusive answers to either.

\paragraph{\textbf{Low Satisfiability of Output Criteria.}}

Our approach generally performs poorly for \gls{qwalk} programs, aligning with the findings from previous studies indicating that continuous-time quantum walks do not typically converge to uniform superpositions~\citep{gerhardt2003continuoustime}. 
Our results suggest that only programs with specific numbers of walks benefit from reductions, indicating that uniformity in the final state vector is rare.
We find some programs in the \gls{var} category that obtained no reductions from \cref{alg:reduction_algorithm}. 
Programs like the quantum conditional phase flip and most integer addition programs failed to achieve any reduction from our approach, suggesting issues with satisfying the output criteria \cref{eq:real_uniform_state_vector}.
State vector inspections reveal that this is due to complex amplitudes, meaning they are not eigenstates of mixed Hadamard basis, as our approach requires.

Thus, these are two examples of low satisfiability of the output criteria where reduction is unsuccessful. 
One where uniformity is not satisfied and the other where the amplitudes are not real numbers.

\subsection{Mutant Sensitivity}

We observed that the Default approach hardly detected phase flip faults.
By hardly, we mean that this score is minuscule, possibly due to false positives.
This is expected as phase flip faults to the final state vector cause flipped signs in the probability amplitudes. 
Consequently, we cannot discover such Z faults, even if we were to perform an infinite number of measurements in the computational basis.

\paragraph{\textbf{X Fault vs. Z Fault Detection Tradeoff.}}

While the Z gate acts as a bit flip gate in our reduced basis, allowing phase flip fault discovery, the X gate also modifies its behavior. 
We can see from the transformations $X\ket{+}=\ket{+}$ and $X\ket{-}=-\ket{-}$ that X may act as a phase flip gate in the mixed Hadamard basis. 
Thus, the X and Z gates switch roles, such that in the mixed Hadamard basis, the X gate acts as Z and Z acts as X. 
This explains why we may observe a reduced detectability of X mutants in our evaluation. 
The \gls{gs} category was the only one that did not experience such an increase but rather increased detection of X faults. 
However, we can partially attribute this to the nature of the program category. 
Due to the large reductions that occur for \gls{gs}, there is a smaller number of states in the reduced \gls{ps} when compared to the other categories' \glspl{ps}.
Thereby, by performing an X gate fault, a basis state not present in the \gls{ps} is created with a high probability.

Furthermore, implied by the X fault vs. Z fault tradeoff, our approach could improve fault detection by combining measurements in both the computational basis and projective measurements in the mixed Hadamard basis. 
This combination would ensure that the role switch between X and Z gates does not result in missed faults for either gate.

\subsection{\textbf{Greedy vs. Random}}

The Greedy approach can get stuck in local minima, such as $[\mathbbm{1}, H,\mathbbm{1}]$ demonstrated in our example (see \cref{fig:graph_examples}). Considering this, one might expect the Random approach to outperform Greedy when sufficient samples are available to explore the search space of bases. 
In our experimental setup, we observed that the Greedy approach generally outperformed the Random approach when given the same number of samples, even in programs with fewer qubits.

\paragraph{\textbf{Curse of Dimensionality.}}
The search runtime for Random is comparable to Greedy, both scaling as $\mathcal{O}(\#Qubits^2)$, but given the exponential growth of the search space ($2^{\#Qubits}$), Random struggles with the curse of dimensionality~\citep{Vervliet_cod,Marimont_cod}.
This could limit Random's effectiveness, particularly as the number of qubits increases, as finding a good basis becomes less likely.
While increasing the number of random searches might improve the outcomes, the exponentially larger search space makes significant improvements unlikely without incurring substantial runtime overhead.

\paragraph{\textbf{Some Bases are Only Discovered by Random.}}

The probabilistic nature of the Random approach allows it to occasionally find bases that are not reachable by Greedy, which terminates once no further reductions are identified. 
This characteristic can be advantageous in scenarios where the search landscape contains multiple viable pathways to a reduction, as shown in \textbf{Example A} in the appendix~\citep{oldfield_2024_11191215}.
However, the random approach samples from an exponentially increasing search space of bases. 
Thus, Random occasionally selects a basis that adds more runtime than it saves due to a high number of Hadamard gates. 
This is not the case for Greedy, except for some outliers. Greedy, thus, for higher qubit counts of \gls{gs}, converges consistently to a basis that results in an improved test runtime over Default, while Random is considerably less consistent for high qubit counts.

\subsection{Practical Aspects for \gls{qst}.}

Here, we discuss the practical considerations of our approach and results.

\paragraph{\textbf{Comparison of Greedy Heuristics with Other Approaches in Classical Software Testing.}}

Greedy algorithms have historically been favored for their simplicity and low computational cost, particularly in tasks like test case prioritization and minimization in classical software testing~\citep{Elbaum_2002}. 
As the field matured, more advanced methods, such as genetic algorithms, were introduced to handle increasingly complex search spaces~\citep{Hemmati_2010}.
Similarly, in quantum software testing (\gls{qst}), where the field is still developing, Greedy offers a practical starting point since we mainly deal with small circuits with limited search spaces (e.g., test inputs). 
Its balance of simplicity and efficiency provides a solid foundation, with room for future exploration of more sophisticated methods as the field evolves.

In our specific approach, the search space is relatively simple, which allows Greedy to be both effective and efficient. 
While advanced heuristics might be required for larger or more complex spaces, Greedy provides a practical solution with minimal computational overhead in this case.

\paragraph{\textbf{Quantum-Specific Bugs as a Result of Faults.}}

While our approach evaluates seeded gate faults, it is important to recognize that they manifest as bugs, as introduced in \cref{sec:background}, depending on the program category, i.e., \gls{grov}, \gls{gs}, \gls{qwalk}, and \gls{var}. 
Each fault type (X, Z, and $R_y$) causes bugs that impact the program's behavior, as introduced in \cref{sec:study_subjects}.

For instance, in a Grover search, if the correct element is $1001$ and a bit flip (X fault) occurs on the third qubit, the algorithm would incorrectly find $1011$. 
Thus, for Grover, X faults primarily lead to incorrect search results. 
This is reflected in our evaluation of RQ3, where we observe a high mutation score of 96.9\% for X faults in Grover using the Default approach (\cref{tab:rq3_summary_mutation_score}), indicating frequent incorrect search elements.

Similarly, $R_y$ faults impact the probability of finding the correct element. 
In our results, the mutation score for $R_y$ faults in Grover is 87.3\% with the Default approach, corresponding to incorrect likelihoods rather than incorrect elements.

This relationship between faults and bugs extends to other program types, such as graph states. 
Each graph state program represents a specific graph (as described in \cref{sec:study_subjects}). 
A Z fault, such as a phase flip, alters the graph representation, leading to incorrect graphs. 
This is evident in our RQ3 evaluation, where Z faults in graph states produced a high mutation score (91.9\% with the Default approach), corresponding to incorrect graph representations.

\paragraph{\textbf{Obtaining the Default \gls{ps}.}}

As our reduction approach assumes that we have access to the state vector of the \gls{sut}, serving as a program's \HLIGHT{Default \gls{ps}}, we now discuss how we obtain it in practice.
We see two approaches to obtain a correct final state vector. 
The first is through experimental methods such as quantum state tomography~\citep{QCQIBook}, which reconstructs the density matrix of a given program. 
Although full state tomography is resource-intensive and requires exponentially many measurements, more efficient quantum state tomography remains an active research field \cite{Efficient_State_Tomography,Eisert_2020,Baumgratz_2013,Motka_2014,Torlai_2018}. 
Weakly entangled states, such as those we consider, may allow for efficient state tomography using matrix-product state tomography~\citep{Efficient_State_Tomography,Schmale_2022}.
The second is through mathematical methods of analytical calculation from probability amplitude expressions. 
Such as the known transformation formula of quantum Fourier transform or amplitude amplification~\citep{QCQIBook}.

\paragraph{ \textbf{Is Greedy Better than Random?} }

While Greedy has an inbuilt stopping criteria for the number of objective function executions, Random requires this as input.
The Greedy approach searches more specific paths through the basis search space, while Random may explore more parts of the space.
The potential drawbacks of applying the Greedy approach as opposed to Random are (1) the vulnerability of Greedy getting stuck in local minima and (2) the inability to find certain bases. 
We could alleviate (1) through techniques like Basin Hopping~\citep{BasinHopping}, incorporating random basis changes to find previously unavailable search paths. 
(2), however, could result in the Greedy approach being unable to detect certain faults if the undiscovered basis is required to detect a particular fault.
The potential drawbacks of Random as opposed to Greedy are (1) the curse of dimensionality from exponentially growing search spaces and (2) requiring a specified number of random samples to be input into the algorithm.
We can alleviate (2), however, by specifying a formula for the number of searches, such as applying the theoretical runtime formula of the Greedy approach (see appendix~\citep{oldfield_2024_11191215}). 

Thus, while our results show that both Greedy and Random often perform similarly, the Greedy algorithm's tendency to find higher reductions more consistently for higher qubit counts suggests it may be more reliable for practical applications. 
This is because more qubits, rather than fewer, will be required by future quantum programs to obtain quantum advantage~\citep{Dalzell_2020,ichikawa2023comprehensive}.

\paragraph{\textbf{Reduction-Circuit.}}
We see a practical limitation of our approach in how we perform the reduction. 
While our reduction approach avoids the exponential computation of the inverse matrix problem by performing reductions in the reduction-circuit, we assume that the circuit can be initialized to the \HLIGHT{Default \gls{ps}}. 
In our experiments, we initialize the state vector directly for experimental simplicity. 
For a real device, gate instructions create this initialization instead. 
However, we argue that this threat is mitigated by the same gate instructions that gave us the \HLIGHT{Default \gls{ps}} in the first place. 
Thus, if we can obtain a \HLIGHT{Default \gls{ps}}, we can also initialize it. 

\paragraph{\textbf{Applications of Specification Reduction in \gls{qst}.}}

While our approach, as demonstrated in our evaluation, can be applied independently within \gls{qst}, the \textit{Reduction} component from \cref{fig:approach_diagram} can also be used as a stand-alone module in other QST approaches. 
This is particularly useful when exponential sampling for test assessments faces scalability issues, requiring more efficient and effective testing, especially enabling phase flip fault detection.
Most \gls{qst} approaches developed in the last few years have relied on quantum program sampling together with some form of test oracle and, in some of their evaluations, have also inserted phase flip faults in their mutation analysis in their evaluation.
For instance, the input-output coverage approach of \citet{quitoASE21tool} or the mutation testing approach of \citet{Muskit} could directly benefit.
These approaches also make use of a similar type of test oracle, assessing the probability distributions, meaning phase flip faults would go undetected with a high likelihood.
Other approaches that do not utilize program specification-based oracles, such as the metamorphic testing-based approach of \citet{QMorph}, could still benefit, as they utilize sampling to compare the distributions.
In this case, one could also benefit from projective measurements to perform faster distribution sampling comparisons, as suggested by our approach.

\paragraph{\textbf{Integration with Development Environments and Hardware Platforms.}}

Although our approach is grounded in theoretical foundations and is independent of specific programming languages or quantum hardware backends, we want to discuss some technical aspects. We have implemented our evaluation using IBM's Qiskit~\citep{QiskitCite}, but other popular programming environments, such as Rigetti's Forest~\citep{joshua_combes_2019_3455848} or Google's Cirq~\citep{cirq_developers_2023_10247207}, are equally applicable. The primary implementation differences across these environments are syntactical, as the core functionality we use is generally available in most quantum software development kits.

On the hardware side, aside from noise, which we address in \cref{sec:threats_to_validity}, our approach is designed to run on any quantum backend based on the gate model for universal quantum computing~\citep{QCQIBook}. This includes systems such as IBM's Q system one and two, Google's Sycamore, Quantinuum's H-series, and Rigetti's Aspen or Aanka. However, our approach is not suitable for non-gate-model systems like those developed by D-Wave, which are currently specialized for solving optimization problems using quantum annealing~\citep{kadowaki1998quantum}.

\section{Related Work}

This section discusses related work.

\paragraph{\textbf{Abstraction of Quantum Circuits}}

Efficient methods for designing and automating software are essential in quantum software engineering to harness the benefits of quantum algorithms. 
\citet{wille2022decision} address this by using decision diagrams to abstract quantum concepts like circuits and state vectors, simplifying quantum design automation. 
Our approach complements theirs by advancing the abstraction of quantum program specifications through basis dependence, and we propose that it can be further enhanced by the high-level abstractions from \citet{wille2022decision}, potentially representing our specifications as diagrams. 
These diagrams could improve automation and enable the extension of our specifications to other general bases beyond the mixed Hadamard basis. 
Unlike decision diagrams primarily used to represent the quantum states and operations, our method integrates basis dependence directly into program specifications, focusing on enhancing \gls{qst} efficiency.
Our experimental evaluation maintains a foundational level of abstraction by integrating basis dependence into program specifications, demonstrating how this abstract concept experimentally enhances \gls{qst}.

\paragraph{\textbf{Projective Measurements}}

\citet{Proq} introduce an approach called Proq, where they define projective measurement-based runtime assertions and prove theoretical results about their efficiency and effectiveness in testing. 
While they do not provide an experimental evaluation of fault detection, they do present an implementation strategy for their assertions, along with a case study on realistic quantum algorithms such as Shor's and Harrow–Hassidim–Lloyd algorithms.

As with Proq, on the theoretical side, our approach also considers projective measurements, specifically using mixed Hadamard bases (though not limited to this), while Proq includes projective measurements for more general states, such as maximally entangled states. 
Proq uses knowledge of the quantum algorithms to derive assertions, for example, by mathematically calculating the state vector at various points in the circuit.

Similarly, we derive our \HLIGHT{Default \gls{ps}} based on algorithm-specific knowledge, just as Proq does. However, on the experimental side, our approach goes further by incorporating program specifications as a high-level software testing component to guide \gls{qst} in practical implementations.

In addition, while Proq focuses primarily on the theoretical application of projective measurements in quantum software testing, our work provides empirical evidence demonstrating their effectiveness and efficiency in real-world \gls{qst} scenarios.

\paragraph{\textbf{Other \gls{qst} Methodologies}}

In quantum symbolic execution, \citet{nan2022quantum} exploits quantum superposition to provide multiple inputs to a quantum program, effectively reducing the quantum resource requirements. 
Although their approach does not detect end-circuit phase flip faults, as X basis measurements are required, they achieve a large input coverage, whereas we focus on larger output coverage using a single input per subject.

\citet{QMorph} utilize metamorphic testing by comparing the statistical distributions between a source quantum program and its transformed version through metamorphic relations. 
This method facilitates high-level handling of programming bugs at the language-specific statement level and requires less knowledge about the quantum algorithm, making it easily applicable. 
However, while their approach is easily applicable, to detect circuit faults efficiently, particularly phase flip faults, a metamorphic relationship that incorporates quantum-specific algorithmic knowledge may be necessary. 
In contrast, our use of projective measurements enhances both the efficiency of sampling and the detection of phase flip faults, addressing this gap.

\section{Conclusions and Future Directions}
\label{sec:conclusion}

\glsresetall{}

In the growing field of quantum computing, efficient and effective quantum software testing is essential.
To this aim, we proposed an approach to reduce quantum program specifications to perform projective measurements in mixed Hadamard bases. 
We empirically evaluated our approach and found that reduction is highly efficient and effective, with the Greedy approach outperforming the Random baseline in terms of average reduction rate and runtime.
Specifically, reductions were most effective in the \gls{grov} and \gls{gs} categories, demonstrating high and consistent reduction rates, whereas \gls{qwalk} and \gls{var} categories showed significantly poorer performance. 

Regarding the impact on \gls{qst} efficiency, our reduction approach enhanced testing efficiency compared to the Default approach, where only computational basis measurements are performed, with the Greedy method slightly more favorable in \gls{gs} programs for larger specifications. 
The greatest improvements were noted in \gls{grov} and \gls{gs} categories, highlighting the influence of program characteristics on reduction success. 
However, the correlation between the degree of reduction and actual runtime improvements was Moderate, indicating that reductions do not uniformly predict efficiency improvements.

Regarding the effectiveness of \gls{qst}, applying reduced specifications significantly improved mutation scores, particularly through enhanced detection of phase flip faults. 
Although the performance between reduction methods was generally comparable, the Greedy approach showed a slight advantage in the \gls{gs} category. 
The overall correlation between reduction and mutation score effectiveness was Negligible, driven by program characteristics and the inverse relationship of detecting X and Z faults. 
These results underline the potential and limitations of reduction approaches in enhancing the efficiency and effectiveness of \gls{qst}, providing valuable insights into the dependency of performance gains on specific program characteristics and reduction strategies.


Going forward, we envision three promising directions future research should explore.
\begin{inparaenum}
    \item Methods should be developed to empirically obtain program specifications through techniques like state tomography or mathematically by inspecting algorithms, which would streamline the reduction process.
    \item The reduction algorithm could be generalized to incorporate other bases beyond the mixed Hadamard, such as including Y gates for states with complex amplitudes or CNOT gates for maximally entangled states.
    This would introduce a more complex search space, possibly requiring advanced search techniques like genetic algorithms, as the Greedy approach may be limited by local minima. 
    As our approach allowed for the detection of bit flip faults from the Z gate, a future approach would expand our approach further to detect S gate faults by including Y gates in the search space.
    This specific gate choice will allow program additional specification reductions in quantum programs such as quantum Fourier transform.
    \item Combining different measurement bases to improve fault detection, such as using both the computational and mixed Hadamard bases or computational basis and Y gate basis for S gate faults. 
    This combined approach would address the X-Z fault detection tradeoff or other tradeoffs of this nature, ensuring both bit flip and phase flip errors or other faults are effectively detected.
\end{inparaenum}

\begin{acks}
This work is supported by the Qu-Test project (\texttt{299827}) funded by the Research Council of Norway (RCN) and has benefited from the \glsentryfull{ex3}, which is supported by the RCN project \texttt{270053}.
S. Ali also acknowledges the support from Simula's internal strategic project on quantum software engineering and the \textit{Quantum Hub initiative} (OsloMet).
\end{acks}

\bibliographystyle{ACM-Reference-Format}
\bibliography{biblio}

\end{document}


\definecolor{lightgray}{gray}{0.9}

\title{Faster and Better Quantum Software Testing through Specification Reduction and Projective Measurements}
\subtitle{Appendix}


\author{Noah H. Oldfield}
\orcid{0000-0002-9059-0694}
\affiliation{%
  \institution{Simula Research Laboratory and University of Oslo}
  \city{Oslo}
  \country{Norway}}
\email{noah@simula.no}

\author{Christoph Laaber}
\orcid{0000-0001-6817-331X}
\affiliation{%
  \institution{Simula Research Laboratory}
  \city{Oslo}
  \country{Norway}}
\email{laaber@simula.no}

\author{Tao Yue}
\orcid{0000-0003-3262-5577}
\affiliation{%
  \institution{Simula Research Laboratory}
  \city{Oslo}
  \country{Norway}}
\email{taoyue@gmail.com}

\author{Shaukat Ali}
\orcid{0000-0002-9979-3519}
\affiliation{%
  \institution{Simula Research Laboratory}
  \city{Oslo}
  \country{Norway}}
\affiliation{%
  \institution{Oslo Metropolitan University}
  \city{Oslo}
  \country{Norway}}
\email{shaukat@simula.no}

\maketitle

\section{Introduction}
This appendix provides supplementary details for the TOSEM paper titled "Faster and Better Quantum Software Testing through Specification Reduction and Projective Measurements."

\section{Example A}

Given the state vector:

\begin{equation}
    \frac{1}{\sqrt{8}}\Big( \ket{00000} + \ket{00010} + \ket{00100} - \ket{00110} + \ket{01000} + \ket{01010} - \ket{01100} + \ket{01110} \Big)
\end{equation}

The following reduced state can not be found with the Greedy approach:

\begin{equation}
    \frac{1}{\sqrt{4}}\Big( \ket{0+++0} + \ket{0+-+0} + \ket{0-+-0} - \ket{0---0} \Big)    \label{eq:not_possible_to_find_greedy}
\end{equation}

\paragraph{\textbf{Demonstration:}}
Let's look at the reduced state resulting from a Hadamard gate applied to the second, third and fourth qubits: 

\begin{align}
 \ket{\psi}_{ihiii} &=   \frac{1}{2}\Big( \ket{0+000} + \ket{0+010} + \ket{0-100} - \ket{0-110}  \Big) \\
 \ket{\psi}_{iihii} &=   \frac{1}{2}\Big( \ket{00+00} + \ket{00-10} + \ket{01-00} + \ket{01+10}  \Big) \\
  \ket{\psi}_{iiihi} &=   \frac{1}{2}\Big( \ket{000+0} + \ket{001-0} + \ket{010+0} - \ket{011-0}  \Big)
\end{align}

None of these states are possible to find with the Greedy algorithm, as we can only progress with Greedy further with either a reduction or an increase in the number of basis states. 
Thus, neither options can lead to finding the state \cref{eq:not_possible_to_find_greedy}, because only reduction is allowed further.

\section{Theoretical Runtime of Reduction Algorithm}
\label{sec:runtime_complexity}

We quantify the theoretical runtime of \cref{alg:reduction_algorithm} by the number of basis transformations required to reduce a given \gls{ps}.
An exhaustive search to find the global minimum of the rank, ($N_{\texttt{psPrevious}}$ from \cref{alg:reduction_algorithm}), will at most need to sample all possible basis transformations, one for each choice of the values $x_0,x_1,\cdots,x_{n-1}$, in the basis transformation \cref{eq:basis_transform} ($n=\#\texttt{Qubits})$. 
Therefore, exhaustive search requires $\mathcal{O}(2^n)$ basis transformations to find the global minimum.

Our Greedy reduction algorithm, finds a local or global minimum with $\mathcal{O}(n^2)$ number of basis transformations. 
We demonstrate this theoretical runtime by considering a few iterations of \cref{alg:reduction_algorithm}. 
In the first iteration of the \HLIGHT{Basis Change} stage, there are $n$ possible positions for the single-qubit Hadamard gate. 
If we find a reduction and proceed to the next iteration, there are now $n-1$ possible transformations, as we fixed the gate in the first iteration. 
Subsequently, if we find another reduction, there are $n-2$ possibilities, and so on until there is only one final choice for the last Hadamard gate. 
Therefore, the upper bound of the number of basis transformations for the Greedy algorithm is equal to the series of consecutive natural numbers:

\begin{equation}
    f(n)=n + (n-1) + (n-2) + \cdots + 1 = \frac{n(n+1)}{2}
    \label{eq:Greedy_runtime_formula}
\end{equation}

In \cref{eq:Greedy_runtime_formula}, we compute the maximum number of objective function evaluations $f(n)$ required by our Greedy approach in order to obtain a reduction.

\section{Detailed Sampling Strategies}

Here, we provide additional details sampling strategies for our 4 program categories: \gls{grov}, \gls{qwalk}, \gls{var} and \gls{gs}.

\subsection{\gls{grov}
}
In order to successfully apply the reflection about the mean operation of the Grover operator, the set $M$, defining the number of matching entries in the search space, must be of size $M< 2^n/2$, where $n$ is the number of qubits \cite{GroversAlgo}.
We generate program variations by first specifying a range of qubit values, and then for a fixed number of qubits, we specify a set of output states between $0$ and $N-1$, with $N=2^n$, which are to be amplified. 
We then iterate over set sizes $M$ up to $M=N-2$ with an increment determined by $N-2/2$ divided by the number of program variations within each qubit. 
We exclude $N-1$ because setting the boundary at $N-1$ makes the increment an integer, avoiding rounding. 
We then randomly select output states in the set as we iterate over different set sizes. 
When the qubit and input set are specified, we maximize the output criterion \cref{eq:real_uniform_state_vector} by performing repeated iterations of the Grover operator until the probability amplitudes of states outside the input set reach an order of magnitude of $10^{-4}$. 
When $M$ is very small compared to $N$, the optimal number of iterations is approximately given by $\pi/4\sqrt{N/M}$ \cite{GroversAlgo}.

\subsection{\gls{gs}}

Graph states are representations of a graph $G=(V,E)$ where the set of vertices $V$ of the graph are encoded as quantum states \cite{Hein_2004, PhysRevA.73.022334}. 
The quantum program for creating a graph state for a given graph works by initializing all states to zero. 
Then, we apply a Hadamard gate to every qubit in the register, yielding a uniform superposition. 
Lastly, a controlled Z gate is applied between all pairs of vertices $(a,b)$ that connect to an edge from the set $E$ for all edges. 
Due to this program structure, the \gls{gs} programs are low depth, but the combination of Hadamard and controlled Z gates always results in $2^{n-2}$ number of output states, which makes \gls{gs} programs a reliable sample of programs with a high number of output states. 

Our sampling strategy for generating \gls{gs} program variations consists of defining the type of graph of our graph state program by defining the set of edges $E$ provided to the program. 
Given a qubit count $n$, the set of edges consists of pairs of indices $(k,l)$ where $k\neq l$ and both are selected from the set $\{0,1,\cdots, n-1\}$. 
With respect to our testing runtime constraints and to maximize the output criterion \cref{eq:computational_basis}, we select one type of graph, namely the ring graph, which consists of the edge set $E=\{(0,1), (1,2), \cdots, (n-1, 0)\}$.

\subsection{\gls{qwalk}}

Quantum walks are the quantum counterpart of classical random walks, utilizing principles of superposition and interference. 
Unlike classical walks that model probabilistic movement in space, quantum walks operate with a "walker" in a superposed state, moving through positions based on a quantum coin's state.
The walker's position is represented by a set of qubits, with each qubit encoding different potential positions. 
An additional qubit, the quantum coin, is manipulated using a Hadamard gate to create a superposition state. 
This state influences the walker's movement, controlled through gates such as $CNOT$ and multi-controlled $CNOT$.
The walker's movement in a quantum walk is stepwise, determined by the coin qubit's state. 
At the end of the walk, measuring the qubits yields the probability distribution of the walker's positions. 
Quantum walks enable efficient exploration of complex structures and offer advantages in computation and simulation over some classical methods \cite{Venegas_Andraca_2012_Quantum_Walk}.

We select quantum walk programs for qubits ranging from 5 to 8. Then, for varying degrees of walks, the final state vector spreads out more or less, leading to a varying saturation of the output criterion \cref{eq:real_uniform_state_vector}. 
Thus, we select program variations within each qubit number automatically by varying the number of walks performed in the respective program.

\subsection{\gls{var}}

We select programs from the \gls{var} category by visually inspecting the final state vectors of the programs using the QCengine API from the original source of the programs~\citep{johnston2019programming}. 
The programs are denoted by \textit{Example $x/y$} where $x$ and $y$ refers to example number $y$ of chapter $x$. 
From the $52$ programs of the program source, we select a sample of programs that maximize the output criterion \cref{eq:real_uniform_state_vector} after we apply the following exclusion criteria. 
We exclude programs that are simple demonstrational programs, such as the program Example 2-1, which simulates a random bit. Thus, we exclude the programs Examples: 2-1, 2-2, 2-3 and 3-1. 
Next, we exclude programs that only contain a single output state prior to measurement, as at least two output states are required for reduction. 
This excludes the programs Examples: 3-4, 5-6, 10-2, 12-2, 12-4, 14-GT, 14-BV and 14-S. Next, we exclude programs for runtime constraints as they contain a relatively high number of qubits combined with a high depth, such as Example 4-2 with 30 qubits. We exclude the following programs Example: 4-2, 11-2, 11-4, 11-6, 12-1. 
After performing these exclusions, we select a sample of programs where the output criterion is maximized, yielding the programs that we list in \cref{tab:study_subjects_var}. 
For each program, we generate 4 program variations by varying the input state between 0-3. 
For further program descriptions, we refer to the source \cite{honarvar2020property}.

\section{Results}

\textit{As a technical note, we run Experiments 1 and 2 with Qiskit version 0.45.1.}

Here, we show detailed tables of our results for RQ1, RQ2, and RQ3.
All tables contain pairwise statistical tests between the Greedy approach and Random baseline for a given row, which are described in the full paper.

\subsection{RQ1 Tables}

\begin{table}[]
    \centering
    \caption{Reduction rate results for the \textbf{\gls{grov}} program category.}
\label{tab:rq1_full_reduction_table_grov}
\resizebox{\columnwidth}{!}{%
\begin{tabular}{rlrr>{\columncolor{lightgray}}r>{\columncolor{lightgray}}rrrl}
\toprule
\textbf{ID} & \textbf{Category} & \textbf{\#Qubits} & \textbf{Depth} & \textbf{Greedy [\%]} &    \textbf{Random [\%]} & \textbf{p-value} & $\bf{\hat{A}_{12}}$ & \textbf{Magnitude} \\
\midrule
 \textbf{0} &     \textbf{\gls{grov}} &        \textbf{6} &   \textbf{178} & \textbf{72.4 $\pm$ 9.6} & \textbf{56.0 $\pm$ 1.8} & \textbf{3.6e-27} &          \textbf{0.873} &       \textbf{(L)} \\
          1 &              \gls{grov} &                 6 &             30 &          65.6 $\pm$ 0.0 &          65.6 $\pm$ 0.0 &              1.0 &                     0.5 &                (N) \\
          2 &              \gls{grov} &                 6 &            121 &          68.8 $\pm$ 0.0 &          66.4 $\pm$ 5.9 &          1.1e-04 &                    0.57 &                (N) \\
 \textbf{3} &     \textbf{\gls{grov}} &        \textbf{6} &    \textbf{42} & \textbf{60.9 $\pm$ 6.6} & \textbf{58.8 $\pm$ 4.6} & \textbf{4.6e-07} &          \textbf{0.698} &       \textbf{(M)} \\
          4 &              \gls{grov} &                 6 &             48 &          60.8 $\pm$ 7.9 &          59.5 $\pm$ 8.5 &          0.29283 &                   0.538 &                (N) \\
          5 &              \gls{grov} &                 6 &             54 &          56.9 $\pm$ 5.9 &          57.2 $\pm$ 5.1 &          0.98945 &                   0.499 &                (N) \\
 \textbf{6} &     \textbf{\gls{grov}} &        \textbf{6} &    \textbf{66} & \textbf{59.4 $\pm$ 0.0} & \textbf{55.7 $\pm$ 5.6} & \textbf{1.1e-13} &           \textbf{0.72} &       \textbf{(M)} \\
\bottomrule
\end{tabular}%
}
\end{table}

\begin{table}[]
    \centering
    \caption{Reduction rate results for the \textbf{\gls{qwalk}} program category.}
    \label{tab:rq1_full_reduction_table_qwalk}
\resizebox{\columnwidth}{!}{%
\begin{tabular}{rlrr>{\columncolor{lightgray}}r>{\columncolor{lightgray}}rrrl}
\toprule
\textbf{ID} & \textbf{Category} & \textbf{\#Qubits} & \textbf{Depth} & \textbf{Greedy [\%]} &     \textbf{Random [\%]} & \textbf{p-value} & $\bf{\hat{A}_{12}}$ & \textbf{Magnitude} \\
\midrule
          0 &             \gls{qwalk} &                 3 &             14 &          50.0 $\pm$ 0.0 &          43.0 $\pm$ 17.4 &          1.1e-04 &                    0.57 &                (N) \\
 \textbf{1} &    \textbf{\gls{qwalk}} &        \textbf{3} &    \textbf{21} & \textbf{50.0 $\pm$ 0.0} & \textbf{34.0 $\pm$ 23.4} & \textbf{7.5e-10} &           \textbf{0.66} &       \textbf{(S)} \\
          2 &             \gls{qwalk} &                 3 &             28 &           0.0 $\pm$ 0.0 &            0.0 $\pm$ 0.0 &              1.0 &                     0.5 &                (N) \\
          3 &             \gls{qwalk} &                 3 &             35 &           0.0 $\pm$ 0.0 &            0.0 $\pm$ 0.0 &              1.0 &                     0.5 &                (N) \\
          4 &             \gls{qwalk} &                 3 &             42 &          50.0 $\pm$ 0.0 &          46.5 $\pm$ 12.8 &           0.0073 &                   0.535 &                (N) \\
 \textbf{5} &    \textbf{\gls{qwalk}} &        \textbf{3} &    \textbf{49} & \textbf{50.0 $\pm$ 0.0} & \textbf{30.0 $\pm$ 24.6} & \textbf{1.8e-12} &            \textbf{0.7} &       \textbf{(M)} \\
          6 &             \gls{qwalk} &                 3 &             56 &           0.0 $\pm$ 0.0 &            0.0 $\pm$ 0.0 &              1.0 &                     0.5 &                (N) \\
\bottomrule
\end{tabular}%
}
\end{table}

\begin{table}[]
    \centering
    \caption{Reduction rate results for the \textbf{\gls{var}} program category.}
    \label{tab:rq1_full_reduction_table_var}
\resizebox{\columnwidth}{!}{%
\begin{tabular}{rlrr>{\columncolor{lightgray}}r>{\columncolor{lightgray}}rrrl}
\toprule
\textbf{ID} & \textbf{Category} & \textbf{\#Qubits} & \textbf{Depth} &  \textbf{Greedy [\%]} &     \textbf{Random [\%]} & \textbf{p-value} & $\bf{\hat{A}_{12}}$ & \textbf{Magnitude} \\
\midrule
          0 &               \gls{var} &                 2 &              6 &           25.0 $\pm$ 0.0 &           25.0 $\pm$ 0.0 &              1.0 &                     0.5 &                (N) \\
          1 &               \gls{var} &                 2 &              6 &           25.0 $\pm$ 0.0 &           25.0 $\pm$ 0.0 &              1.0 &                     0.5 &                (N) \\
          2 &               \gls{var} &                 2 &              7 &           25.0 $\pm$ 0.0 &           25.0 $\pm$ 0.0 &              1.0 &                     0.5 &                (N) \\
          3 &               \gls{var} &                 2 &              7 &           25.0 $\pm$ 0.0 &           25.0 $\pm$ 0.0 &              1.0 &                     0.5 &                (N) \\
 \textbf{4} &      \textbf{\gls{var}} &        \textbf{3} &     \textbf{3} &  \textbf{75.0 $\pm$ 0.0} & \textbf{69.2 $\pm$ 10.6} & \textbf{3.7e-07} &          \textbf{0.615} &       \textbf{(S)} \\
 \textbf{5} &      \textbf{\gls{var}} &        \textbf{3} &     \textbf{4} &  \textbf{75.0 $\pm$ 0.0} & \textbf{69.2 $\pm$ 10.6} & \textbf{3.7e-07} &          \textbf{0.615} &       \textbf{(S)} \\
 \textbf{6} &      \textbf{\gls{var}} &        \textbf{3} &     \textbf{3} &  \textbf{75.0 $\pm$ 0.0} & \textbf{68.0 $\pm$ 11.3} & \textbf{1.3e-08} &           \textbf{0.64} &       \textbf{(S)} \\
\bottomrule
\end{tabular}%
}
\end{table}

\begin{table}[]
    \centering
    \caption{Reduction rate results for the \textbf{\gls{gs}} program category.}
    \label{tab:rq1_full_reduction_table_gs}
\resizebox{\columnwidth}{!}{%
\begin{tabular}{rlrr>{\columncolor{lightgray}}r>{\columncolor{lightgray}}rrrl}
\toprule
\textbf{ID} & \textbf{Category} & \textbf{\#Qubits} & \textbf{Depth} &  \textbf{Greedy [\%]} &     \textbf{Random [\%]} & \textbf{p-value} & $\bf{\hat{A}_{12}}$ & \textbf{Magnitude} \\
\midrule
 \textbf{0} &       \textbf{\gls{gs}} &        \textbf{3} &     \textbf{5} &  \textbf{50.0 $\pm$ 0.0} & \textbf{29.5 $\pm$ 24.7} & \textbf{8.0e-13} &          \textbf{0.705} &       \textbf{(M)} \\
 \textbf{1} &       \textbf{\gls{gs}} &        \textbf{4} &     \textbf{6} &  \textbf{50.0 $\pm$ 0.0} & \textbf{35.5 $\pm$ 22.8} & \textbf{6.3e-09} &          \textbf{0.645} &       \textbf{(S)} \\
 \textbf{2} &       \textbf{\gls{gs}} &        \textbf{5} &     \textbf{7} & \textbf{67.8 $\pm$ 11.4} & \textbf{59.5 $\pm$ 16.2} & \textbf{7.7e-05} &          \textbf{0.639} &       \textbf{(S)} \\
 \textbf{3} &       \textbf{\gls{gs}} &        \textbf{6} &     \textbf{8} &  \textbf{75.0 $\pm$ 0.0} & \textbf{68.2 $\pm$ 11.2} & \textbf{2.5e-08} &          \textbf{0.635} &       \textbf{(S)} \\
          4 &                \gls{gs} &                 7 &              9 &           80.2 $\pm$ 6.2 &           80.0 $\pm$ 6.9 &           0.9353 &                   0.503 &                (N) \\
 \textbf{5} &       \textbf{\gls{gs}} &        \textbf{8} &    \textbf{10} &  \textbf{86.0 $\pm$ 4.1} &  \textbf{79.4 $\pm$ 6.7} & \textbf{1.0e-13} &          \textbf{0.756} &       \textbf{(L)} \\
 \textbf{6} &       \textbf{\gls{gs}} &        \textbf{9} &    \textbf{11} &  \textbf{89.6 $\pm$ 3.0} &  \textbf{83.9 $\pm$ 5.7} & \textbf{3.9e-15} &          \textbf{0.762} &       \textbf{(L)} \\
\bottomrule
\end{tabular}%
}
\end{table}

\begin{table}[tbp]
    \centering
\caption{Reduction runtime results for the \gls{grov} program category.}
\label{tab:rq1_full_runtime_table_grov}
\resizebox{\columnwidth}{!}{%
\begin{tabular}{rlrr>{\columncolor{lightgray}}r>{\columncolor{lightgray}}r>{\columncolor{lightgray}}r>{\columncolor{lightgray}}rrrl}
\toprule
\textbf{ID} & \textbf{Category} & \textbf{\#Qubits} & \textbf{Depth} & \textbf{\#T} & \textbf{f(\#Q)} &    \textbf{Greedy [ms]} &       \textbf{Random [ms]} & \textbf{p-value} & $\bf{\hat{A}_{12}}$ & \textbf{Magnitude} \\
\midrule
          0 &              \gls{grov} &                 6 &            178 &           18 &              21 &           313.2 $\pm$ 25.7 &          313.2 $\pm$ 25.0 &          0.91342 &                   0.505 &                (N) \\
 \textbf{1} &     \textbf{\gls{grov}} &        \textbf{6} &    \textbf{30} &  \textbf{15} &     \textbf{21} &    \textbf{99.9 $\pm$ 5.0} &  \textbf{103.4 $\pm$ 5.1} & \textbf{2.3e-26} &          \textbf{0.065} &       \textbf{(L)} \\
 \textbf{2} &     \textbf{\gls{grov}} &        \textbf{6} &   \textbf{121} &  \textbf{15} &     \textbf{21} &  \textbf{196.2 $\pm$ 13.8} & \textbf{200.2 $\pm$ 13.8} & \textbf{1.3e-23} &           \textbf{0.09} &       \textbf{(L)} \\
 \textbf{3} &     \textbf{\gls{grov}} &        \textbf{6} &    \textbf{42} &  \textbf{18} &     \textbf{21} &   \textbf{122.3 $\pm$ 7.4} &  \textbf{124.0 $\pm$ 6.9} & \textbf{1.7e-05} &          \textbf{0.324} &       \textbf{(M)} \\
 \textbf{4} &     \textbf{\gls{grov}} &        \textbf{6} &    \textbf{48} &  \textbf{18} &     \textbf{21} &  \textbf{120.5 $\pm$ 11.5} &  \textbf{121.2 $\pm$ 8.0} & \textbf{0.00112} &          \textbf{0.367} &       \textbf{(S)} \\
 \textbf{5} &     \textbf{\gls{grov}} &        \textbf{6} &    \textbf{54} &  \textbf{18} &     \textbf{21} &   \textbf{119.8 $\pm$ 8.7} &  \textbf{122.3 $\pm$ 7.9} & \textbf{4.5e-06} &          \textbf{0.312} &       \textbf{(M)} \\
 \textbf{6} &     \textbf{\gls{grov}} &        \textbf{6} &    \textbf{66} &  \textbf{15} &     \textbf{21} &   \textbf{128.3 $\pm$ 7.4} &  \textbf{132.2 $\pm$ 7.5} & \textbf{5.7e-26} &          \textbf{0.069} &       \textbf{(L)} \\
\bottomrule
\end{tabular}
}
\end{table}

\begin{table}[tbp]
    \centering
\caption{Reduction runtime results for the \textbf{\gls{qwalk}} program category.}
\label{tab:rq1_full_runtime_table_qwalk}
\resizebox{\columnwidth}{!}{%
\begin{tabular}{rlrr>{\columncolor{lightgray}}r>{\columncolor{lightgray}}r>{\columncolor{lightgray}}r>{\columncolor{lightgray}}rrrl}
\toprule
\textbf{ID} & \textbf{Category} & \textbf{\#Qubits} & \textbf{Depth} & \textbf{\#T} & \textbf{f(\#Q)} &    \textbf{Greedy [ms]} &       \textbf{Random [ms]} & \textbf{p-value} & $\bf{\hat{A}_{12}}$ & \textbf{Magnitude} \\
\midrule
 \textbf{0} &    \textbf{\gls{qwalk}} &        \textbf{3} &    \textbf{14} &   \textbf{5} &      \textbf{6} &   \textbf{31.4 $\pm$ 3.8} &  \textbf{31.7 $\pm$ 1.8} & \textbf{1.6e-09} &          \textbf{0.253} &       \textbf{(L)} \\
 \textbf{1} &    \textbf{\gls{qwalk}} &        \textbf{3} &    \textbf{21} &   \textbf{5} &      \textbf{6} &   \textbf{34.7 $\pm$ 2.3} &  \textbf{35.2 $\pm$ 1.3} & \textbf{4.5e-13} &          \textbf{0.204} &       \textbf{(L)} \\
 \textbf{2} &    \textbf{\gls{qwalk}} &        \textbf{3} &    \textbf{28} &   \textbf{3} &      \textbf{6} &   \textbf{31.4 $\pm$ 3.4} &  \textbf{32.8 $\pm$ 1.7} & \textbf{2.6e-28} &          \textbf{0.048} &       \textbf{(L)} \\
 \textbf{3} &    \textbf{\gls{qwalk}} &        \textbf{3} &    \textbf{35} &   \textbf{3} &      \textbf{6} &   \textbf{34.4 $\pm$ 2.4} &  \textbf{36.5 $\pm$ 2.7} & \textbf{2.0e-31} &          \textbf{0.023} &       \textbf{(L)} \\
 \textbf{4} &    \textbf{\gls{qwalk}} &        \textbf{3} &    \textbf{42} &   \textbf{5} &      \textbf{6} &   \textbf{44.5 $\pm$ 2.9} &  \textbf{45.2 $\pm$ 2.2} & \textbf{1.9e-14} &          \textbf{0.186} &       \textbf{(L)} \\
 \textbf{5} &    \textbf{\gls{qwalk}} &        \textbf{3} &    \textbf{49} &   \textbf{5} &      \textbf{6} &   \textbf{48.1 $\pm$ 2.4} &  \textbf{48.4 $\pm$ 3.1} & \textbf{0.01445} &            \textbf{0.4} &       \textbf{(S)} \\
 \textbf{6} &    \textbf{\gls{qwalk}} &        \textbf{3} &    \textbf{56} &   \textbf{3} &      \textbf{6} &   \textbf{44.7 $\pm$ 3.9} &  \textbf{46.8 $\pm$ 8.3} & \textbf{7.4e-25} &          \textbf{0.079} &       \textbf{(L)} \\
\bottomrule
\end{tabular}
}
\end{table}

\begin{table}[tbp]
    \centering
\caption{Reduction runtime results for the \textbf{\gls{var}} program category.}
\label{tab:rq1_full_runtime_table_var}
\resizebox{\columnwidth}{!}{%
\begin{tabular}{rlrr>{\columncolor{lightgray}}r>{\columncolor{lightgray}}r>{\columncolor{lightgray}}r>{\columncolor{lightgray}}rrrl}
\toprule
\textbf{ID} & \textbf{Category} & \textbf{\#Qubits} & \textbf{Depth} & \textbf{\#T} & \textbf{f(\#Q)} &    \textbf{Greedy [ms]} &       \textbf{Random [ms]} & \textbf{p-value} & $\bf{\hat{A}_{12}}$ & \textbf{Magnitude} \\
\midrule
 \textbf{0} &      \textbf{\gls{var}} &        \textbf{2} &     \textbf{6} &   \textbf{3} &      \textbf{3} &  \textbf{24.4 $\pm$ 33.2} &  \textbf{22.8 $\pm$ 25.5} & \textbf{7.1e-27} &          \textbf{0.939} &       \textbf{(L)} \\
 \textbf{1} &      \textbf{\gls{var}} &        \textbf{2} &     \textbf{6} &   \textbf{3} &      \textbf{3} &   \textbf{21.8 $\pm$ 4.0} &   \textbf{20.7 $\pm$ 2.8} & \textbf{1.1e-25} &          \textbf{0.929} &       \textbf{(L)} \\
 \textbf{2} &      \textbf{\gls{var}} &        \textbf{2} &     \textbf{7} &   \textbf{3} &      \textbf{3} &   \textbf{21.3 $\pm$ 4.4} &   \textbf{20.4 $\pm$ 0.6} & \textbf{2.3e-15} &          \textbf{0.824} &       \textbf{(L)} \\
 \textbf{3} &      \textbf{\gls{var}} &        \textbf{2} &     \textbf{7} &   \textbf{3} &      \textbf{3} &   \textbf{21.4 $\pm$ 0.6} &   \textbf{20.7 $\pm$ 3.5} & \textbf{9.4e-28} &          \textbf{0.947} &       \textbf{(L)} \\
 \textbf{4} &      \textbf{\gls{var}} &        \textbf{3} &     \textbf{3} &   \textbf{6} &      \textbf{6} &   \textbf{28.9 $\pm$ 2.7} &   \textbf{28.1 $\pm$ 0.8} & \textbf{3.1e-09} &          \textbf{0.743} &       \textbf{(L)} \\
 \textbf{5} &      \textbf{\gls{var}} &        \textbf{3} &     \textbf{4} &   \textbf{6} &      \textbf{6} &   \textbf{29.8 $\pm$ 1.3} &   \textbf{29.3 $\pm$ 1.0} & \textbf{0.00555} &          \textbf{0.614} &       \textbf{(S)} \\
 \textbf{6} &      \textbf{\gls{var}} &        \textbf{3} &     \textbf{3} &   \textbf{6} &      \textbf{6} &   \textbf{30.2 $\pm$ 5.4} &   \textbf{29.4 $\pm$ 2.8} & \textbf{4.2e-07} &          \textbf{0.707} &       \textbf{(M)} \\
\bottomrule
\end{tabular}
}
\end{table}

\begin{table}[tbp]
    \centering
\caption{Reduction runtime results for the \textbf{\gls{gs}} program category.}
\label{tab:rq1_full_runtime_table_gs}
\resizebox{\columnwidth}{!}{%
\begin{tabular}{rlrr>{\columncolor{lightgray}}r>{\columncolor{lightgray}}r>{\columncolor{lightgray}}r>{\columncolor{lightgray}}rrrl}
\toprule
\textbf{ID} & \textbf{Category} & \textbf{\#Qubits} & \textbf{Depth} & \textbf{\#T} & \textbf{f(\#Q)} &    \textbf{Greedy [ms]} &       \textbf{Random [ms]} & \textbf{p-value} & $\bf{\hat{A}_{12}}$ & \textbf{Magnitude} \\
\midrule
 \textbf{0} &       \textbf{\gls{gs}} &        \textbf{3} &     \textbf{5} &   \textbf{5} &      \textbf{6} &       \textbf{24.3 $\pm$ 0.4} &       \textbf{24.9 $\pm$ 1.3} & \textbf{1.6e-10} &          \textbf{0.238} &       \textbf{(L)} \\
 \textbf{1} &       \textbf{\gls{gs}} &        \textbf{4} &     \textbf{6} &   \textbf{7} &     \textbf{10} &       \textbf{31.8 $\pm$ 1.8} &       \textbf{33.7 $\pm$ 2.1} & \textbf{2.7e-29} &           \textbf{0.04} &       \textbf{(L)} \\
 \textbf{2} &       \textbf{\gls{gs}} &        \textbf{5} &     \textbf{7} &  \textbf{12} &     \textbf{15} &       \textbf{47.6 $\pm$ 5.6} &       \textbf{50.7 $\pm$ 5.1} & \textbf{7.0e-11} &          \textbf{0.233} &       \textbf{(L)} \\
 \textbf{3} &       \textbf{\gls{gs}} &        \textbf{6} &     \textbf{8} &  \textbf{15} &     \textbf{21} &       \textbf{65.0 $\pm$ 2.1} &       \textbf{69.3 $\pm$ 2.7} & \textbf{4.4e-31} &          \textbf{0.025} &       \textbf{(L)} \\
 \textbf{4} &       \textbf{\gls{gs}} &        \textbf{7} &     \textbf{9} &  \textbf{22} &     \textbf{28} &      \textbf{92.7 $\pm$ 10.0} &      \textbf{99.1 $\pm$ 10.4} & \textbf{5.3e-08} &          \textbf{0.277} &       \textbf{(M)} \\
 \textbf{5} &       \textbf{\gls{gs}} &        \textbf{8} &    \textbf{10} &  \textbf{26} &     \textbf{36} &      \textbf{136.8 $\pm$ 9.3} &      \textbf{146.1 $\pm$ 9.7} & \textbf{2.8e-17} &          \textbf{0.154} &       \textbf{(L)} \\
\bottomrule
\end{tabular}
}
\end{table}

In \cref{tab:rq1_full_reduction_table_grov,tab:rq1_full_reduction_table_qwalk,tab:rq1_full_reduction_table_var,tab:rq1_full_reduction_table_gs,tab:rq1_full_runtime_table_grov,tab:rq1_full_runtime_table_qwalk,tab:rq1_full_runtime_table_var,tab:rq1_full_runtime_table_gs}, we provide the reduction rate and runtime results for RQ1 by category and by qubit count.
The \textbf{Greedy} and \textbf{Random} columns depict the reduction rate percentage and the reduction runtime in milliseconds.
The three rightmost columns show the pairwise statistical \gls{mwu} tests between the Greedy and Random approach where $\hat{A}_{12}$ is the Vargha-Delanay effect size along with its nominal effect size magnitude category.
If a result is significant, rows are made bold.
In the runtime tables, we also include two extra columns.
The \textbf{\#T} column, shows the number of objective function evaluations used experimentally by the Greedy approach.
Next, the \textbf{f(\#Qubits)} column shows the theoretical maximum number of objective function calls given by \cref{eq:Greedy_runtime_formula} with $\#Qubits=n$.

\subsection{RQ2 Tables}

\begin{table}[tbp]
    \centering
    \caption{Testing runtime results for the \textbf{\gls{grov}} program category.}
    \label{tab:rq2_full_test_runtime_table_grov}
\resizebox{\columnwidth}{!}{%
\begin{tabular}{rlrr>{\columncolor{lightgray}}r>{\columncolor{lightgray}}r>{\columncolor{lightgray}}rrrrrrrrrrrrr}
& & & & & & & \textbf{DGR} & \multicolumn{3}{c}{\textbf{DG}} & \multicolumn{3}{c}{\textbf{DR}} & \multicolumn{3}{c}{\textbf{GR}} \\
\cmidrule(lr){8-9} \cmidrule(lr){9-11} \cmidrule(lr){12-14} \cmidrule(lr){15-17}
\textbf{ID} & \textbf{Category} & \textbf{\#Qubits} & \textbf{Depth} & \textbf{Default [s]} & \textbf{Greedy [s]} & \textbf{Random [s]} & \textbf{p-value} & \textbf{p-value} & $\bf{\hat{A}_{12}}$ & \textbf{Magnitude} & \textbf{p-value} & $\bf{\hat{A}_{12}}$ & \textbf{Magnitude} & \textbf{p-value} & $\bf{\hat{A}_{12}}$ & \textbf{Magnitude} \\
\midrule
          0 &              \gls{grov} &                 6 &            178 &      77.6 $\pm$ 17.8 &          5.2 $\pm$ 7.2 &     14.8 $\pm$ 16.0 &             1.9e-170 &            4.2e-124 &                      0.956 &                   (L) &            4.7e-124 &                      0.956 &                   (L) &                 1.0 &                       0.33 &                   (M) \\
          1 &              \gls{grov} &                 6 &             30 &       12.9 $\pm$ 2.3 &          1.7 $\pm$ 2.0 &       1.7 $\pm$ 2.0 &             7.2e-176 &            9.0e-134 &                      0.973 &                   (L) &            8.7e-134 &                      0.973 &                   (L) &             0.83744 &                      0.481 &                   (N) \\
          2 &              \gls{grov} &                 6 &            121 &       44.5 $\pm$ 8.0 &          4.1 $\pm$ 5.9 &       4.3 $\pm$ 5.8 &             1.1e-177 &            4.8e-135 &                      0.976 &                   (L) &            4.8e-135 &                      0.976 &                   (L) &             0.93558 &                      0.471 &                   (N) \\
          3 &              \gls{grov} &                 6 &             42 &       16.8 $\pm$ 3.2 &          4.0 $\pm$ 2.9 &       2.1 $\pm$ 2.7 &             6.3e-182 &            3.0e-136 &                      0.978 &                   (L) &            2.5e-136 &                      0.978 &                   (L) &             3.1e-08 &                      0.604 &                   (S) \\
          4 &              \gls{grov} &                 6 &             48 &       16.9 $\pm$ 3.9 &          2.7 $\pm$ 3.5 &       2.6 $\pm$ 3.4 &             1.3e-164 &            2.8e-125 &                      0.958 &                   (L) &            2.8e-125 &                      0.958 &                   (L) &             0.11332 &                      0.523 &                   (N) \\
          5 &              \gls{grov} &                 6 &             54 &       17.5 $\pm$ 3.5 &          3.3 $\pm$ 3.8 &       3.2 $\pm$ 3.8 &             1.1e-169 &            6.2e-129 &                      0.964 &                   (L) &            5.8e-129 &                      0.964 &                   (L) &             0.02287 &                      0.538 &                   (N) \\
          6 &              \gls{grov} &                 6 &             66 &       21.8 $\pm$ 4.7 &          2.7 $\pm$ 3.8 &       2.6 $\pm$ 3.7 &             5.6e-168 &            6.6e-128 &                      0.963 &                   (L) &            8.9e-128 &                      0.962 &                   (L) &             0.13709 &                      0.521 &                   (N) \\
          7 &              \gls{grov} &                 6 &             72 &       22.2 $\pm$ 4.9 &          1.8 $\pm$ 2.5 &       2.4 $\pm$ 3.2 &             1.1e-170 &            5.5e-129 &                      0.964 &                   (L) &            6.9e-129 &                      0.964 &                   (L) &             0.99993 &                      0.427 &                   (N) \\
          8 &              \gls{grov} &                 6 &             78 &       21.8 $\pm$ 4.6 &          2.8 $\pm$ 3.5 &       2.7 $\pm$ 3.7 &             5.1e-173 &            1.1e-131 &                       0.97 &                   (L) &            1.2e-131 &                       0.97 &                   (L) &             0.21497 &                      0.515 &                   (N) \\
          9 &              \gls{grov} &                 6 &            125 &       35.3 $\pm$ 5.8 &          0.7 $\pm$ 0.5 &       1.3 $\pm$ 1.0 &             6.7e-181 &            1.3e-136 &                      0.978 &                   (L) &            1.8e-136 &                      0.978 &                   (L) &             0.99998 &                      0.422 &                   (S) \\
         10 &              \gls{grov} &                 6 &             90 &       25.5 $\pm$ 5.6 &          6.1 $\pm$ 4.3 &       6.0 $\pm$ 4.2 &             2.9e-169 &            7.4e-129 &                      0.964 &                   (L) &            7.5e-129 &                      0.964 &                   (L) &             0.68266 &                      0.491 &                   (N) \\
         11 &              \gls{grov} &                 7 &             57 &      69.8 $\pm$ 12.3 &          3.8 $\pm$ 5.2 &     10.3 $\pm$ 11.8 &             3.3e-183 &            3.8e-135 &                      0.976 &                   (L) &            5.6e-135 &                      0.976 &                   (L) &                 1.0 &                      0.352 &                   (S) \\
         12 &              \gls{grov} &                 7 &             36 &       35.2 $\pm$ 6.8 &          3.7 $\pm$ 5.3 &       3.8 $\pm$ 5.4 &             2.2e-174 &            1.6e-132 &                      0.971 &                   (L) &            1.6e-132 &                      0.971 &                   (L) &             0.95366 &                      0.468 &                   (N) \\
         13 &              \gls{grov} &                 7 &             48 &       41.5 $\pm$ 7.5 &          2.9 $\pm$ 4.1 &       2.9 $\pm$ 4.0 &             1.6e-177 &            5.0e-135 &                      0.976 &                   (L) &            5.3e-135 &                      0.976 &                   (L) &             0.15817 &                      0.519 &                   (N) \\
         14 &              \gls{grov} &                 7 &            205 &     155.0 $\pm$ 27.9 &        17.6 $\pm$ 26.5 &     18.3 $\pm$ 28.0 &             6.0e-177 &            8.0e-134 &                      0.973 &                   (L) &            9.8e-134 &                      0.973 &                   (L) &             0.99975 &                      0.433 &                   (N) \\
         15 &              \gls{grov} &                 7 &            492 &     340.7 $\pm$ 73.0 &        33.6 $\pm$ 53.8 &     39.5 $\pm$ 58.8 &             4.8e-166 &            2.0e-126 &                       0.96 &                   (L) &            2.0e-126 &                       0.96 &                   (L) &             0.69569 &                       0.49 &                   (N) \\
         16 &              \gls{grov} &                 7 &             96 &      67.1 $\pm$ 15.1 &         7.4 $\pm$ 11.6 &      7.7 $\pm$ 12.0 &             6.6e-163 &            4.6e-124 &                      0.956 &                   (L) &            4.8e-124 &                      0.956 &                   (L) &             0.77371 &                      0.486 &                   (N) \\
         17 &              \gls{grov} &                 7 &            108 &      66.5 $\pm$ 11.5 &          5.1 $\pm$ 7.9 &      7.4 $\pm$ 11.2 &             1.9e-178 &            5.3e-135 &                      0.976 &                   (L) &            5.6e-135 &                      0.976 &                   (L) &             0.99951 &                      0.437 &                   (N) \\
         18 &              \gls{grov} &                 7 &            120 &      86.4 $\pm$ 15.4 &         8.2 $\pm$ 11.9 &     11.4 $\pm$ 17.1 &             2.7e-176 &            8.8e-134 &                      0.973 &                   (L) &            9.6e-134 &                      0.973 &                   (L) &             0.99077 &                      0.455 &                   (N) \\
         19 &              \gls{grov} &                 7 &            132 &      82.8 $\pm$ 20.0 &         8.3 $\pm$ 13.0 &      9.4 $\pm$ 15.5 &             3.7e-158 &            1.6e-120 &                      0.949 &                   (L) &            1.5e-120 &                      0.949 &                   (L) &             0.58905 &                      0.496 &                   (N) \\
         20 &              \gls{grov} &                 7 &            144 &      91.8 $\pm$ 15.2 &         9.0 $\pm$ 13.8 &     12.0 $\pm$ 17.8 &             4.8e-180 &            2.9e-136 &                      0.978 &                   (L) &            3.2e-136 &                      0.978 &                   (L) &             0.99927 &                      0.439 &                   (N) \\
         21 &              \gls{grov} &                 7 &            156 &      81.4 $\pm$ 16.3 &         9.1 $\pm$ 14.3 &     10.6 $\pm$ 15.5 &             4.6e-171 &            4.4e-130 &                      0.967 &                   (L) &            4.6e-130 &                      0.967 &                   (L) &             0.92731 &                      0.472 &                   (N) \\
         22 &              \gls{grov} &                 8 &             79 &     200.6 $\pm$ 38.2 &        12.4 $\pm$ 20.0 &     14.0 $\pm$ 20.5 &             6.6e-173 &            2.6e-131 &                      0.969 &                   (L) &            2.7e-131 &                      0.969 &                   (L) &             0.98047 &                       0.46 &                   (N) \\
         23 &              \gls{grov} &                 8 &             54 &      97.6 $\pm$ 18.8 &         7.3 $\pm$ 12.2 &      7.3 $\pm$ 12.4 &             1.4e-172 &            2.7e-131 &                      0.969 &                   (L) &            2.8e-131 &                      0.969 &                   (L) &             0.83588 &                      0.481 &                   (N) \\
         24 &              \gls{grov} &                 8 &             84 &     124.1 $\pm$ 18.8 &        11.0 $\pm$ 18.9 &     11.9 $\pm$ 19.9 &             2.2e-182 &            1.0e-138 &                      0.982 &                   (L) &            1.0e-138 &                      0.982 &                   (L) &             0.62297 &                      0.494 &                   (N) \\
         25 &              \gls{grov} &                 8 &            114 &     171.2 $\pm$ 29.5 &        13.6 $\pm$ 22.7 &     16.5 $\pm$ 27.8 &             2.1e-177 &            5.7e-135 &                      0.976 &                   (L) &            5.5e-135 &                      0.976 &                   (L) &             0.45543 &                      0.502 &                   (N) \\
         26 &              \gls{grov} &                 8 &            144 &     192.9 $\pm$ 39.1 &        18.8 $\pm$ 28.7 &     25.0 $\pm$ 37.7 &             2.7e-171 &            3.6e-129 &                      0.965 &                   (L) &            4.6e-129 &                      0.965 &                   (L) &             0.99999 &                      0.419 &                   (S) \\
         27 &              \gls{grov} &                 8 &            174 &     218.0 $\pm$ 39.5 &        20.8 $\pm$ 33.7 &     20.8 $\pm$ 33.8 &             2.0e-174 &            1.7e-132 &                      0.971 &                   (L) &            1.7e-132 &                      0.971 &                   (L) &             0.03615 &                      0.535 &                   (N) \\
         28 &              \gls{grov} &                 8 &            204 &     297.1 $\pm$ 52.0 &        23.8 $\pm$ 38.4 &     32.7 $\pm$ 52.0 &             3.3e-176 &            9.6e-134 &                      0.973 &                   (L) &            9.7e-134 &                      0.973 &                   (L) &             0.98504 &                      0.458 &                   (N) \\
         29 &              \gls{grov} &                 8 &            234 &     316.8 $\pm$ 51.2 &        27.0 $\pm$ 45.2 &     30.5 $\pm$ 50.3 &             4.0e-179 &            3.2e-136 &                      0.978 &                   (L) &            3.2e-136 &                      0.978 &                   (L) &             0.79822 &                      0.484 &                   (N) \\
         30 &              \gls{grov} &                 8 &            264 &     338.2 $\pm$ 50.7 &        35.0 $\pm$ 58.1 &     32.6 $\pm$ 54.8 &             1.6e-182 &            1.0e-138 &                      0.982 &                   (L) &            1.0e-138 &                      0.982 &                   (L) &             0.11062 &                      0.524 &                   (N) \\
         31 &              \gls{grov} &                 8 &            294 &     372.8 $\pm$ 46.6 &        40.3 $\pm$ 60.3 &     39.1 $\pm$ 64.4 &             1.8e-187 &            1.6e-142 &                      0.989 &                   (L) &            1.6e-142 &                      0.989 &                   (L) &             0.27613 &                      0.511 &                   (N) \\
         32 &              \gls{grov} &                 8 &            324 &     428.7 $\pm$ 67.1 &        49.6 $\pm$ 79.3 &     49.4 $\pm$ 78.1 &             7.6e-181 &            1.8e-137 &                       0.98 &                   (L) &            1.8e-137 &                       0.98 &                   (L) &             0.11948 &                      0.523 &                   (N) \\
         33 &              \gls{grov} &                 8 &            354 &     462.1 $\pm$ 88.2 &        43.7 $\pm$ 75.1 &     46.5 $\pm$ 80.4 &             1.8e-172 &            2.8e-131 &                      0.969 &                   (L) &            2.8e-131 &                      0.969 &                   (L) &             0.60036 &                      0.495 &                   (N) \\
         34 &              \gls{grov} &                 9 &            101 &     521.0 $\pm$ 97.1 &        15.8 $\pm$ 21.4 &     20.7 $\pm$ 31.6 &             3.7e-174 &            1.6e-132 &                      0.971 &                   (L) &            1.6e-132 &                      0.971 &                   (L) &             0.27209 &                      0.512 &                   (N) \\
         35 &              \gls{grov} &                 9 &            166 &    570.6 $\pm$ 101.0 &        37.4 $\pm$ 67.6 &     47.7 $\pm$ 85.2 &             4.1e-179 &            3.2e-136 &                      0.978 &                   (L) &            3.2e-136 &                      0.978 &                   (L) &             0.79871 &                      0.484 &                   (N) \\
         36 &              \gls{grov} &                 9 &            144 &     498.5 $\pm$ 88.5 &        36.7 $\pm$ 62.2 &     38.7 $\pm$ 66.7 &             4.3e-179 &            3.2e-136 &                      0.978 &                   (L) &            3.3e-136 &                      0.978 &                   (L) &             0.76164 &                      0.486 &                   (N) \\
         37 &              \gls{grov} &                 9 &            204 &    617.6 $\pm$ 121.6 &        49.0 $\pm$ 83.6 &     53.7 $\pm$ 89.9 &             1.6e-172 &            2.8e-131 &                      0.969 &                   (L) &            2.8e-131 &                      0.969 &                   (L) &             0.76378 &                      0.486 &                   (N) \\
         38 &              \gls{grov} &                 9 &            264 &    827.5 $\pm$ 163.6 &       61.8 $\pm$ 113.0 &    68.3 $\pm$ 120.9 &             5.7e-171 &            4.6e-130 &                      0.967 &                   (L) &            4.6e-130 &                      0.967 &                   (L) &             0.13181 &                      0.522 &                   (N) \\
         39 &              \gls{grov} &                 9 &            324 &    968.9 $\pm$ 152.7 &       83.0 $\pm$ 140.5 &    86.8 $\pm$ 156.0 &             1.0e-180 &            1.8e-137 &                       0.98 &                   (L) &            1.8e-137 &                       0.98 &                   (L) &             0.56393 &                      0.497 &                   (N) \\
         40 &              \gls{grov} &                 9 &            384 &   1089.0 $\pm$ 158.8 &      104.0 $\pm$ 181.9 &   106.2 $\pm$ 190.5 &             1.9e-182 &            1.0e-138 &                      0.982 &                   (L) &            1.0e-138 &                      0.982 &                   (L) &             0.80667 &                      0.483 &                   (N) \\
         41 &              \gls{grov} &                 9 &            444 &   1284.6 $\pm$ 179.9 &      111.6 $\pm$ 196.8 &   120.0 $\pm$ 220.2 &             2.2e-182 &            1.0e-138 &                      0.982 &                   (L) &            1.0e-138 &                      0.982 &                   (L) &             0.59202 &                      0.496 &                   (N) \\
         42 &              \gls{grov} &                 9 &            504 &   1436.5 $\pm$ 247.4 &      135.2 $\pm$ 227.5 &   156.8 $\pm$ 257.9 &             5.4e-176 &            9.9e-134 &                      0.973 &                   (L) &            9.9e-134 &                      0.973 &                   (L) &             0.94605 &                      0.469 &                   (N) \\
         43 &              \gls{grov} &                 9 &            564 &   1537.1 $\pm$ 263.9 &      121.9 $\pm$ 210.0 &   143.0 $\pm$ 239.5 &             3.0e-176 &            9.8e-134 &                      0.973 &                   (L) &            9.9e-134 &                      0.973 &                   (L) &             0.98856 &                      0.456 &                   (N) \\
         44 &              \gls{grov} &                 9 &            624 &   1724.9 $\pm$ 327.6 &      136.1 $\pm$ 265.8 &   140.4 $\pm$ 279.0 &             7.0e-171 &            4.6e-130 &                      0.967 &                   (L) &            4.7e-130 &                      0.967 &                   (L) &             0.71139 &                      0.489 &                   (N) \\
         45 &              \gls{grov} &                 9 &            684 &   1895.1 $\pm$ 289.4 &      196.9 $\pm$ 304.6 &   199.7 $\pm$ 332.6 &             4.8e-179 &            3.3e-136 &                      0.978 &                   (L) &            3.3e-136 &                      0.978 &                   (L) &              0.5949 &                      0.495 &                   (N) \\
\bottomrule
\end{tabular}%
}
\end{table}

\begin{table}[tbp]
    \centering
    \caption{Testing runtime results for the \textbf{\gls{qwalk}} program category.}
    \label{tab:rq2_full_test_runtime_table_qwalk}
\resizebox{\columnwidth}{!}{%
\begin{tabular}{rlrr>{\columncolor{lightgray}}r>{\columncolor{lightgray}}r>{\columncolor{lightgray}}rrrrrrrrrrrrr}
& & & & & & & \textbf{DGR} & \multicolumn{3}{c}{\textbf{DG}} & \multicolumn{3}{c}{\textbf{DR}} & \multicolumn{3}{c}{\textbf{GR}} \\
\cmidrule(lr){8-9} \cmidrule(lr){9-11} \cmidrule(lr){12-14} \cmidrule(lr){15-17}
\textbf{ID} & \textbf{Category} & \textbf{\#Qubits} & \textbf{Depth} & \textbf{Default [s]} & \textbf{Greedy [s]} & \textbf{Random [s]} & \textbf{p-value} & \textbf{p-value} & $\bf{\hat{A}_{12}}$ & \textbf{Magnitude} & \textbf{p-value} & $\bf{\hat{A}_{12}}$ & \textbf{Magnitude} & \textbf{p-value} & $\bf{\hat{A}_{12}}$ & \textbf{Magnitude} \\
\midrule
          0 &             \gls{qwalk} &                 3 &             14 &        0.4 $\pm$ 0.1 &          0.1 $\pm$ 0.1 &       0.1 $\pm$ 0.1 &             2.2e-156 &            3.6e-121 &                      0.947 &                   (L) &            3.2e-118 &                      0.943 &                   (L) &              0.0795 &                      0.527 &                   (N) \\
          1 &             \gls{qwalk} &                 3 &             21 &        0.3 $\pm$ 0.1 &          0.1 $\pm$ 0.1 &       0.1 $\pm$ 0.1 &             3.2e-129 &             1.2e-99 &                      0.905 &                   (L) &             1.2e-98 &                      0.904 &                   (L) &             0.05597 &                       0.53 &                   (N) \\
          2 &             \gls{qwalk} &                 3 &             28 &        0.2 $\pm$ 0.1 &          0.1 $\pm$ 0.1 &       0.1 $\pm$ 0.1 &              0.24595 &             0.05989 &                       0.53 &                   (N) &             0.10246 &                      0.524 &                   (N) &             0.67428 &                      0.491 &                   (N) \\
          3 &             \gls{qwalk} &                 3 &             35 &        0.3 $\pm$ 0.2 &          0.2 $\pm$ 0.2 &       0.2 $\pm$ 0.2 &              2.1e-08 &             1.5e-05 &                       0.58 &                   (S) &             3.1e-09 &                      0.612 &                   (S) &             0.08183 &                      0.527 &                   (N) \\
          4 &             \gls{qwalk} &                 3 &             42 &        1.0 $\pm$ 0.2 &          0.3 $\pm$ 0.2 &       0.2 $\pm$ 0.2 &             1.8e-158 &            6.7e-120 &                      0.947 &                   (L) &            1.9e-120 &                      0.948 &                   (L) &             3.8e-04 &                      0.565 &                   (N) \\
          5 &             \gls{qwalk} &                 3 &             49 &        0.5 $\pm$ 0.2 &          0.2 $\pm$ 0.1 &       0.2 $\pm$ 0.1 &             1.4e-119 &             1.0e-91 &                      0.889 &                   (L) &             4.2e-92 &                       0.89 &                   (L) &             0.40572 &                      0.505 &                   (N) \\
          6 &             \gls{qwalk} &                 3 &             56 &        0.2 $\pm$ 0.1 &          0.2 $\pm$ 0.1 &       0.2 $\pm$ 0.1 &              0.11308 &             0.08203 &                      0.527 &                   (N) &             0.10554 &                      0.524 &                   (N) &             0.01319 &                      0.543 &                   (N) \\
          7 &             \gls{qwalk} &                 3 &             63 &        0.5 $\pm$ 0.3 &          0.3 $\pm$ 0.3 &       0.3 $\pm$ 0.3 &              1.2e-12 &             1.5e-07 &                      0.598 &                   (S) &             2.3e-12 &                      0.633 &                   (S) &             0.00157 &                      0.557 &                   (N) \\
          8 &             \gls{qwalk} &                 3 &             70 &        1.5 $\pm$ 0.4 &          0.4 $\pm$ 0.3 &       0.4 $\pm$ 0.3 &             9.6e-154 &            1.0e-116 &                      0.941 &                   (L) &            4.1e-118 &                      0.943 &                   (L) &              0.2666 &                      0.512 &                   (N) \\
          9 &             \gls{qwalk} &                 3 &             77 &        0.8 $\pm$ 0.3 &          0.3 $\pm$ 0.2 &       0.3 $\pm$ 0.2 &             8.3e-119 &             2.2e-90 &                      0.886 &                   (L) &             3.3e-91 &                      0.888 &                   (L) &             0.00728 &                      0.547 &                   (N) \\
         10 &             \gls{qwalk} &                 3 &             84 &        0.4 $\pm$ 0.2 &          0.3 $\pm$ 0.2 &       0.3 $\pm$ 0.2 &              0.00743 &             0.02075 &                      0.539 &                   (N) &             0.00277 &                      0.553 &                   (N) &             0.03519 &                      0.535 &                   (N) \\
         11 &             \gls{qwalk} &                 3 &             91 &        0.6 $\pm$ 0.4 &          0.4 $\pm$ 0.4 &       0.4 $\pm$ 0.4 &              0.00155 &              0.0039 &                      0.551 &                   (N) &             2.2e-04 &                      0.567 &                   (N) &             0.36425 &                      0.507 &                   (N) \\
         12 &             \gls{qwalk} &                 3 &             98 &        2.0 $\pm$ 0.4 &          0.5 $\pm$ 0.4 &       0.5 $\pm$ 0.4 &             8.7e-162 &            7.0e-123 &                      0.952 &                   (L) &            1.0e-122 &                      0.952 &                   (L) &             6.7e-04 &                      0.561 &                   (N) \\
         13 &             \gls{qwalk} &                 4 &             18 &        0.3 $\pm$ 0.2 &          0.2 $\pm$ 0.2 &       0.2 $\pm$ 0.2 &              1.7e-05 &             7.9e-05 &                      0.573 &                   (N) &             7.3e-06 &                      0.583 &                   (S) &             0.70417 &                       0.49 &                   (N) \\
         14 &             \gls{qwalk} &                 4 &             27 &        0.5 $\pm$ 0.4 &          0.3 $\pm$ 0.3 &       0.4 $\pm$ 0.3 &              3.6e-12 &             9.1e-12 &                      0.629 &                   (S) &             6.3e-09 &                       0.61 &                   (S) &             0.89609 &                      0.476 &                   (N) \\
         15 &             \gls{qwalk} &                 4 &             36 &        1.2 $\pm$ 0.8 &          0.4 $\pm$ 0.4 &       0.5 $\pm$ 0.5 &              2.1e-52 &             2.6e-41 &                      0.758 &                   (L) &             1.7e-40 &                      0.755 &                   (L) &             0.09555 &                      0.525 &                   (N) \\
         16 &             \gls{qwalk} &                 4 &             45 &        0.8 $\pm$ 0.6 &          0.5 $\pm$ 0.5 &       0.5 $\pm$ 0.5 &              1.1e-10 &             1.8e-10 &                      0.621 &                   (S) &             2.3e-08 &                      0.605 &                   (S) &             0.32079 &                      0.509 &                   (N) \\
         17 &             \gls{qwalk} &                 4 &             54 &        1.6 $\pm$ 1.2 &          0.6 $\pm$ 0.6 &       0.6 $\pm$ 0.6 &              2.7e-52 &             4.0e-40 &                      0.754 &                   (L) &             7.2e-41 &                      0.757 &                   (L) &             0.00848 &                      0.546 &                   (N) \\
         18 &             \gls{qwalk} &                 4 &             63 &        1.0 $\pm$ 0.8 &          0.7 $\pm$ 0.7 &       0.7 $\pm$ 0.7 &              7.2e-13 &             9.2e-12 &                      0.629 &                   (S) &             2.7e-10 &                      0.619 &                   (S) &             0.16875 &                      0.518 &                   (N) \\
         19 &             \gls{qwalk} &                 4 &             72 &        2.0 $\pm$ 1.5 &          0.7 $\pm$ 0.8 &       0.7 $\pm$ 0.8 &              2.9e-49 &             1.6e-35 &                      0.738 &                   (L) &             2.6e-40 &                      0.755 &                   (L) &              0.0026 &                      0.554 &                   (N) \\
         20 &             \gls{qwalk} &                 4 &             81 &        1.3 $\pm$ 1.1 &          0.7 $\pm$ 0.8 &       0.7 $\pm$ 0.9 &              3.8e-18 &             2.3e-16 &                      0.656 &                   (S) &             7.0e-14 &                      0.642 &                   (S) &             0.29837 &                       0.51 &                   (N) \\
         21 &             \gls{qwalk} &                 4 &             90 &        1.3 $\pm$ 1.0 &          0.5 $\pm$ 0.5 &       0.5 $\pm$ 0.5 &              8.2e-32 &             2.6e-24 &                      0.694 &                   (M) &             6.1e-25 &                      0.697 &                   (M) &             4.1e-04 &                      0.564 &                   (N) \\
         22 &             \gls{qwalk} &                 4 &             99 &        3.0 $\pm$ 2.0 &          0.6 $\pm$ 0.5 &       0.6 $\pm$ 0.5 &              9.1e-82 &             5.4e-60 &                      0.813 &                   (L) &             2.0e-63 &                      0.822 &                   (L) &             7.1e-06 &                      0.583 &                   (S) \\
         23 &             \gls{qwalk} &                 4 &            108 &        1.7 $\pm$ 1.1 &          0.6 $\pm$ 0.6 &       0.6 $\pm$ 0.5 &              9.4e-53 &             3.3e-40 &                      0.754 &                   (L) &             7.0e-40 &                      0.753 &                   (L) &             1.4e-05 &                       0.58 &                   (S) \\
         24 &             \gls{qwalk} &                 4 &            117 &        3.4 $\pm$ 2.4 &          1.1 $\pm$ 1.2 &       1.1 $\pm$ 1.2 &              6.0e-63 &             4.3e-49 &                      0.782 &                   (L) &             7.4e-49 &                      0.782 &                   (L) &             0.31182 &                      0.509 &                   (N) \\
         25 &             \gls{qwalk} &                 4 &            126 &        1.7 $\pm$ 1.3 &          1.1 $\pm$ 1.3 &       1.2 $\pm$ 1.3 &              2.8e-06 &             1.9e-05 &                      0.579 &                   (S) &             1.8e-06 &                      0.589 &                   (S) &             0.50036 &                        0.5 &                   (N) \\
         26 &             \gls{qwalk} &                 5 &             22 &        0.5 $\pm$ 0.3 &          0.3 $\pm$ 0.3 &       0.3 $\pm$ 0.3 &              8.9e-13 &             2.4e-10 &                       0.62 &                   (S) &             1.2e-11 &                      0.629 &                   (S) &             0.50713 &                        0.5 &                   (N) \\
         27 &             \gls{qwalk} &                 5 &             33 &        0.8 $\pm$ 0.5 &          0.5 $\pm$ 0.5 &       0.5 $\pm$ 0.5 &              4.6e-13 &             1.5e-09 &                      0.614 &                   (S) &             5.4e-13 &                      0.637 &                   (S) &             0.61007 &                      0.495 &                   (N) \\
         28 &             \gls{qwalk} &                 5 &             44 &        1.8 $\pm$ 1.0 &          0.9 $\pm$ 1.1 &       0.9 $\pm$ 1.1 &              1.0e-23 &             5.8e-16 &                      0.654 &                   (S) &             3.5e-22 &                      0.685 &                   (M) &             0.06513 &                      0.529 &                   (N) \\
         29 &             \gls{qwalk} &                 5 &             55 &        1.6 $\pm$ 1.3 &          1.0 $\pm$ 1.3 &       1.0 $\pm$ 1.3 &              3.7e-10 &             9.2e-08 &                        0.6 &                   (S) &             5.1e-10 &                      0.618 &                   (S) &             0.20521 &                      0.516 &                   (N) \\
         30 &             \gls{qwalk} &                 5 &             66 &        4.3 $\pm$ 2.1 &          2.3 $\pm$ 2.3 &       2.3 $\pm$ 2.3 &              5.5e-31 &             1.3e-19 &                      0.673 &                   (M) &             2.2e-30 &                      0.719 &                   (M) &             0.43587 &                      0.503 &                   (N) \\
         31 &             \gls{qwalk} &                 5 &             77 &        5.3 $\pm$ 2.6 &          1.6 $\pm$ 2.0 &       2.0 $\pm$ 2.6 &              5.6e-85 &             2.2e-74 &                      0.851 &                   (L) &             3.9e-55 &                        0.8 &                   (L) &             0.99627 &                      0.448 &                   (N) \\
         32 &             \gls{qwalk} &                 5 &             88 &        9.2 $\pm$ 1.7 &          1.9 $\pm$ 2.5 &       1.9 $\pm$ 2.4 &             1.1e-172 &            2.3e-131 &                      0.969 &                   (L) &            2.5e-131 &                      0.969 &                   (L) &             0.09474 &                      0.525 &                   (N) \\
         33 &             \gls{qwalk} &                 5 &             99 &        6.2 $\pm$ 3.5 &          1.9 $\pm$ 2.5 &       2.1 $\pm$ 2.7 &              1.2e-79 &             2.4e-63 &                      0.823 &                   (L) &             3.7e-59 &                      0.811 &                   (L) &             0.95303 &                      0.468 &                   (N) \\
         34 &             \gls{qwalk} &                 5 &            110 &        7.5 $\pm$ 4.1 &          1.6 $\pm$ 2.1 &       1.6 $\pm$ 2.1 &              3.4e-98 &             2.4e-75 &                      0.853 &                   (L) &             2.5e-75 &                      0.853 &                   (L) &             0.07096 &                      0.528 &                   (N) \\
         35 &             \gls{qwalk} &                 5 &            121 &       10.3 $\pm$ 4.0 &          2.6 $\pm$ 3.4 &       2.6 $\pm$ 3.4 &             3.2e-133 &            2.5e-101 &                      0.911 &                   (L) &            2.0e-101 &                      0.911 &                   (L) &             0.00711 &                      0.547 &                   (N) \\
         36 &             \gls{qwalk} &                 5 &            132 &       13.3 $\pm$ 2.8 &          2.4 $\pm$ 3.0 &       2.4 $\pm$ 3.0 &             8.1e-168 &            1.1e-127 &                      0.962 &                   (L) &            1.1e-127 &                      0.962 &                   (L) &             0.09462 &                      0.525 &                   (N) \\
         37 &             \gls{qwalk} &                 5 &            143 &       10.3 $\pm$ 5.0 &          5.9 $\pm$ 5.7 &       6.1 $\pm$ 5.8 &              7.2e-25 &             2.3e-15 &                      0.651 &                   (S) &             2.3e-25 &                      0.699 &                   (M) &             0.40798 &                      0.504 &                   (N) \\
         38 &             \gls{qwalk} &                 5 &            154 &       15.2 $\pm$ 3.4 &          3.5 $\pm$ 4.6 &       3.6 $\pm$ 4.8 &             1.2e-160 &            7.0e-123 &                      0.953 &                   (L) &            5.2e-121 &                       0.95 &                   (L) &             0.99199 &                      0.454 &                   (N) \\
\bottomrule
\end{tabular}%
}
\end{table}

\begin{table}[tbp]
    \centering
    \caption{Testing runtime results for the \textbf{\gls{var}} program category.}
    \label{tab:rq2_full_test_runtime_table_var}
\resizebox{\columnwidth}{!}{%
\begin{tabular}{rlrr>{\columncolor{lightgray}}r>{\columncolor{lightgray}}r>{\columncolor{lightgray}}rrrrrrrrrrrrr}
& & & & & & & \textbf{DGR} & \multicolumn{3}{c}{\textbf{DG}} & \multicolumn{3}{c}{\textbf{DR}} & \multicolumn{3}{c}{\textbf{GR}} \\
\cmidrule(lr){8-9} \cmidrule(lr){9-11} \cmidrule(lr){12-14} \cmidrule(lr){15-17}
\textbf{ID} & \textbf{Category} & \textbf{\#Qubits} & \textbf{Depth} & \textbf{Default [s]} & \textbf{Greedy [s]} & \textbf{Random [s]} & \textbf{p-value} & \textbf{p-value} & $\bf{\hat{A}_{12}}$ & \textbf{Magnitude} & \textbf{p-value} & $\bf{\hat{A}_{12}}$ & \textbf{Magnitude} & \textbf{p-value} & $\bf{\hat{A}_{12}}$ & \textbf{Magnitude} \\
\midrule
          0 &               \gls{var} &                 2 &              6 &        0.3 $\pm$ 0.0 &          0.1 $\pm$ 0.1 &       0.1 $\pm$ 0.1 &             4.6e-145 &            3.5e-147 &                      0.993 &                   (L) &             1.1e-74 &                      0.849 &                   (L) &             0.99989 &                      0.429 &                   (N) \\
          1 &               \gls{var} &                 2 &              6 &        0.3 $\pm$ 0.0 &          0.1 $\pm$ 0.1 &       0.1 $\pm$ 0.1 &             2.0e-121 &            8.7e-112 &                      0.931 &                   (L) &             4.6e-71 &                      0.842 &                   (L) &             0.99998 &                      0.422 &                   (S) \\
          2 &               \gls{var} &                 2 &              7 &        0.3 $\pm$ 0.0 &          0.1 $\pm$ 0.1 &       0.1 $\pm$ 0.1 &             7.0e-130 &            2.6e-113 &                      0.934 &                   (L) &             2.0e-83 &                      0.871 &                   (L) &             0.99951 &                      0.437 &                   (N) \\
          3 &               \gls{var} &                 2 &              7 &        0.3 $\pm$ 0.0 &          0.1 $\pm$ 0.1 &       0.1 $\pm$ 0.1 &             1.0e-136 &            1.5e-115 &                      0.938 &                   (L) &             1.5e-93 &                      0.894 &                   (L) &             0.90879 &                      0.474 &                   (N) \\
          4 &               \gls{var} &                 3 &              3 &        0.1 $\pm$ 0.1 &          0.1 $\pm$ 0.0 &       0.1 $\pm$ 0.0 &              1.4e-37 &             2.3e-35 &                      0.736 &                   (M) &             5.8e-25 &                      0.697 &                   (M) &             0.49374 &                        0.5 &                   (N) \\
          5 &               \gls{var} &                 3 &              4 &        0.2 $\pm$ 0.1 &          0.1 $\pm$ 0.0 &       0.1 $\pm$ 0.0 &              1.6e-45 &             1.6e-39 &                      0.751 &                   (L) &             3.1e-32 &                      0.726 &                   (M) &             0.80405 &                      0.484 &                   (N) \\
          6 &               \gls{var} &                 3 &              3 &        0.2 $\pm$ 0.1 &          0.1 $\pm$ 0.0 &       0.1 $\pm$ 0.0 &              2.8e-45 &             2.7e-39 &                       0.75 &                   (L) &             1.0e-31 &                      0.724 &                   (M) &             0.94248 &                       0.47 &                   (N) \\
          7 &               \gls{var} &                 3 &              4 &        0.2 $\pm$ 0.1 &          0.1 $\pm$ 0.0 &       0.1 $\pm$ 0.1 &              9.1e-44 &             8.3e-41 &                      0.754 &                   (L) &             2.3e-28 &                      0.711 &                   (M) &               0.866 &                      0.479 &                   (N) \\
          8 &               \gls{var} &                 3 &              4 &        0.4 $\pm$ 0.1 &          0.1 $\pm$ 0.1 &       0.1 $\pm$ 0.1 &             5.3e-193 &            1.2e-150 &                        1.0 &                   (L) &            1.9e-143 &                      0.989 &                   (L) &             0.93925 &                       0.47 &                   (N) \\
          9 &               \gls{var} &                 3 &              4 &        0.5 $\pm$ 0.1 &          0.1 $\pm$ 0.1 &       0.1 $\pm$ 0.1 &             6.2e-196 &            1.7e-150 &                        1.0 &                   (L) &            2.4e-148 &                      0.998 &                   (L) &             0.53582 &                      0.498 &                   (N) \\
         10 &               \gls{var} &                 3 &              5 &        0.5 $\pm$ 0.1 &          0.1 $\pm$ 0.1 &       0.1 $\pm$ 0.1 &             1.4e-194 &            1.3e-150 &                        1.0 &                   (L) &            2.7e-146 &                      0.994 &                   (L) &             0.36691 &                      0.507 &                   (N) \\
         11 &               \gls{var} &                 3 &              5 &        0.5 $\pm$ 0.0 &          0.1 $\pm$ 0.1 &       0.1 $\pm$ 0.1 &             7.0e-198 &            5.4e-152 &                        1.0 &                   (L) &            1.7e-149 &                      0.999 &                   (L) &             0.99805 &                      0.445 &                   (N) \\
         12 &               \gls{var} &                 4 &             69 &        1.3 $\pm$ 0.9 &          0.3 $\pm$ 0.2 &       0.4 $\pm$ 0.3 &              9.0e-67 &             3.6e-50 &                      0.785 &                   (L) &             1.2e-43 &                      0.766 &                   (L) &                 1.0 &                      0.349 &                   (S) \\
         13 &               \gls{var} &                 4 &             70 &        8.8 $\pm$ 1.1 &          0.3 $\pm$ 0.2 &       0.7 $\pm$ 0.6 &             1.3e-206 &            8.5e-150 &                        1.0 &                   (L) &            5.0e-149 &                        1.0 &                   (L) &                 1.0 &                      0.302 &                   (M) \\
         14 &               \gls{var} &                 4 &             70 &        8.7 $\pm$ 0.9 &          0.3 $\pm$ 0.2 &       0.7 $\pm$ 0.5 &             2.9e-203 &            6.1e-150 &                        1.0 &                   (L) &            4.9e-149 &                        1.0 &                   (L) &                 1.0 &                      0.338 &                   (S) \\
         15 &               \gls{var} &                 4 &             70 &        8.4 $\pm$ 0.5 &          0.3 $\pm$ 0.2 &       0.7 $\pm$ 0.6 &             6.8e-208 &            6.6e-150 &                        1.0 &                   (L) &            5.0e-149 &                        1.0 &                   (L) &                 1.0 &                       0.29 &                   (M) \\
         16 &               \gls{var} &                 4 &             11 &        0.9 $\pm$ 0.2 &          0.1 $\pm$ 0.1 &       0.2 $\pm$ 0.1 &             6.5e-167 &            2.9e-128 &                       0.96 &                   (L) &            3.7e-127 &                       0.96 &                   (L) &             0.54041 &                      0.498 &                   (N) \\
         17 &               \gls{var} &                 4 &             11 &        0.9 $\pm$ 0.2 &          0.1 $\pm$ 0.1 &       0.2 $\pm$ 0.1 &             6.6e-166 &            1.9e-127 &                      0.958 &                   (L) &            2.4e-126 &                      0.958 &                   (L) &             0.03778 &                      0.534 &                   (N) \\
         18 &               \gls{var} &                 4 &             12 &        0.9 $\pm$ 0.3 &          0.1 $\pm$ 0.1 &       0.2 $\pm$ 0.1 &             3.7e-143 &            1.6e-110 &                      0.926 &                   (L) &            5.5e-109 &                      0.925 &                   (L) &             0.12775 &                      0.522 &                   (N) \\
         19 &               \gls{var} &                 4 &             12 &        0.9 $\pm$ 0.2 &          0.1 $\pm$ 0.1 &       0.2 $\pm$ 0.1 &             6.1e-169 &            1.8e-129 &                      0.963 &                   (L) &            1.4e-128 &                      0.963 &                   (L) &             0.10277 &                      0.524 &                   (N) \\
         20 &               \gls{var} &                 5 &             11 &        0.4 $\pm$ 0.3 &          0.2 $\pm$ 0.2 &       0.2 $\pm$ 0.2 &              1.9e-36 &             1.0e-27 &                      0.708 &                   (M) &             8.1e-22 &                      0.683 &                   (M) &                 1.0 &                      0.365 &                   (S) \\
         21 &               \gls{var} &                 5 &             11 &        0.4 $\pm$ 0.3 &          0.2 $\pm$ 0.2 &       0.2 $\pm$ 0.2 &              4.9e-36 &             1.5e-27 &                      0.708 &                   (M) &             2.1e-23 &                      0.691 &                   (M) &                 1.0 &                      0.381 &                   (S) \\
         22 &               \gls{var} &                 5 &             12 &        0.4 $\pm$ 0.3 &          0.2 $\pm$ 0.2 &       0.2 $\pm$ 0.2 &              1.1e-34 &             5.7e-27 &                      0.706 &                   (M) &             2.1e-24 &                      0.695 &                   (M) &                 1.0 &                      0.405 &                   (S) \\
         23 &               \gls{var} &                 5 &             12 &        0.4 $\pm$ 0.3 &          0.2 $\pm$ 0.2 &       0.2 $\pm$ 0.2 &              3.8e-30 &             6.7e-24 &                      0.693 &                   (M) &             5.7e-20 &                      0.675 &                   (M) &                 1.0 &                      0.401 &                   (S) \\
         24 &               \gls{var} &                 6 &              8 &        0.3 $\pm$ 0.1 &          0.1 $\pm$ 0.1 &       0.2 $\pm$ 0.1 &              2.2e-73 &             6.1e-53 &                      0.792 &                   (L) &             4.5e-56 &                      0.802 &                   (L) &                 1.0 &                       0.38 &                   (S) \\
         25 &               \gls{var} &                 6 &              8 &        0.3 $\pm$ 0.1 &          0.1 $\pm$ 0.1 &       0.1 $\pm$ 0.1 &             3.5e-185 &            3.8e-143 &                      0.987 &                   (L) &            1.2e-133 &                      0.971 &                   (L) &                 1.0 &                      0.379 &                   (S) \\
         26 &               \gls{var} &                 6 &              8 &        0.4 $\pm$ 0.1 &          0.1 $\pm$ 0.1 &       0.2 $\pm$ 0.1 &              2.0e-73 &             7.0e-54 &                      0.795 &                   (L) &             1.8e-55 &                        0.8 &                   (L) &                 1.0 &                      0.387 &                   (S) \\
         27 &               \gls{var} &                 6 &              8 &        0.3 $\pm$ 0.1 &          0.1 $\pm$ 0.1 &       0.2 $\pm$ 0.1 &              3.8e-64 &             4.4e-47 &                      0.775 &                   (L) &             2.5e-45 &                       0.77 &                   (L) &                 1.0 &                      0.362 &                   (S) \\
         28 &               \gls{var} &                 6 &             14 &        0.3 $\pm$ 0.0 &          0.1 $\pm$ 0.1 &       0.2 $\pm$ 0.1 &             3.7e-185 &            1.7e-149 &                      0.998 &                   (L) &            8.3e-111 &                      0.928 &                   (L) &                 1.0 &                      0.286 &                   (M) \\
         29 &               \gls{var} &                 6 &             14 &        0.3 $\pm$ 0.0 &          0.1 $\pm$ 0.1 &       0.1 $\pm$ 0.1 &             3.1e-189 &            2.1e-149 &                      0.996 &                   (L) &            1.8e-125 &                      0.955 &                   (L) &                 1.0 &                      0.319 &                   (M) \\
         30 &               \gls{var} &                 6 &             14 &        0.3 $\pm$ 0.0 &          0.1 $\pm$ 0.1 &       0.2 $\pm$ 0.1 &              7.5e-53 &             8.6e-38 &                      0.745 &                   (L) &             3.7e-36 &                      0.739 &                   (L) &                 1.0 &                      0.349 &                   (S) \\
         31 &               \gls{var} &                 6 &             14 &        0.3 $\pm$ 0.0 &          0.1 $\pm$ 0.1 &       0.2 $\pm$ 0.1 &             5.6e-187 &            2.3e-151 &                        1.0 &                   (L) &            1.3e-110 &                      0.927 &                   (L) &                 1.0 &                      0.278 &                   (M) \\
         32 &               \gls{var} &                 6 &             11 &        0.2 $\pm$ 0.2 &          0.2 $\pm$ 0.2 &       0.2 $\pm$ 0.2 &              1.1e-08 &             0.15011 &                       0.52 &                   (N) &             0.82369 &                      0.482 &                   (N) &                 1.0 &                      0.345 &                   (S) \\
         33 &               \gls{var} &                 6 &             12 &        0.2 $\pm$ 0.2 &          0.2 $\pm$ 0.2 &       0.2 $\pm$ 0.2 &              7.1e-07 &             0.02029 &                      0.539 &                   (N) &             0.12465 &                      0.522 &                   (N) &                 1.0 &                      0.365 &                   (S) \\
         34 &               \gls{var} &                 6 &             12 &        0.2 $\pm$ 0.2 &          0.2 $\pm$ 0.2 &       0.2 $\pm$ 0.2 &              1.2e-05 &             0.07803 &                      0.527 &                   (N) &             0.35595 &                      0.507 &                   (N) &                 1.0 &                      0.376 &                   (S) \\
         35 &               \gls{var} &                 6 &             12 &        0.2 $\pm$ 0.2 &          0.2 $\pm$ 0.2 &       0.2 $\pm$ 0.2 &              3.4e-07 &             4.4e-04 &                      0.564 &                   (N) &             0.00893 &                      0.546 &                   (N) &                 1.0 &                      0.384 &                   (S) \\
         36 &               \gls{var} &                 7 &             41 &        1.3 $\pm$ 1.1 &          0.3 $\pm$ 0.1 &       0.8 $\pm$ 0.6 &              7.2e-43 &             9.2e-28 &                      0.709 &                   (M) &             1.0e-06 &                      0.591 &                   (S) &                 1.0 &                      0.252 &                   (L) \\
         37 &               \gls{var} &                 7 &             41 &        1.2 $\pm$ 1.0 &          0.6 $\pm$ 0.6 &       1.7 $\pm$ 1.3 &              5.9e-46 &             4.4e-12 &                      0.631 &                   (S) &                 1.0 &                      0.329 &                   (M) &                 1.0 &                      0.236 &                   (L) \\
         38 &               \gls{var} &                 7 &             41 &        1.3 $\pm$ 1.1 &          0.6 $\pm$ 0.6 &       0.7 $\pm$ 0.7 &              2.5e-24 &             4.0e-20 &                      0.675 &                   (M) &             5.6e-17 &                       0.66 &                   (S) &             0.99992 &                      0.427 &                   (N) \\
         39 &               \gls{var} &                 7 &             41 &        1.4 $\pm$ 1.1 &          0.3 $\pm$ 0.1 &       0.8 $\pm$ 0.6 &              6.4e-48 &             1.4e-30 &                      0.719 &                   (M) &             1.5e-09 &                      0.614 &                   (S) &                 1.0 &                      0.241 &                   (L) \\
         40 &               \gls{var} &                 8 &             17 &        1.1 $\pm$ 1.1 &          0.4 $\pm$ 0.3 &       0.7 $\pm$ 0.6 &              2.6e-16 &             4.3e-11 &                      0.625 &                   (S) &             1.5e-04 &                      0.569 &                   (N) &                 1.0 &                       0.35 &                   (S) \\
         41 &               \gls{var} &                 8 &             17 &        1.2 $\pm$ 1.1 &          0.4 $\pm$ 0.3 &       0.7 $\pm$ 0.7 &              3.7e-22 &             1.8e-16 &                      0.657 &                   (S) &             3.4e-07 &                      0.596 &                   (S) &                 1.0 &                      0.345 &                   (S) \\
         42 &               \gls{var} &                 8 &             17 &        1.2 $\pm$ 1.2 &          0.5 $\pm$ 0.6 &       0.6 $\pm$ 0.6 &              1.8e-15 &             1.5e-12 &                      0.634 &                   (S) &             1.3e-07 &                      0.599 &                   (S) &                 1.0 &                      0.392 &                   (S) \\
         43 &               \gls{var} &                 8 &             17 &        1.2 $\pm$ 1.1 &          0.4 $\pm$ 0.3 &       0.7 $\pm$ 0.7 &              3.3e-20 &             8.0e-14 &                      0.642 &                   (S) &             3.4e-07 &                      0.596 &                   (S) &                 1.0 &                      0.343 &                   (S) \\
\bottomrule
\end{tabular}%
}
\end{table}

\begin{table}[tbp]
    \centering
    \caption{Testing runtime results for the \textbf{\gls{gs}} program category.}
    \label{tab:rq2_full_test_runtime_table_gs}
\resizebox{\columnwidth}{!}{%
\begin{tabular}{rlrr>{\columncolor{lightgray}}r>{\columncolor{lightgray}}r>{\columncolor{lightgray}}rrrrrrrrrrrrr}
& & & & & & & \textbf{DGR} & \multicolumn{3}{c}{\textbf{DG}} & \multicolumn{3}{c}{\textbf{DR}} & \multicolumn{3}{c}{\textbf{GR}} \\
\cmidrule(lr){8-9} \cmidrule(lr){9-11} \cmidrule(lr){12-14} \cmidrule(lr){15-17}
\textbf{ID} & \textbf{Category} & \textbf{\#Qubits} & \textbf{Depth} & \textbf{Default [s]} & \textbf{Greedy [s]} & \textbf{Random [s]} & \textbf{p-value} & \textbf{p-value} & $\bf{\hat{A}_{12}}$ & \textbf{Magnitude} & \textbf{p-value} & $\bf{\hat{A}_{12}}$ & \textbf{Magnitude} & \textbf{p-value} & $\bf{\hat{A}_{12}}$ & \textbf{Magnitude} \\
\midrule
          0 &                \gls{gs} &                 3 &              5 &        0.1 $\pm$ 0.0 &          0.1 $\pm$ 0.0 &       0.1 $\pm$ 0.0 &              5.5e-36 &             2.1e-19 &                      0.671 &                   (M) &             3.9e-30 &                      0.717 &                   (M) &             3.7e-11 &                      0.624 &                   (S) \\
          1 &                \gls{gs} &                 4 &              6 &        0.2 $\pm$ 0.1 &          0.1 $\pm$ 0.1 &       0.1 $\pm$ 0.1 &             5.2e-121 &             1.8e-93 &                      0.891 &                   (L) &             4.1e-91 &                      0.887 &                   (L) &             1.6e-04 &                      0.569 &                   (N) \\
          2 &                \gls{gs} &                 5 &              7 &        0.5 $\pm$ 0.2 &          0.1 $\pm$ 0.1 &       0.2 $\pm$ 0.2 &             2.4e-127 &            5.5e-100 &                      0.906 &                   (L) &             1.0e-89 &                      0.885 &                   (L) &                 1.0 &                       0.39 &                   (S) \\
          3 &                \gls{gs} &                 6 &              8 &        1.2 $\pm$ 0.3 &          0.2 $\pm$ 0.2 &       0.2 $\pm$ 0.2 &             1.3e-153 &            2.3e-116 &                       0.94 &                   (L) &            4.9e-117 &                      0.941 &                   (L) &             0.00193 &                      0.555 &                   (N) \\
          4 &                \gls{gs} &                 7 &              9 &        2.7 $\pm$ 0.7 &          0.3 $\pm$ 0.4 &       0.3 $\pm$ 0.3 &             3.5e-161 &            6.3e-121 &                      0.949 &                   (L) &            5.2e-121 &                      0.949 &                   (L) &             6.2e-08 &                      0.601 &                   (S) \\
          5 &                \gls{gs} &                 8 &             10 &        6.0 $\pm$ 1.7 &          0.4 $\pm$ 0.4 &       0.6 $\pm$ 0.7 &             1.1e-144 &            1.7e-110 &                      0.929 &                   (L) &            2.5e-110 &                      0.929 &                   (L) &             0.86791 &                      0.479 &                   (N) \\
          6 &                \gls{gs} &                 9 &             11 &       13.3 $\pm$ 3.6 &          0.8 $\pm$ 0.9 &       1.0 $\pm$ 1.0 &             3.8e-153 &            4.8e-115 &                      0.938 &                   (L) &            6.8e-115 &                      0.938 &                   (L) &                 1.0 &                      0.402 &                   (S) \\
          7 &                \gls{gs} &                10 &             12 &       29.5 $\pm$ 6.8 &          1.0 $\pm$ 1.0 &       2.0 $\pm$ 2.1 &             3.5e-162 &            4.0e-123 &                      0.953 &                   (L) &            5.2e-123 &                      0.953 &                   (L) &             0.99748 &                      0.446 &                   (N) \\
          8 &                \gls{gs} &                11 &             13 &      64.5 $\pm$ 13.3 &          2.0 $\pm$ 2.1 &       4.1 $\pm$ 4.5 &             6.1e-168 &            6.4e-128 &                      0.962 &                   (L) &            9.3e-128 &                      0.962 &                   (L) &             0.89935 &                      0.475 &                   (N) \\
          9 &                \gls{gs} &                12 &             14 &     142.3 $\pm$ 23.8 &          3.1 $\pm$ 2.4 &     10.1 $\pm$ 10.2 &             1.4e-177 &            4.1e-135 &                      0.976 &                   (L) &            5.3e-135 &                      0.976 &                   (L) &             0.87336 &                      0.478 &                   (N) \\
         10 &                \gls{gs} &                13 &             15 &     306.8 $\pm$ 57.7 &          6.4 $\pm$ 5.0 &     17.9 $\pm$ 20.2 &             3.9e-173 &            2.5e-131 &                      0.969 &                   (L) &            2.6e-131 &                      0.969 &                   (L) &             0.00555 &                      0.549 &                   (N) \\
         11 &                \gls{gs} &                14 &             16 &    682.3 $\pm$ 110.4 &        13.8 $\pm$ 10.6 &     48.6 $\pm$ 46.4 &             8.5e-210 &            2.8e-136 &                      0.978 &                   (L) &            3.1e-136 &                      0.978 &                   (L) &                 1.0 &                      0.157 &                   (L) \\
         12 &                \gls{gs} &                15 &             17 &   1587.0 $\pm$ 303.5 &        25.7 $\pm$ 12.2 &   120.8 $\pm$ 101.5 &             3.5e-208 &            1.6e-132 &                      0.971 &                   (L) &            1.6e-132 &                      0.971 &                   (L) &                 1.0 &                      0.138 &                   (L) \\
         13 &                \gls{gs} &                16 &             18 &  3670.9 $\pm$ 1059.6 &        54.2 $\pm$ 24.4 &   285.9 $\pm$ 215.1 &             2.9e-202 &            4.6e-124 &                      0.956 &                   (L) &            4.7e-124 &                      0.956 &                   (L) &                 1.0 &                      0.111 &                   (L) \\
\bottomrule
\end{tabular}%
}
\end{table}

In \cref{tab:rq2_full_test_runtime_table_grov,tab:rq2_full_test_runtime_table_qwalk,tab:rq2_full_test_runtime_table_var,tab:rq2_full_test_runtime_table_gs}, we depict the testing runtimes for RQ2 by program category and qubit count.
The gray columns \textbf{Default}, \textbf{Greedy} and \textbf{Random} show the testing runtimes in seconds for the respective approach.
In the \textbf{DGR} column, we show the p-value resulting from a Kruskal-Wallis test between the three approaches where \textbf{D} refers to Default, \textbf{G} to Greedy and \textbf{R} to Random.
Following this, in the \textbf{DG}, \textbf{DR} and \textbf{GR} columns, we show the pairwise statistical test results, consisting of the p-value, effect size and effect magnitude category for the comparisons between the respective approaches.

\subsection{RQ3 Tables}

\begin{table}[tbp]
    \centering
\caption{Mutation scores for the \textbf{\gls{grov}} program category. Table 1/3.}
\label{tab:rq4_full_mutation_score_grov_p1}
\resizebox{\columnwidth}{!}{%
\begin{tabular}{rllrr>{\columncolor{lightgray}}r>{\columncolor{lightgray}}r>{\columncolor{lightgray}}rrrrlrllrll}
\toprule
& & & & & & & & \textbf{DGR} & \multicolumn{3}{c}{\textbf{DG}} & \multicolumn{3}{c}{\textbf{DR}} & \multicolumn{3}{c}{\textbf{GR}} \\
\cmidrule(lr){9-9} \cmidrule(lr){10-12} \cmidrule(lr){13-15} \cmidrule(lr){16-18}
\textbf{ID} & Mutant Type & \textbf{Category} & \textbf{\#Qubits} & \textbf{Depth} & \textbf{Default [\%]} & \textbf{Greedy [\%]} & \textbf{Random [\%]} & \textbf{p-value}  & \textbf{p-value}  & $\bf{\hat{A}_{12}}$  & \textbf{Magnitude}  & \textbf{p-value}  & $\bf{\hat{A}_{12}}$  & \textbf{Magnitude}  & \textbf{p-value}  & $\bf{\hat{A}_{12}}$  & \textbf{Magnitude}  \\
\midrule
          0 &         all &              \gls{grov} &                 6 &            178 &        71.6 $\pm$ 6.3 &          78.7 $\pm$ 5.9 &       73.1 $\pm$ 8.7 &              6.7e-04 &             0.99994 &                      0.222 &                   (L) &             0.81456 &                      0.436 &                   (N) &             0.00584 &                      0.683 &                   (M) \\
          0 &           X &              \gls{grov} &                 6 &            178 &       100.0 $\pm$ 0.0 &         56.7 $\pm$ 21.7 &      54.4 $\pm$ 28.3 &              1.5e-12 &             3.5e-12 &                      0.967 &                   (L) &             6.4e-11 &                      0.933 &                   (L) &             0.39383 &                      0.519 &                   (N) \\
          0 &           Z &              \gls{grov} &                 6 &            178 &         0.0 $\pm$ 0.0 &         48.9 $\pm$ 21.0 &      34.4 $\pm$ 23.9 &              1.3e-12 &                 1.0 &                      0.017 &                   (L) &                 1.0 &                      0.117 &                   (L) &              0.0123 &                      0.654 &                   (S) \\
\bottomrule
\end{tabular}
}
\end{table}
\begin{table}[tbp]
    \centering
\caption{Mutation scores for the \textbf{\gls{grov}} program category. Table 2/3.}
\label{tab:rq4_full_mutation_score_grov_p2}
\resizebox{\columnwidth}{!}{%
\begin{tabular}{rllrr>{\columncolor{lightgray}}r>{\columncolor{lightgray}}r>{\columncolor{lightgray}}rrrrlrllrll}
\toprule
& & & & & & & & \textbf{DGR} & \multicolumn{3}{c}{\textbf{DG}} & \multicolumn{3}{c}{\textbf{DR}} & \multicolumn{3}{c}{\textbf{GR}} \\
\cmidrule(lr){9-9} \cmidrule(lr){10-12} \cmidrule(lr){13-15} \cmidrule(lr){16-18}
\textbf{ID} & Mutant Type & \textbf{Category} & \textbf{\#Qubits} & \textbf{Depth} & \textbf{Default [\%]} & \textbf{Greedy [\%]} & \textbf{Random [\%]} & \textbf{p-value}  & \textbf{p-value}  & $\bf{\hat{A}_{12}}$  & \textbf{Magnitude}  & \textbf{p-value}  & $\bf{\hat{A}_{12}}$  & \textbf{Magnitude}  & \textbf{p-value}  & $\bf{\hat{A}_{12}}$  & \textbf{Magnitude}  \\
\midrule
         21 &         all &              \gls{grov} &                 7 &            156 &        69.3 $\pm$ 7.1 &         78.7 $\pm$ 10.3 &       75.3 $\pm$ 8.4 &              4.7e-04 &             0.99985 &                      0.233 &                   (L) &             0.99733 &                      0.297 &                   (M) &             0.06618 &                      0.611 &                   (S) \\
         21 &           X &              \gls{grov} &                 7 &            156 &       100.0 $\pm$ 0.0 &         57.8 $\pm$ 31.5 &      70.0 $\pm$ 26.8 &              4.7e-09 &             2.3e-09 &                      0.883 &                   (L) &             5.0e-08 &                      0.833 &                   (L) &             0.94031 &                       0.39 &                   (S) \\
         21 &           Z &              \gls{grov} &                 7 &            156 &        3.3 $\pm$ 10.2 &         50.0 $\pm$ 28.7 &      23.3 $\pm$ 23.4 &              4.2e-09 &                 1.0 &                      0.103 &                   (L) &             0.99995 &                       0.26 &                   (L) &             2.1e-04 &                      0.752 &                   (L) \\
\bottomrule
\end{tabular}
}
\end{table}
\begin{table}[tbp]
    \centering
\caption{Mutation scores for the \textbf{\gls{grov}} program category. Table 3/3.}
\label{tab:rq4_full_mutation_score_grov_p3}
\resizebox{\columnwidth}{!}{%
\begin{tabular}{rllrr>{\columncolor{lightgray}}r>{\columncolor{lightgray}}r>{\columncolor{lightgray}}rrrrlrllrll}
\toprule
& & & & & & & & \textbf{DGR} & \multicolumn{3}{c}{\textbf{DG}} & \multicolumn{3}{c}{\textbf{DR}} & \multicolumn{3}{c}{\textbf{GR}} \\
\cmidrule(lr){9-9} \cmidrule(lr){10-12} \cmidrule(lr){13-15} \cmidrule(lr){16-18}
\textbf{ID} & Mutant Type & \textbf{Category} & \textbf{\#Qubits} & \textbf{Depth} & \textbf{Default [\%]} & \textbf{Greedy [\%]} & \textbf{Random [\%]} & \textbf{p-value}  & \textbf{p-value}  & $\bf{\hat{A}_{12}}$  & \textbf{Magnitude}  & \textbf{p-value}  & $\bf{\hat{A}_{12}}$  & \textbf{Magnitude}  & \textbf{p-value}  & $\bf{\hat{A}_{12}}$  & \textbf{Magnitude}  \\
\midrule
         42 &         all &              \gls{grov} &                 9 &            504 &        79.3 $\pm$ 2.7 &          78.0 $\pm$ 9.8 &       76.4 $\pm$ 9.1 &               0.1657 &             0.20394 &                      0.558 &                   (N) &             0.01493 &                      0.649 &                   (S) &             0.25439 &                      0.549 &                   (N) \\
         42 &           X &              \gls{grov} &                 9 &            504 &       100.0 $\pm$ 0.0 &         57.8 $\pm$ 31.5 &      63.3 $\pm$ 33.2 &              1.9e-08 &             2.3e-09 &                      0.883 &                   (L) &             1.5e-07 &                      0.817 &                   (L) &             0.74449 &                      0.453 &                   (N) \\
         42 &           Z &              \gls{grov} &                 9 &            504 &         1.1 $\pm$ 6.1 &         35.6 $\pm$ 26.2 &      26.7 $\pm$ 25.4 &              6.6e-08 &                 1.0 &                      0.144 &                   (L) &                 1.0 &                      0.198 &                   (L) &             0.07539 &                      0.601 &                   (S) \\
\bottomrule
\end{tabular}
}
\end{table}

\begin{table}[tbp]
    \centering
\caption{Mutation scores for the \textbf{\gls{qwalk}} program category. Table 1/2.}
\label{tab:rq4_full_mutation_score_qwalk_p1}
\resizebox{\columnwidth}{!}{%
\begin{tabular}{rllrr>{\columncolor{lightgray}}r>{\columncolor{lightgray}}r>{\columncolor{lightgray}}rrrrlrllrll}
\toprule
& & & & & & & & \textbf{DGR} & \multicolumn{3}{c}{\textbf{DG}} & \multicolumn{3}{c}{\textbf{DR}} & \multicolumn{3}{c}{\textbf{GR}} \\
\cmidrule(lr){9-9} \cmidrule(lr){10-12} \cmidrule(lr){13-15} \cmidrule(lr){16-18}
\textbf{ID} & Mutant Type & \textbf{Category} & \textbf{\#Qubits} & \textbf{Depth} & \textbf{Default [\%]} & \textbf{Greedy [\%]} & \textbf{Random [\%]} & \textbf{p-value}  & \textbf{p-value}  & $\bf{\hat{A}_{12}}$  & \textbf{Magnitude}  & \textbf{p-value}  & $\bf{\hat{A}_{12}}$  & \textbf{Magnitude}  & \textbf{p-value}  & $\bf{\hat{A}_{12}}$  & \textbf{Magnitude}  \\
\midrule
          0 &         all &             \gls{qwalk} &                 3 &             14 &       26.7 $\pm$ 10.1 &          72.0 $\pm$ 7.9 &       71.8 $\pm$ 9.5 &              7.9e-14 &                 1.0 &                        0.0 &                   (L) &                 1.0 &                        0.0 &                   (L) &             0.45469 &                      0.509 &                   (N) \\
          0 &           X &             \gls{qwalk} &                 3 &             14 &       32.2 $\pm$ 28.3 &         71.1 $\pm$ 25.9 &      61.1 $\pm$ 26.4 &              6.8e-06 &                 1.0 &                      0.175 &                   (L) &             0.99978 &                       0.25 &                   (L) &              0.0608 &                      0.608 &                   (S) \\
          0 &           Z &             \gls{qwalk} &                 3 &             14 &        3.3 $\pm$ 10.2 &         28.9 $\pm$ 21.0 &      33.3 $\pm$ 29.0 &              4.8e-07 &                 1.0 &                      0.177 &                   (L) &                 1.0 &                      0.188 &                   (L) &             0.65521 &                      0.473 &                   (N) \\
\bottomrule
\end{tabular}
}
\end{table}
\begin{table}[tbp]
    \centering
\caption{Mutation scores for the \textbf{\gls{qwalk}} program category. Table 2/2.}
\label{tab:rq4_full_mutation_score_qwalk_p2}
\resizebox{\columnwidth}{!}{%
\begin{tabular}{rllrr>{\columncolor{lightgray}}r>{\columncolor{lightgray}}r>{\columncolor{lightgray}}rrrrlrllrll}
\toprule
& & & & & & & & \textbf{DGR} & \multicolumn{3}{c}{\textbf{DG}} & \multicolumn{3}{c}{\textbf{DR}} & \multicolumn{3}{c}{\textbf{GR}} \\
\cmidrule(lr){9-9} \cmidrule(lr){10-12} \cmidrule(lr){13-15} \cmidrule(lr){16-18}
\textbf{ID} & Mutant Type & \textbf{Category} & \textbf{\#Qubits} & \textbf{Depth} & \textbf{Default [\%]} & \textbf{Greedy [\%]} & \textbf{Random [\%]} & \textbf{p-value}  & \textbf{p-value}  & $\bf{\hat{A}_{12}}$  & \textbf{Magnitude}  & \textbf{p-value}  & $\bf{\hat{A}_{12}}$  & \textbf{Magnitude}  & \textbf{p-value}  & $\bf{\hat{A}_{12}}$  & \textbf{Magnitude}  \\
\midrule
         20 &         all &             \gls{qwalk} &                 4 &             81 &        64.9 $\pm$ 6.8 &          74.9 $\pm$ 9.5 &       75.6 $\pm$ 8.3 &              4.1e-06 &             0.99997 &                      0.204 &                   (L) &                 1.0 &                      0.163 &                   (L) &             0.50607 &                      0.499 &                   (N) \\
         20 &           X &             \gls{qwalk} &                 4 &             81 &       100.0 $\pm$ 0.0 &         76.7 $\pm$ 26.5 &      76.7 $\pm$ 26.5 &              7.3e-06 &             2.3e-06 &                      0.767 &                   (L) &             2.3e-06 &                      0.767 &                   (L) &             0.50323 &                        0.5 &                   (N) \\
         20 &           Z &             \gls{qwalk} &                 4 &             81 &         2.2 $\pm$ 8.5 &         31.1 $\pm$ 21.3 &      31.1 $\pm$ 21.3 &              1.2e-08 &                 1.0 &                      0.144 &                   (L) &                 1.0 &                      0.144 &                   (L) &             0.50337 &                        0.5 &                   (N) \\
\bottomrule
\end{tabular}
}
\end{table}

\begin{table}[tbp]
    \centering
\caption{Mutation scores for the \textbf{\gls{var}} program category. Table 1/3.}
\label{tab:rq4_full_mutation_score_var_p1}
\resizebox{\columnwidth}{!}{%
\begin{tabular}{rllrr>{\columncolor{lightgray}}r>{\columncolor{lightgray}}r>{\columncolor{lightgray}}rrrrlrllrll}
\toprule
& & & & & & & & \textbf{DGR} & \multicolumn{3}{c}{\textbf{DG}} & \multicolumn{3}{c}{\textbf{DR}} & \multicolumn{3}{c}{\textbf{GR}} \\
\cmidrule(lr){9-9} \cmidrule(lr){10-12} \cmidrule(lr){13-15} \cmidrule(lr){16-18}
\textbf{ID} & Mutant Type & \textbf{Category} & \textbf{\#Qubits} & \textbf{Depth} & \textbf{Default [\%]} & \textbf{Greedy [\%]} & \textbf{Random [\%]} & \textbf{p-value}  & \textbf{p-value}  & $\bf{\hat{A}_{12}}$  & \textbf{Magnitude}  & \textbf{p-value}  & $\bf{\hat{A}_{12}}$  & \textbf{Magnitude}  & \textbf{p-value}  & $\bf{\hat{A}_{12}}$  & \textbf{Magnitude}  \\
\midrule
          0 &         all &               \gls{var} &                 2 &              6 &         8.2 $\pm$ 6.9 &         71.8 $\pm$ 11.0 &      72.7 $\pm$ 11.3 &              8.5e-14 &                 1.0 &                        0.0 &                   (L) &                 1.0 &                        0.0 &                   (L) &             0.56575 &                      0.488 &                   (N) \\
          0 &           X &               \gls{var} &                 2 &              6 &         1.1 $\pm$ 6.1 &         54.4 $\pm$ 29.7 &      55.6 $\pm$ 29.5 &              2.3e-11 &                 1.0 &                      0.073 &                   (L) &                 1.0 &                      0.072 &                   (L) &             0.57327 &                      0.488 &                   (N) \\
          0 &           Z &               \gls{var} &                 2 &              6 &         0.0 $\pm$ 0.0 &         53.3 $\pm$ 33.4 &      53.3 $\pm$ 33.4 &              8.8e-11 &                 1.0 &                      0.083 &                   (L) &                 1.0 &                      0.083 &                   (L) &             0.50308 &                        0.5 &                   (N) \\
\bottomrule
\end{tabular}
}
\end{table}
\begin{table}[tbp]
    \centering
\caption{Mutation scores for the \textbf{\gls{var}} program category. Table 2/3.}
\label{tab:rq4_full_mutation_score_var_p2}
\resizebox{\columnwidth}{!}{%
\begin{tabular}{rllrr>{\columncolor{lightgray}}r>{\columncolor{lightgray}}r>{\columncolor{lightgray}}rrrrlrllrll}
\toprule
& & & & & & & & \textbf{DGR} & \multicolumn{3}{c}{\textbf{DG}} & \multicolumn{3}{c}{\textbf{DR}} & \multicolumn{3}{c}{\textbf{GR}} \\
\cmidrule(lr){9-9} \cmidrule(lr){10-12} \cmidrule(lr){13-15} \cmidrule(lr){16-18}
\textbf{ID} & Mutant Type & \textbf{Category} & \textbf{\#Qubits} & \textbf{Depth} & \textbf{Default [\%]} & \textbf{Greedy [\%]} & \textbf{Random [\%]} & \textbf{p-value}  & \textbf{p-value}  & $\bf{\hat{A}_{12}}$  & \textbf{Magnitude}  & \textbf{p-value}  & $\bf{\hat{A}_{12}}$  & \textbf{Magnitude}  & \textbf{p-value}  & $\bf{\hat{A}_{12}}$  & \textbf{Magnitude}  \\
\midrule
         21 &         all &               \gls{var} &                 5 &             11 &        56.2 $\pm$ 8.9 &          73.3 $\pm$ 9.9 &       75.8 $\pm$ 6.9 &              3.4e-11 &                 1.0 &                      0.102 &                   (L) &                 1.0 &                       0.02 &                   (L) &             0.76633 &                      0.448 &                   (N) \\
         21 &           X &               \gls{var} &                 5 &             11 &       38.9 $\pm$ 29.1 &         76.7 $\pm$ 26.5 &      84.4 $\pm$ 19.0 &              4.1e-08 &                 1.0 &                       0.18 &                   (L) &                 1.0 &                      0.115 &                   (L) &             0.85525 &                      0.429 &                   (N) \\
         21 &           Z &               \gls{var} &                 5 &             11 &         0.0 $\pm$ 0.0 &         24.4 $\pm$ 26.2 &      13.3 $\pm$ 22.5 &              2.3e-05 &                 1.0 &                      0.233 &                   (L) &             0.99971 &                      0.333 &                   (M) &             0.03515 &                      0.621 &                   (S) \\
\bottomrule
\end{tabular}
}
\end{table}
\begin{table}[tbp]
    \centering
\caption{Mutation scores for the \textbf{\gls{var}} program category. Table 3/3.}
\label{tab:rq4_full_mutation_score_var_p3}
\resizebox{\columnwidth}{!}{%
\begin{tabular}{rllrr>{\columncolor{lightgray}}r>{\columncolor{lightgray}}r>{\columncolor{lightgray}}rrrrlrllrll}
\toprule
& & & & & & & & \textbf{DGR} & \multicolumn{3}{c}{\textbf{DG}} & \multicolumn{3}{c}{\textbf{DR}} & \multicolumn{3}{c}{\textbf{GR}} \\
\cmidrule(lr){9-9} \cmidrule(lr){10-12} \cmidrule(lr){13-15} \cmidrule(lr){16-18}
\textbf{ID} & Mutant Type & \textbf{Category} & \textbf{\#Qubits} & \textbf{Depth} & \textbf{Default [\%]} & \textbf{Greedy [\%]} & \textbf{Random [\%]} & \textbf{p-value}  & \textbf{p-value}  & $\bf{\hat{A}_{12}}$  & \textbf{Magnitude}  & \textbf{p-value}  & $\bf{\hat{A}_{12}}$  & \textbf{Magnitude}  & \textbf{p-value}  & $\bf{\hat{A}_{12}}$  & \textbf{Magnitude}  \\
\midrule
         41 &         all &               \gls{var} &                 8 &             17 &        67.6 $\pm$ 7.4 &          72.2 $\pm$ 8.6 &      69.3 $\pm$ 10.4 &               0.0626 &             0.98901 &                      0.334 &                   (M) &             0.94113 &                      0.387 &                   (S) &             0.20841 &                      0.559 &                   (N) \\
         41 &           X &               \gls{var} &                 8 &             17 &       48.9 $\pm$ 33.6 &         72.2 $\pm$ 27.8 &      58.9 $\pm$ 29.9 &              0.02108 &             0.99647 &                      0.307 &                   (M) &             0.89221 &                      0.412 &                   (S) &             0.04686 &                       0.62 &                   (S) \\
         41 &           Z &               \gls{var} &                 8 &             17 &        3.3 $\pm$ 10.2 &         23.3 $\pm$ 21.7 &      24.4 $\pm$ 21.3 &              2.4e-05 &             0.99998 &                      0.245 &                   (L) &             0.99999 &                      0.228 &                   (L) &             0.59122 &                      0.485 &                   (N) \\
\bottomrule
\end{tabular}
}
\end{table}

\begin{table}[tbp]
    \centering
\caption{Mutation scores for the \textbf{\gls{gs}} program category.}
\label{tab:rq4_full_mutation_score_gs}
\resizebox{\columnwidth}{!}{%
\begin{tabular}{rllrr>{\columncolor{lightgray}}r>{\columncolor{lightgray}}r>{\columncolor{lightgray}}rrrrlrllrll}
\toprule
& & & & & & & & \textbf{DGR} & \multicolumn{3}{c}{\textbf{DG}} & \multicolumn{3}{c}{\textbf{DR}} & \multicolumn{3}{c}{\textbf{GR}} \\
\cmidrule(lr){9-9} \cmidrule(lr){10-12} \cmidrule(lr){13-15} \cmidrule(lr){16-18}
\textbf{ID} & Mutant Type & \textbf{Category} & \textbf{\#Qubits} & \textbf{Depth} & \textbf{Default [\%]} & \textbf{Greedy [\%]} & \textbf{Random [\%]} & \textbf{p-value}  & \textbf{p-value}  & $\bf{\hat{A}_{12}}$  & \textbf{Magnitude}  & \textbf{p-value}  & $\bf{\hat{A}_{12}}$  & \textbf{Magnitude}  & \textbf{p-value}  & $\bf{\hat{A}_{12}}$  & \textbf{Magnitude}  \\
\midrule
          0 &         all &                \gls{gs} &                 3 &              5 &       44.9 $\pm$ 10.3 &         67.3 $\pm$ 12.8 &      68.9 $\pm$ 13.3 &              8.5e-10 &                 1.0 &                      0.089 &                   (L) &                 1.0 &                      0.081 &                   (L) &             0.67941 &                      0.466 &                   (N) \\
          0 &           X &                \gls{gs} &                 3 &              5 &       67.8 $\pm$ 32.1 &         67.8 $\pm$ 32.1 &      67.8 $\pm$ 32.1 &                  1.0 &             0.50315 &                        0.5 &                   (N) &             0.50315 &                        0.5 &                   (N) &             0.50315 &                        0.5 &                   (N) \\
          0 &           Z &                \gls{gs} &                 3 &              5 &         2.2 $\pm$ 8.5 &         30.0 $\pm$ 26.8 &      30.0 $\pm$ 26.8 &              3.4e-06 &                 1.0 &                      0.208 &                   (L) &                 1.0 &                      0.208 &                   (L) &             0.50314 &                        0.5 &                   (N) \\
\bottomrule
\end{tabular}
}
\end{table}

In \cref{tab:rq4_full_mutation_score_grov_p1,tab:rq4_full_mutation_score_grov_p2,tab:rq4_full_mutation_score_grov_p3,tab:rq4_full_mutation_score_qwalk_p1,tab:rq4_full_mutation_score_qwalk_p2,tab:rq4_full_mutation_score_var_p1,tab:rq4_full_mutation_score_var_p2,tab:rq4_full_mutation_score_var_p3,tab:rq4_full_mutation_score_gs}, we show the mutation score results for RQ3 by mutant, category, and qubit count.
In the gray columns \textbf{Default}, \textbf{Greedy} and \textbf{Random}, we list the mutation scores and to the right of the gray columns we provide the statistical test results as we described for the RQ2 tables.

\FloatBarrier

\bibliographystyle{ACM-Reference-Format}
\bibliography{biblio}